 \documentclass[useAMS,usegraphicx, usenatbib]{mn2e}
\usepackage{aas_macros}

\usepackage{epsfig}
\usepackage{verbatim, amsmath, amsfonts, amssymb,amssymb}
\usepackage[utf8x]{inputenc}
\usepackage[table]{xcolor}
\usepackage{booktabs, longtable}
\usepackage{multirow}
\usepackage{float}
\usepackage{lscape}
\usepackage{graphicx, subfigure}
\usepackage{rotating}

\begin{document}
\title{Kernel PCA for type Ia supernovae photometric classification}

\author[E. E. O. Ishida and R. S. de Souza]
{E. E. O. Ishida$^{1,2}\thanks{e-mail: emilleishida@usp.br (EEOI)}$ and 
R. S. de Souza  $^{3,1,2}$
\\
$^{1}$IAG, Universidade de S\~ao Paulo, Rua do Mat\~ao 1226, Cidade Universit\'aria,
CEP 05508-900, S\~ao Paulo, SP, Brazil\\
$^{2}$Max-Planck-Institut f\"ur Astrophysik, Karl-Schwarzschild-Str. 1, D-85748 Garching, Germany\\
$^{3}$Korea Astronomy and Space Science Institute, Daejeon, 305-348, Republic of Korea 
}

 \date{Accepted -- Received  --}

\pagerange{\pageref{firstpage}--\pageref{lastpage}} \pubyear{2010}

\maketitle
\label{firstpage}

\begin{abstract}
The problem of supernova photometric identification will be extremely important for large surveys in the next decade. In this work, we propose the use of Kernel Principal Component Analysis (KPCA) combined with k = 1 nearest neighbour algorithm (1NN) as a framework
for supernovae (SNe) photometric classification. The method does not rely on information about redshift or local environmental variables, so it is less sensitive to bias than its template fitting counterparts. The classification is entirely based on information within the spectroscopic confirmed sample and each new light curve is classified one at a time. This allows us to update the principal component (PC) parameter space if a new spectroscopic light curve is available while also avoids the need of re-determining it for each individual new classification. We applied the method to different instances of the \textit{Supernova Photometric Classification Challenge} (SNPCC) data set. Our method provide good purity  results in all data sample analysed, when SNR$\geq$5. As a consequence, we can state that if a sample as the post-SNPCC  was available today, we would be able to classify $\approx 15\%$ of the initial data set with purity $\gtrsim$ 90\% (D$_{7}$+SNR3). Results from the original SNPCC sample, reported as a function of redshift, show that our method provides high purity (up to $\approx 97\%$), specially in the range of $0.2\leq z < 0.4$, when compared to results from the SNPCC, while maintaining a moderate figure of merit ($\approx 0.25$). This makes our algorithm ideal for a first approach to an unlabelled  data set or to be used as a complement in increasing the training sample for other algorithms. We also present results for  SNe photometric classification using only pre-maximum epochs, obtaining  63\% purity and 77\% successful classification rates (SNR$\geq$5). In a tougher scenario, considering only SNe with MLCS2k2 fit probability $>$0.1, we demonstrate that KPCA+1NN is able to improve  the classification results up to $>95\%$ (SNR$\geq$3) purity without the need of redshift information. Results are sensitive to the information contained in each light curve, as a consequence, higher quality data points lead to  higher successful classification rates. The method is flexible enough to be applied to other astrophysical transients, as long as a training and a test sample are provided. 
\end{abstract}

\section{Introduction}

Since its discovery  \citep{riess1998,perlmutter1999}, dark energy (DE) has become a big challenge  in theoretical physics and cosmology. 
In order to improve our understanding  about its nature, multiple  observations are used to add better constraints over DE characteristics \citep[e.g.,][]{mantz2010,blake2011,plionis2011}. In special,  large samples of type Ia supernovae (SNe Ia) are being used to measure luminosity distances as a function of redshift in order to constraint cosmological parameters \citep[e.g.,][]{kessler2009c,ishida2011,Benitez2011, conley2011}.  As part of the efforts towards understanding DE, we expect many thousands of SNe candidates  from large photometric surveys, such as the \textit{Large Synoptic Survey Telescope} (LSST) \citep{lsst},  SkyMapper \citep{skymapper} and the \textit{Dark Energy Survey} (DES) \citep{des}. However, with rapidly increasing available data, it is already  impracticable to provide spectroscopical confirmation for all potential SNe Ia discovered in large field imaging surveys. After a great effort in allocating their resources for spectroscopic follow-up, the \textit{SuperNova Legacy Survey} (SNLS)\citep{Astier2006} and the \textit{Sloan Digital Sky Survey} (SDSS)\citep{York2000}, were able to provide confirmation for almost half of their light-curves. These constitute the major SNe Ia samples currently available, but it is very unlikely that their power of spectroscopic follow-up will continue to increase as it did in the last decade \citep{Kessler2010b}. In this context, we do not have much choice left other than develop (or adapt) statistical and computational tools which allow us to perform classification on photometric data alone. Beyond that, such tools should ideally provide a quick and flexible framework, where information from new data may be smoothly added in the pipeline.

Trying to solve this puzzle, in the recent years a good diversity of techniques were applied to the problem of  SNe photometric classification \citep{Poznanski2002,Johnson2006,Sullivan2006,Poznanski2007,kuznetsova2007,Kunz2007,Sako2008,Rodney2009,Gong2010,Falck2010}. 
Most of them use the idea of  template fitting, so  the classification is estimated by comparison between the unlabelled SN  and a  set of confirmed light curve templates. The method starts with the hypothesis that the new, unlabelled, light curve belongs to one of the categories in the template sample. The procedure then continues to determine which category best resembles the characteristic of this new object. It produced good results \citep{Sako2008}, but its final classification rates are highly sensitive to the characteristics of the template sample. 

To overcome such difficulty, \citet{Newling2011,Sako2011} describe different techniques which address a posterior probability to each classification output.  These algorithms produce not a specific type for each SN, but a probability of  belonging to each one of the template classes. Such an improvement allow the user to impose selection cuts on posterior probability and, for example,  use for cosmology only those SNe with a high probability of being Ia.

Another interesting approach proposed by \citet{Kunz2007}, and further developed by \citet{Newling2011b}, takes a somewhat different path. Instead of separating between Ia and non-Ia before the cosmological analysis, they use all the available data. However, the influence of each data point in determining the cosmological parameters is weighted according to their posterior probability (obtained from some classifier like that of \citet{Sako2011}, for example). The method was able to identify the fiducial cosmological parameters in a simulated data set,  although some bias still remains and worth further investigation.

Following a different line of thought, \citet{Richards2011} (hereafter R2012) proposes the use of diffusion maps to translate each light curve into a  low dimensional parameter space. Such space is constructed using the entire sample and, after a suitable representation is found, the label of the spectroscopic sample is revealed. In the final step a random forest classification algorithm is used to assign a label to the photometric light curves, based on their low dimensional distribution when compared to the one from the spectroscopically confirmed SNe.  Results were comparable to  template fitting methods in a simulated data set, but it also showed  large sensitivity to the representativeness between training and test samples.

More recently, \citet{karpenka2012} presented a two-step algorithm where each light curve in the spectroscopic sample is first fitted to a parametric function. The values of parameters found are subsequently used in training a neural network (NN) algorithm. The NN is then applied to the photometric sample and, for each light curve, it returns the probability of being a Ia. Their classification results are overall not depending on redshift distribution and, as other analysis cited before, can be vary significantly depending on the training sample used.

In order to better understand and compare the state of art of photometric classification techniques, \citet{Kessler2010} released the \textit{SuperNova Photometric Classification Challenge} (hereafter, SNPCC). It consisted of a blind sample of $\sim$20.000 SNe light curves, generated using the \textit{SuperNova ANAlysis}\footnote{http://sdssdp62.fnal.gov/sdsssn/SNANA-PUBLIC/} (SNANA) light curve simulator \citep{Kessler2009b}, and designed to mimic data from the DES. Approximately 1000 of these were given with labels, so to represent a spectroscopically confirmed sub-sample. The participants were offered 2 instances of the data, with and without the host galaxy photometric redshift (photo-\textit{z}). Around a dozen entries were submitted to the Challenge and, although none of them obtained an outstanding result when compared to others, it provided a clear picture of what can be done currently and what we should require from future surveys in order to improve photometric classifications. There was also an instance of the data containing only observations before maximum, which aimed at choosing potential spectroscopic follow-up candidates. However, this data set did not received replies from the participants \citep{Kessler2010b}. After the Challenge, the organizers released an updated version of the data, including all labels, bug fixes and other improvements found necessary during the competition\footnote{http://sdssdp62.fnal.gov/sdsssn/SIMGEN\_PUBLIC/}. The works of \citet{Newling2011}, R2012 and \citet{karpenka2012} present detailed results from applying their algorithm to this post-SNPCC\footnote{Nomenclature taken from \citet{Newling2011}.} data.

Given the stimulating activity in the field of SNe photometric classification, and the urgency with which the problem imposes itself, our purpose here is to present an alternative method which \textit{optimizes purity} in the final SNe Ia sample, in order to provide a statistically significant number of photometrically classified SNe Ia for cosmological analysis. Our algorithm uses a machine learning approach, similar in philosophy to the entry of R2012 submitted to the SNPCC. This class of statistical tools has already been  applied to a variety of astronomical topics (for a recent review see \citet{Ball2010}). 
 
We propose the use of \textit{Kernel Principal Component Analysis} (hereafter, KPCA) as a tool to find a suitable low dimension representation of SNe light curves. In constructing this low dimensional space only the spectroscopically confirmed  sample is used.  Each unlabelled light curve is then projected into this space one at a time and a \textit{k-nearest neighbour} (kNN) algorithm  performs the classification. The procedure was applied to the post-SNPCC data set using the entire light curves and also using only pre-maximum observations.  In order to allow a more direct comparison with SNPCC results, we also applied the  algorithm to the complete light curves in the original SNPCC data set\footnote{http://www.hep.anl.gov/SNchallenge/ DES\_BLINDnoHOSTZ.tar.gz}.

Our procedure returns purity levels higher than to top ranked methods reported in the SNPCC. 
The results are sensitive to the spectroscopic sample, but more on the quality of each individual observation than on representativeness between spectroscopic and photometric samples. Assuming that results can only be as good as the input data, we perform classification in sub-samples of SNPCC and post-SNPCC data based on signal to noise ratio (SNR) levels. 

 Considering only light curves with \textit{Multi-color Light Curve Shape} (MLCS2k2) \citep{Jha2007} fit probability, FitProb$>$0.1, we demonstrate that our method is capable of increasing purity and successful classification rates even in a context with only light curves very similar between each other.

The paper is organized as follows: section \ref{sec:PCA} briefly describe linear PCA and its transition to the KPCA formalism. In section \ref{sec:class} we detailed the cross-validation and kNN algorithm used for classification. In section \ref{sec:LC_prep} we present the guidelines to prepare the raw light curve data into a data vector suitable for KPCA. The results applied to a best case scenario simulation, to the post-SNPCC data and comparison with MLCS2k2 fit probability results are shown in section \ref{sec:application}. We report outcomes from applying our method to the original SNPCC data set in section \ref{sec:SNPCC}. Finally, we discuss the results  and future perspectives in section \ref{sec:discuss}. Throughout the text, mainly in section \ref{sec:PCA}, we refer to a few theorems and mathematical statements are made. Those which are most crucial for the development of the KPCA argument are briefly demonstrated in appendix \ref{ap:proof}. A detailed description of results achieve using linear PCA+kNN algorithm is presented in appendix \ref{ap:linearPCA}. Appendix \ref{ap:zcad1SNR5} shows classification rates as a function of redshift and SNR cuts and appendix \ref{ap:d8} displays our achievements when no SNR cuts are applied.  Graphical representation of results from SNPCC data set for all the tests we performed, which can be directly compared to those of \citet{Kessler2010b} are displayed in appendix \ref{ap:SNPCC_comp}. Complete summary tables reporting the number of data points in  different sub-samples of SNPCC and post-SNPCC data and classification results mentioned in the text are shown in appendix {\ref{ap:tables}.
  

\section{Principal Component Analysis}
\label{sec:PCA}

The main goal of PCA is to reduce an initial large number of variables to a smaller set of uncorrelated ones, called \textit{Principal Components} (PCs).  This set of PCs is capable of reproducing as much variance from the original variables as possible. Each of them can be viewed as a composite variable summarizing the original ones, and its eigenvalue indicates how successful this summary is. 
If all variables are highly correlated, one single PC is sufficient to describe the data. If the variables form two or more sets, and correlations are high within sets and low between sets, a second or third PC is needed to summarize the initial variables. PCA solutions with more than one PC are referred to as multi-dimensional solutions. In such cases, the PCs are ordered according to their eigenvalues. The first component is associated with the largest eigenvalue, and accounts for most of the variance, the second accounts for as much as possible of the remaining variance, and so on.

There are a few different ways which lead to the determination of PCs. Particularly, we have already shown that it is possible to derive the PCs beginning from a theoretical description of the likelihood function  \citep[e.g.,][]{ishida2011,ishida2011b}.  

In the present work we are interested in exploring the KPCA and, as a consequence, our description shall be based on dot products. In doing so, the connection between PCA and KPCA occurs almost smoothly. We follow closely \citet{Hofmann2008} and Max Welling's notes \textit{A first encounter with Machine Learning}\footnote{http://www.ics.uci.edu/$\sim$welling/teaching/ICS273Afall11                    /IntroMLBook.pdf}, which the reader is refereed to for a more complete mathematical description of the steps shown here.

\subsection{Linear PCA}
\label{subsec:PCA}

We begin by defining a set of $N$ vectors $G=\{\mathbf{g_{1}, \mathbf{g_{2}}, ... , \mathbf{g_{N}}}\}$, which contains our observational measurements. If $\mathbf{g_{\rm mean}}$ is the vector of mean values of $G$, let $X \in\mathbb{R}^n$ be the set of vectors which holds the centered observations, 
\begin{equation}
\mathbf{x}_{k}=\mathbf{g_k}-\mathbf{g_{\rm mean}}.
\label{eq:mean}
\end{equation}

In order to find the PCs, we shall diagonalize the covariance matrix\footnote{The covariance matrix is traditionally defined as the expectation value of $\mathbf{x}^T\mathbf{x}$. For convenience, we shall address the term \textit{covariance matrix} to the maximum likelihood estimate of the covariance matrix for a finite sample, given by equation (\ref{eq:covmatrix})\citep{Scholkopf1996}.} \footnote{It is also possible to apply PCA to a correlation matrix. This is advised mainly when the data matrix is composed by measurements with different orders of magnitude and/or units.  Since in our particular case all measurements are in the same units (fluxes) and normalized in advance, we shall use the covariance matrix. For a detail discussion on the pros and cons of each case, see \citet{Jollife2002} - section 2.3. }
\begin{equation}
C = \frac{1}{N}\sum_{i=1}^{N}\mathbf{x}_{i}\mathbf{x}_{i}^T.
\label{eq:covmatrix}
\end{equation}
This can be accomplished by solving the eigenvalue equation
\begin{equation}
\lambda_i\mathbf{v}_i = C\mathbf{v}_i,
\label{eq:eigenvalue}
\end{equation}
where $\lambda_i > 0 $ are the eigenvalues and $\mathbf{v}_i\in\mathbb{R}^n$ the eigenvectors of the covariance matrix.

If we consider $V$ the set of eigenvectors of $C$ and $P$ the set of data points projections in $V$, the elements of $P$ will be given by
\begin{equation}
\mathbf{p_{i}}=A_l^T\mathbf{x}_{i},
\end{equation}
where $A_l$ is the matrix formed by the $l$ first PCs as columns. It is possible to show that the elements of $P$ will be uncorrelated, independently of the dimension chosen for matrix $A_l$.

This is where the dimensionality reduction takes place. We can choose the number of PCs that will compose the matrix $A_l$ based on how much of the initial variance we are willing to reproduce in $P$. At the same time, depending on the nature of our data, the spread of the points in the PCs space might also reveal some underlying information, as the existence of two classes of data points, for example.

The main goal of this work is to use PCA to project the data in a sub-space where photometric data vectors associated with different supernova types can be separated. In order to do so, our first step is  to show that it is possible to calculate the projected data points $\in P$ without the need of explicitly defining the eigenvectors $\in V$. This will be important when we consider non-linear correlations in the next sub-section. 

Given that all vectors $\in V$ must lie in the space spanned by the data vectors $\in X$, we can show that (see appendix \ref{ap:proof})
\begin{equation}
\mathbf{v}_a=\sum_{i=1}^N\alpha_i^a\mathbf{x}_{i}, \qquad {\rm with} \qquad \alpha_i^a=\frac{\mathbf{x}_{i}^T \mathbf{v}_a}{N\lambda_a},
\label{eq:vlambda}
\end{equation}
and as a consequence, instead of solving equation (\ref{eq:eigenvalue}) we can also find the elements of $P$ by solving the projected equations\footnote{Equation \ref{eq:projecteEIGV} results from writing each eigenvector as a linear combination of the data vectors. }
\begin{equation}
\mathbf{x}_{i}^T C \mathbf{v}_a=\lambda_a\mathbf{x}_{i}^T\mathbf{v}_a, \quad \forall i,a.
\label{eq:projecteEIGV}
\end{equation}
This leads us to an eigenvalue equation in the form
\begin{equation}
K\mathbf{\alpha}^a=\tilde{\lambda_a}\mathbf{\alpha}^a,
\label{eq:eigenK}
\end{equation}
where 
\begin{equation}
K_{ij}=\mathbf{x}_{i}^T\mathbf{x}_{j},
\end{equation}
and $\tilde{\lambda_a}=N\lambda_a$. Normalizing  $\mathbf{v}_a$, we can also show that $||\alpha^a||=1/\sqrt{N\lambda_a}$.

Finally, consider a test data vector $\mathbf{n}$.  Its projections in the PCs space are given by
\begin{equation}
\mathbf{v}_a^T\mathbf{n}=\sum_{i=1}^N\alpha_i^a\mathbf{x}_{i}^T\mathbf{n}=\sum_{i=1}^N\alpha_i^a K(\mathbf{x}_{i},\mathbf{n}),
\label{eq:featproj}
\end{equation}
where $K(\mathbf{x}_{i},\mathbf{n})=\mathbf{x}_{i}^T\mathbf{n}$.

This demonstration was specifically designed to rely only on the matrix $K$. Although, the classification we aim in this work is not possible in the linear regime. In order to be able to disentangle light curves from different supernovae, we need to perform PCA in a higher dimensional space, where the characteristics we are interested in are linearly correlated. 


\subsection{Kernel Principal Component Analysis}
\label{subsec:KPCA}

KPCA generalizes PCA by first mapping the data non-linearly into a higher dimensional dot product space $\mathbb{F}$ (hereafter, \textit{feature space}):
\begin{eqnarray}
\Phi : \mathbb{R}^n &\rightarrow& \mathbb{F}\nonumber\\
\mathbf{x}&\rightarrow& \Phi(\mathbf{x}),
\end{eqnarray}
where $\Phi$ is a nonlinear function and $\mathbb{F}$ has arbitrary (usually  very large) dimensionality.

The covariance matrix,  $C_{\rm F} \in \mathbb{F}$,  will be defined similarly as 
\begin{equation}
C_{\rm F} = \frac{1}{N}\sum_{i=1}^N\Phi(x_i)\Phi(x_i)^T.
\label{eq:kcovmatrix}
\end{equation}
We assume that $\Phi(\mathbf{x}_i)$ are centred in feature space. We shall come back to this point latter on. 

Consider $\mathbf{v}_{\Phi}^l$ the $l-th$ eigenvector of $C_{\rm F}$ and $\lambda_{\Phi}^l$ its $l-th$ eigenvalue.
Using the same line of argument shown in the previous subsection, we can define an kernel $N\times N$ matrix 
\begin{equation}
K_F(\mathbf{x}_i,\mathbf{x}_j) = (\Phi(\mathbf{x}_i)\cdot\Phi(\mathbf{x}_j)), 
\end{equation}
which allows us to compute the value of dot product in $\mathbb{F}$ without having to carry out the map $\Phi$.  The kernel function has to satisfy the Mercer's theorem to ensure that it is possible to construct a mapping into a space where $K_{\rm F}$ acts as a dot product\footnote{http://ni.cs.tu-berlin.de/lehre/mi-materials/Mercer\_theorem.pdf}. The projection of a new test point, $\mathbf{n}$, is given by
\begin{equation}
(\mathbf{v}_{\Phi}^l\cdot\Phi(\mathbf{n}))=\sum_{i=1}^N\alpha_{\Phi_i}^l K_F(\mathbf{x_i},\mathbf{n}),
\label{eq:featproj2}
\end{equation}
where $\alpha_{\Phi_i}^l$ is defined by the solutions to the eigenvalue equation $N\lambda_{\Phi}\alpha_{\Phi} = K_F\alpha_{\Phi}$.

Finally, it is important to stress that all the arguments shown in this sub-section rely on the assumption that the data are centred in feature space. This is not a direct consequence of using $X$ instead of $G$. Equation (\ref{eq:mean}) is responsible for centring data vectors in $\mathbb{R}^n$, in order to perform centralization in $\mathbb{F}$, we need to construct the kernel matrix using  $\Phi(\pmb{x})-\widetilde{\Phi(\pmb{x})}$. This can also be computed without any information about the function $\Phi$. It is shown in appendix \ref{ap:proof} that the centred kernel matrix, $\widetilde{K_F}$, can be expressed in terms of the non-centered kernel matrix, $K_F$, as
\begin{equation}
\widetilde{K_F}=K_F-1_N K_F-K_F 1_N+1_M K_F 1_N,
\label{eq:KFcent}
\end{equation}
where $(1_N)_{ij}=1/N$. The reader should be aware that we always refer to the centred kernel matrix $\widetilde{K_F}$. However, for the sake of simplicity, the tilde is not used in our notation. 

At this point, we have the tools necessary to compute the centred kernel matrix based on dot products in input space. However, we still need to choose a form for the kernel function $k(\mathbf{x}_i,\mathbf{x}_j):=K_{{F}_{ij}}$. 

In the present work, for the sake of simplicity, we make an \textit{a priori} choice of using  a Gaussian kernel, 
\begin{equation}
k(\mathbf{x}_i,\mathbf{x}_j) = \exp\left[-\frac{\|\mathbf{x}_i-\mathbf{x}_j\|^2}{2\sigma^2}\right],
\label{eq:kernel}
\end{equation}
where the value of $\sigma$ is determined by a cross-validation processes (see subsection \ref{subsec:cross}). Although, it is important to emphasize that there is extensive literature on how to choose the appropriate kernel for each particular data set at hand \citep{Lanckriet2004,Zang2006}. To compare the analysis between different kernel choices is out of the scope of this work. As our goal is to focus on the KPCA procedure itself, we are using the standard kernel choice. An analysis of performances from different kernel choices within the KPCA framework should certain be topic of future research.


\section{ Classification}
\label{sec:class}

By virtue of what was presented so far, we have a set of centred data points, $X$, and a kernel function, $k(\mathbf{x}_i,\mathbf{x}_j)$. This allows us to calculate the kernel matrix in feature space, $K_F$, and its corresponding eigenvalues, $\pmb{\alpha}_{\Phi}$. Using equation (\ref{eq:featproj}), we can obtain the projection of each data point in the eigenvectors of $C_{\rm F}$. 

From now on, we will work in the space spanned by these eigenvectors. More precisely, we will look for a 2-dimensional sub-space of $\mathbf{v}_{\Phi}$, which can optimize our ability to separate the projected data in 2 different classes (namely Ia and non-Ia supernovae). We chose to keep this sub-space bi-dimensional in order to avoid over-fitting to the particular data set we are analysing. 

The procedure describe  before is now applied to two different instances of our data. A data set suitable for the analysis we present here must be composed of two sub-samples. For one of them we have the appropriate label for each data point (we know which class they belong to), from now on this sub-set will be called \textit{training sample}. For the other sub-sample (hereafter \textit{test sample}) the labels are not available, and we want to classify them based on our previous knowledge about the training sample. 

In a first moment, we will concentrate our efforts in the training sample. Its projections in a certain pair  of PCs are calculated through equation (\ref{eq:featproj}). Given that labels of data in  this sample are known, we can calculate projections in different PCs and  determine which  PC pair better translates the initial light curves into a separable point configuration. 

\subsection{The k-Nearest Neighbor algorithm} 
\label{subsec:kNN}

Our choice of which subspace of $\mathbf{v}_{\Phi}$ is more adequate for a specific data situation will be balanced by how well we can classify the training sample using the \textit{k-Nearest Neighbor algorithm} (kNN). 

kNN is one of the most simple classification algorithms and it has been proved efficient in low dimension parameter spaces, ($\rm{dim}\leq 10$, for a further discussion on kNN performance in higher dimensions see \citet{Beyer1999}). The method begins with the training sample organized as $q_i=(x_i,y_i)$, where $x_i$ is the $i-th$ data vector and $y_i$ its label, and a definition of distance between 2 data vectors $d(x_i,x_j)$. Given a new unlabelled test point $q_t(x_t, )$, the algorithm computes the distance between $x_t$ and all the other points in the training sample, $d(x_t,\mathbf{x})$, ordering them from lower to higher distance. The labels of the first $k$ data vectors (the ones closer to $x_t$) are counted as votes in the definition of $y_t$. Finally, $y_t$ is set as the label with highest number of votes. Given this last voting characteristic, kNN is many times refereed to as a type of \textit{majority vote classifier} \citep{James1998}.

Throughout our analysis, we used an Euclidean distance metric and order $k=1$. As this is the first attempt in applying KPCA to the photometric problem, we chose to be bounded by the \textit{Bayes error rate} (hereafter, BER). The BER is defined as the error rate resulting from the best possible classifier. It can be shown that, in the limit of large samples, the error rate of a $k=1$ nearest neighbour algorithm is never larger than $2\times$ BER (for a scratch of the proof see \citet{Ripley1996}, page 195). From now on, this will be refereed to as 1NN algorithm (nearest neighbour with $k=1$).

So far we described how to define a convenient 2-dimensional space where our data points will be separated in Ia and non-Ia populations (sub-section \ref{subsec:KPCA}) and a classification tool that allows us to add a label to a new, unlabelled data point (subsection \ref{subsec:kNN}).  However, we still need to define which pair of PCs of the feature space better maps our data. 
This is done in the next sub-section.

\subsection{Cross-validation}
\label{subsec:cross}

The main idea behind the cross-validation procedure is to remove from the training  sample a random set of $M$ data points, $T^{\rm{out}}$. The remaining part of the training sample is given as input in some classifier algorithm and used to classify the points in $T^{\rm{out}}$. In this way, we can measure the success rate of the classifier over different random choices of $T^{\rm{out}}$ and also compare results from different classifiers given the same training and $T^{\rm{out}}$ sets (for a complete review on cross-validation methods see \citet{Arlot2010}).

The the number of points in $T^{\rm{out}}$ is a free parameter and  must be defined based on the clustering characteristics of the given data set. Here we chose the most classical exhaustive data splitting procedure, sometimes called \textit{Leave One Out} (LOO) algorithm. As the name states, we construct $N$ sub-samples $T^{\rm{out}}$, each one containing only one data point, $M=1$. The training sample is then cross-validated and the performance judged by the average number of correct classifications. 

Data exhaustive algorithms like LOO have a larger variance in the final results, although, they are highly recommended for avoiding biases regarding local data clustering and some non-uniform geometrical distribution of data points in a given parameter space\footnote{http://www.public.asu.edu/$\sim$ltang9/papers/ency-cross-validation.pdf}. 

\subsubsection{The algorithm}

In the context of KPCA, we used LOO and 1NN algorithms to decide the appropriate pair of PCs and  value of $\sigma$ (equation (\ref{eq:kernel})) for each data set.

\begin{figure}
\centering
\includegraphics[trim = 0mm 4mm 4mm 18mm, clip, width=1\columnwidth]{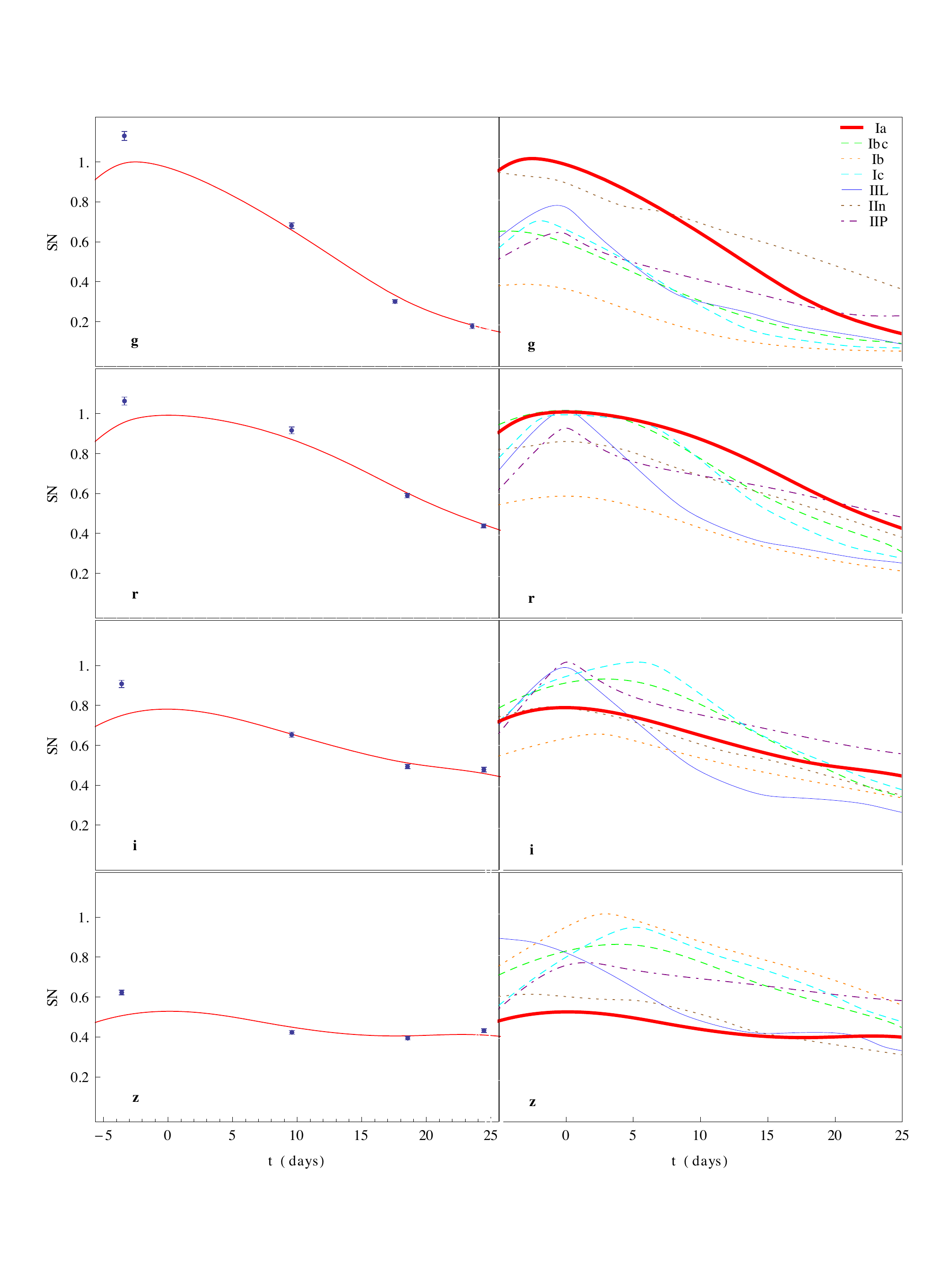}
\caption{Normalized light curves from SIM1. \textbf{Left}: SNe Ia light curve. The plot shows the flux measurements (blue dots) and fitted spline function (red curve), normalized as explained in the text. \textbf{Right}: Example of normalized light curves functions for Ia (red thick), Ib (green dashed), Ibc (orange short-dashed), Ic (cyan dashed), IIL (blue thin), IIn (brown short-dashed) and IIP (purple dot-dashed), according to SNANA classification. The panels from top to bottom run over the DES filters $\{g,r,i,z\}$. The horizontal axis is in units of days since maximum brightness in $r$ band.}
\label{fig:diff_LC}
\end{figure}

The next trick question to answer is: which PCs we should test with the algorithms described before? Obviously there is a high number of vectors in $\mathbf{v}_{\Phi}$ and it would not be possible to test all available pairs. Fortunately, we can make use of the fact that the firsts eigenvectors $\mathbf{v}_{\Phi}$ (those with larger eigenvalues) represent directions of greater data variance in feature space. Although we cannot visualize such vectors, it is easy to confirm that the magnitude of data points projections in $\mathbf{v}_{\Phi}^l$  become very similar to each other for higher $l$. In other words, the smaller eigenvalues correspond to PCs carrying mostly noise, so their projections will, in average, be very similar, and meaningless \citep{Scholkopf1996}. For classification purposes, one expects that the PC pair tailored to provide geometrical separation of the data projection into classes will be among the PCs with higher eigenvalues. 

For the case studied here, we restrict ourselves to testing the first 5 PCs in a first round and extend the search to other PCs only if the classification success rate do not monotonically decrease with the use of higher PCs. In the same line of thought, we start our search with $\sigma \in \{0.1,2.0\}$ in a grid with steps of 0.1 and make this interval wider only if the results do not converge after a first round of evaluations.

The cross-validation algorithm we used is better summarized as:
\begin{enumerate}
\item Pick a PC pair, $\{\rm{PC}_{\rm{A}},\rm{PC}_{\rm{B}}\}$.
\item Define a grid of values for parameter $\sigma$, $\sigma\in\{\sigma_{\rm{min}},\sigma_{\rm{max}}\}$.
\item Pick a value from the above grid, $\sigma_{\rm test}$.
\item Cross validate the training sample using the KCPA projections in the chosen PCs, 1NN and LOO algorithms. 
\item Calculate the average classification success rate for $\{\sigma_{\rm test},{\rm PC}_{\rm A}, {\rm PC}_{\rm B}\}$.
\item Repeat steps (ii) to (v) 10 times. If the average number of successful classifications monotonically decreases in the upper and lower boundaries of $\sigma$, go to step (vii). If not, repeat steps (ii) to (vi) until they do. 
\item Repeat steps (i) to (vi) for all pairs of $\{{\rm A},{\rm B}\}\in \{1,5\}$.
\item If the average number of successful classifications monotonically decreases when using higher PCs, go to step (ix). Otherwise, consider $\{{\rm A},{\rm B}\}\in \{1,10\}$ and repeat steps (i) to (viii).
\item Choose for $\{\sigma,{\rm PC}_{\rm A},{\rm PC}_{\rm B}\}$, values corresponding to the largest average number of successful classifications. 
\end{enumerate}

Once the cross-validation is completed, we use the resulting parameter values to calculate the training sample projections in PC space. We can finally use 1NN algorithm to assign a label to each data point in the test sample.  The final procedure of classifying the test sample is called KPCA+1NN algorithm throughout the text.

The framework described so far can be applied to any set of astrophysical objects, as long as we have a training and a test sample. The cross-validation procedure is performed only in the training sample and each point in the test sample is classified at a time. This avoids running the whole machinery again every time one new point is added to the test sample, and prevent us from introducing misleading data as part of the features to be mapped by the PCs. However, the parameter space composed by the PC pair and value of $\sigma$ can always be updated if we have at hand new data points whose types are known. Only then it is necessary to re-run the cross-validation process.

From now on we focus on the problem of photometrically classifying SNe Ia as a practical example, although the exact same steps could be applied for any transient with observable light curves. In the next section, we describe how the light curve data should be prepare before we try to classify them.

\begin{table}
\caption{Description of the light curve selection cuts. The SNe were required at least one observation in $t\leq t_{\rm low}$, one in $t\geq t_{\rm up}$ and at least 3 observations satisfying a given SNR requirement in each filter in order to be included in any of the data sets analysed in this work. These selection cuts were applied for training and test samples within a specific data set.}
\centering
\begin{tabular}{c | c | c | c |}
\cline{2-4}
 & $t_{\rm low}$ & $t_{\rm up}$ & $\Delta$ \\
\hline
\multicolumn{1}{|c|}{$D_1$} & \multirow{2}{*}{-3} & \multirow{2}{*}{+24} & 1\\
\cline{1-1} \cline{4-4}
\multicolumn{1}{|c|}{$D_2$} & & &3\\
\hline
\multicolumn{1}{|c|}{$D_3$} &\multirow{2}{*}{0} & \multirow{2}{*}{+15} & 1\\
\cline{1-1} \cline{4-4}
\multicolumn{1}{|c|}{$D_4$} & & &3\\
\hline
\multicolumn{1}{|c|}{$D_5$} &\multirow{2}{*}{-10} & \multirow{2}{*}{0} & 1\\
\cline{1-1} \cline{4-4}
\multicolumn{1}{|c|}{$D_6$} & & &3\\
\hline
\multicolumn{1}{|c|}{$D_7$} &\multirow{2}{*}{-3} & \multirow{2}{*}{+45} & 1\\
\cline{1-1} \cline{4-4}
\multicolumn{1}{|c|}{$D_8$} & & &3\\
\hline
\end{tabular}
\label{tab:cad}
\end{table}


\section{Light curve preparation}
\label{sec:LC_prep}

In case we have \textit{b} different filters, the observational data available from the $l-th$ SN can be arranged as
$\mathbf{F}^l=\{F^l_1,...,F^l_b\}$. Considering the $i-th$ filter, $\left(\mathbf{F}^l\right)_{i}=\{\{t^l_{i1},F^l_{i1},\sigma^l_{{\rm F}i1}\},...,\{t^l_{ie},F^l_{ie},\sigma^l_{{\rm F}ie}\}\}$. In our notation, the $t^l_{ij}$ correspond to the $j-th$ observation epoch (in MJD), $F^l_{ij}$ is the measured flux at $t^l_{ij}$, $\sigma^l_{{\rm F}ij}$ is the error in flux measurement and $e$ is the total number of observation epochs in filter $i$.

Our next task is to translate the time of each observation from MJD to the time since maximum brightness in a particular filter. Which filter shall be used as a reference does not have much influence in the final result. The ideal is to choose a band where the ability to determine the time of peak brightness is greater, and use that reference band for all SN in the sample.  The time of maximum brightness in our reference band for the $l-th$ SN is addressed as $t^l_{\rm max}$. As a result, we obtain data points in a particular filter $i$ as $F^l_i=\{\{\left(t_{\rm max}^l\right)_{i1},F^l_{i1},\sigma^l_{i\rm F1}\},...,\{\left(t_{\rm max}^l\right)_{ie},F^l_{ie},\sigma^l_{i\rm Fe}\}\}$, where $\left(t_{\rm max}^l\right)_{ij}=t^l_{ij}-t^l_{\rm max}$.

\begin{figure}
\centering
\includegraphics[scale=0.65]{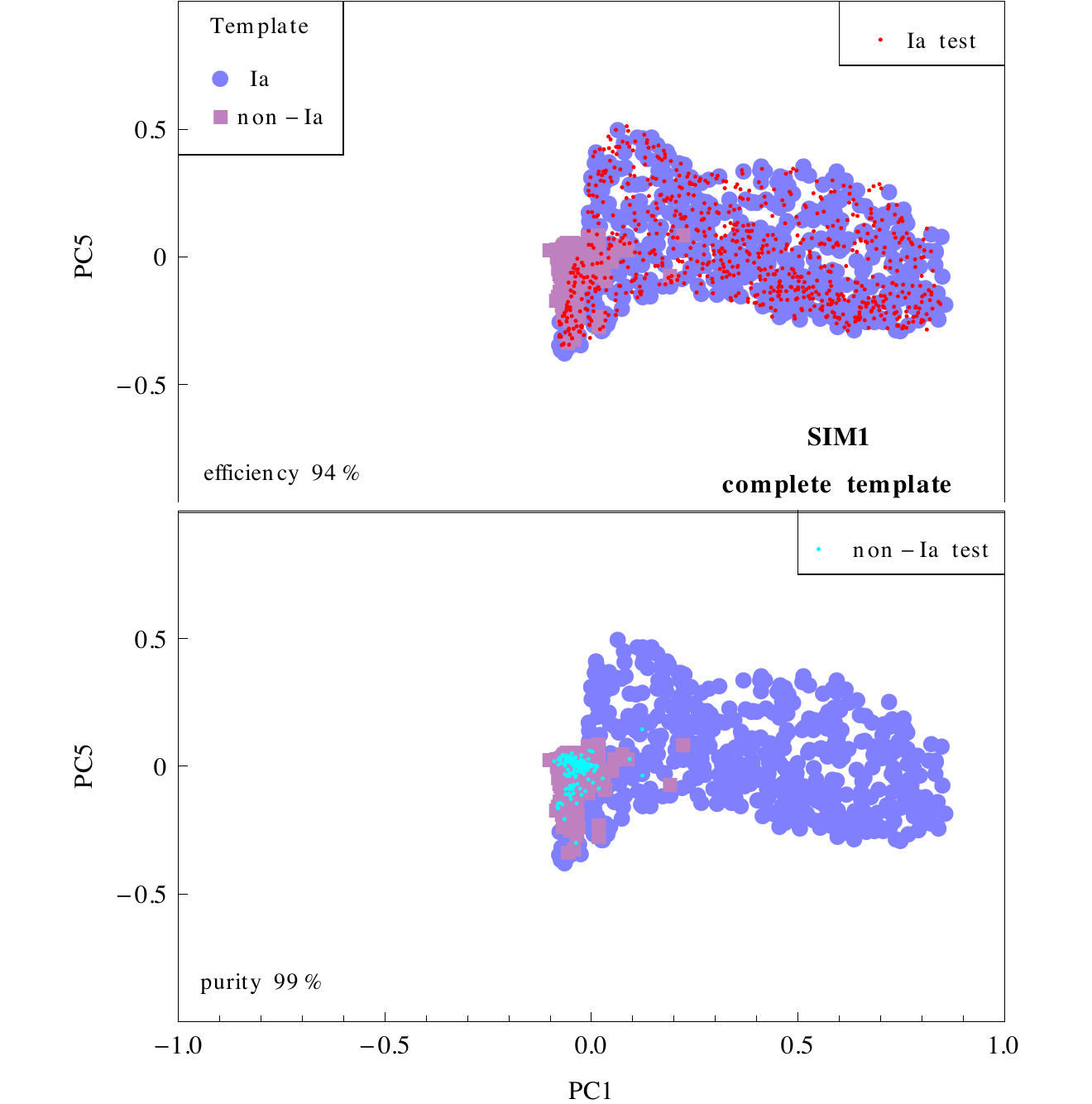}
\caption{Classification results from SIM1. Blue circles (Ia) and purple squares (non-Ia) represent the geometrical \textit{locus} defined by the training sample. \textbf{Top}: Red dots correspond to SNe Ia in the test sample. \textbf{Bottom}: Cyan dots correspond to non-Ia SNe in the test sample. The plot also shows calculated values for eff$_{A}$ and  pur. }
\label{fig:resSIM1}
\end{figure}

We must also deal with the fact that, in a real situation,  the  input from observations consists in some non-uniform sampling of the light curve in various (most cases more than 3) different filters for each SNe. Although, it is necessary to translate such information into a grid equally spaced in time. This is done by using a cubic regression spline fit for each light curve. The spline fit was chosen based on its ability to fit non-uniform functions in a  parameter independent manner. As a consequence, we have a smooth light curve function for  each SNe and filter. 

As a final step, we must keep the light curve functions within a reasonable range (so to avoid divergence in the exponent of equation (\ref{eq:kernel}) due to very bright or dim sources, for example). This is done through the normalization of the light curve functions by the maximum flux measured in all filters for a particular SN. In our notation ${S_N}^{l}_i(t)$ corresponds to the normalized fitted light curve for the $l-th$ SN in filter $i$. The use of the same normalization factor for all filters for a given SN ensures that the colour and shape of each light curves are preserved. 

We now use the $\mathbf{S_N}^l=\{{S_N}^l_1,...,{S_N}^l_b\}$ functions in order to construct our initial data matrix, $\mathbf{G}$, composed by $N$ rows and $M$ columns. Each row contains all information available for a single SN and each column contains the flux measurements in a specific observation epoch and filter. The difference in time since maximum brightness between 2 successive columns of $\mathbf{G}$ is defined as $\Delta$ and for the purposes of this work it is kept constant. However, we do address the analysis with different values for $\Delta$ later on. The lowest and highest observation epoch since $t^l_{\rm max}$ is referred to as $t_{\rm low}$ and $t_{\rm up}$, respectively.

\begin{table}
\centering
\caption{Mean values and standard deviations of residual between the simulated and derived date of peak brightness in each band. The values were obtained through analysis of SNe Ia  light curves in the training samples of SIM1 and SNPCC.}
\begin{tabular} {c c c c c}
& & SIM1 & & SNPCC\\
\cline{1-1} \cline{3-3} \cline{5-5}
filter & &$\Delta t^{\rm max} \pm \sigma_{\Delta t^{\rm max}}$& &$\Delta t^{\rm max} \pm \sigma_{\Delta t^{\rm max}}$\\
\cline{1-1} \cline{3-3} \cline{5-5}
g & &$-3.7 \pm 3.5$& &$1.2 \pm 27.1$\\
r & &$-0.1 \pm 2.6$& &$0.9 \pm 8.2$\\
i & &$1.9 \pm 2.8$ & &$2.3 \pm 9.2$\\
z & &$1.2 \pm 3.6$& & $3.4 \pm 8.4$\\
\cline{1-1} \cline{3-3} \cline{5-5}
\end{tabular}
\label{tab:wfilt}
\end{table}

Throughout this work, we took the conservative approach of not extrapolating functions ${S_N}^{l}_i(t)$ outside the time domain covered by the data. In other words, we only considered classifiable those  SN which have at least one observation epoch $t\leq t_{\rm low}$ and at least one epoch $t\geq t_{\rm up}$, in all available filters. The values of $t_{\rm low}$ and $t_{\rm up}$  must be chosen so to include the largest possible number of SNe and, at the same time, to probe an interval of the light curve which posses information enough to satisfy our classification purposes. We applied the algorithm considering values of $t_{\rm low}$ and $t_{\rm up}$ shown in table \ref{tab:cad}. The demand that this sampling must be fulfilled in all filters could be relaxed, leading to an interesting study about the importance and role of each frequency band. We leave that for a future work, focusing our efforts in data points for which information is available  in all bands.

Joining the previous ingredients, light curves from the $l-th$ SN sampled between $t_{\rm low}$ and $t_{\rm up}$ in steps of length $\Delta$ are stored in a single row of $\mathbf{G}$, sequentially for $b$ different filters. We can now use equations (\ref{eq:mean}) and (\ref{eq:kernel}) to calculate the centred data vectors and kernel matrix, respectively. 


\section{Application}
\label{sec:application}

\subsection{Data sets}

We applied the procedure described so far to different samples taken from the post-SNPCC data set.
The post-SNPCC consisted of $\approx$20.000 SNe light curves, simulated according to DES specifications and using the SNANA light curve simulator. This large set is subdivided in 2 sub-samples: a small spectroscopically confirmed one of 1103 light-curves (training) and a photometric sample of 20216 light curves (test). The role of the training sample was to mimic, in SNe types, proportions and data quality, a spectroscopically confirmed subset available for a survey like DES. After the challenge results were released, the organizers made public an updated version of the simulated data set (post-SNPCC), which was used in most of this work. This updated data set is quite different from the one used in the challenge itself (SNPCC), due to a few bug fixes and other improvements aimed to a more realistic representation of the data expected for DES. As a consequence, its results should not be compared to those of the SNPCC. 
A detailed analysis of our findings from the post-SNPCC faced to others published after the challenge, which use the same data set (namely, \citet{Newling2011}, R2012 and \citet{karpenka2012}), is presented in section \ref{sec:discuss}.

\begin{figure}
\centering
\includegraphics[scale=0.65]{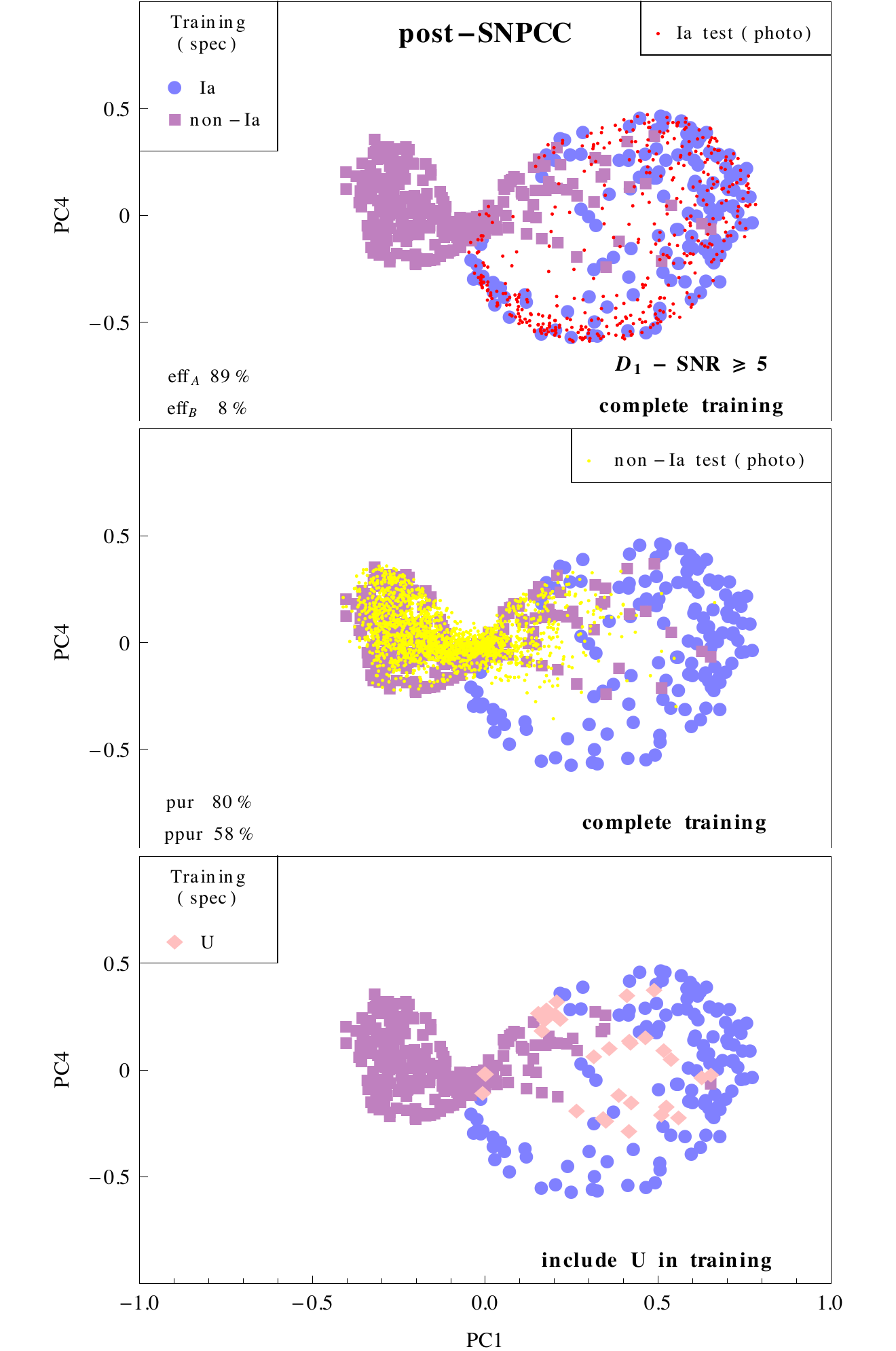}
\caption{Classification results from post-SNPCC, $D_1$+SNR5 data set. The training sample is represented by the blue circles (Ia), purple squares (non-Ia) and pink diamonds (untyped). \textbf{Top}: SNe Ia from the  test sample (red dots) are superimposed to the complete training set divided in Ia and non-Ia. \textbf{Middle}: Non-Ia SNe test sample (yellow dots) are superimposed to the complete training set as in the upper panel. \textbf{Bottom}: Training set points including $U$ as a possible classification type.}
\label{fig:D1SNR5}
\end{figure}
\begin{figure}
\centering
\includegraphics[trim = 0mm 0mm 0mm 3.5mm, clip, width=1\columnwidth]{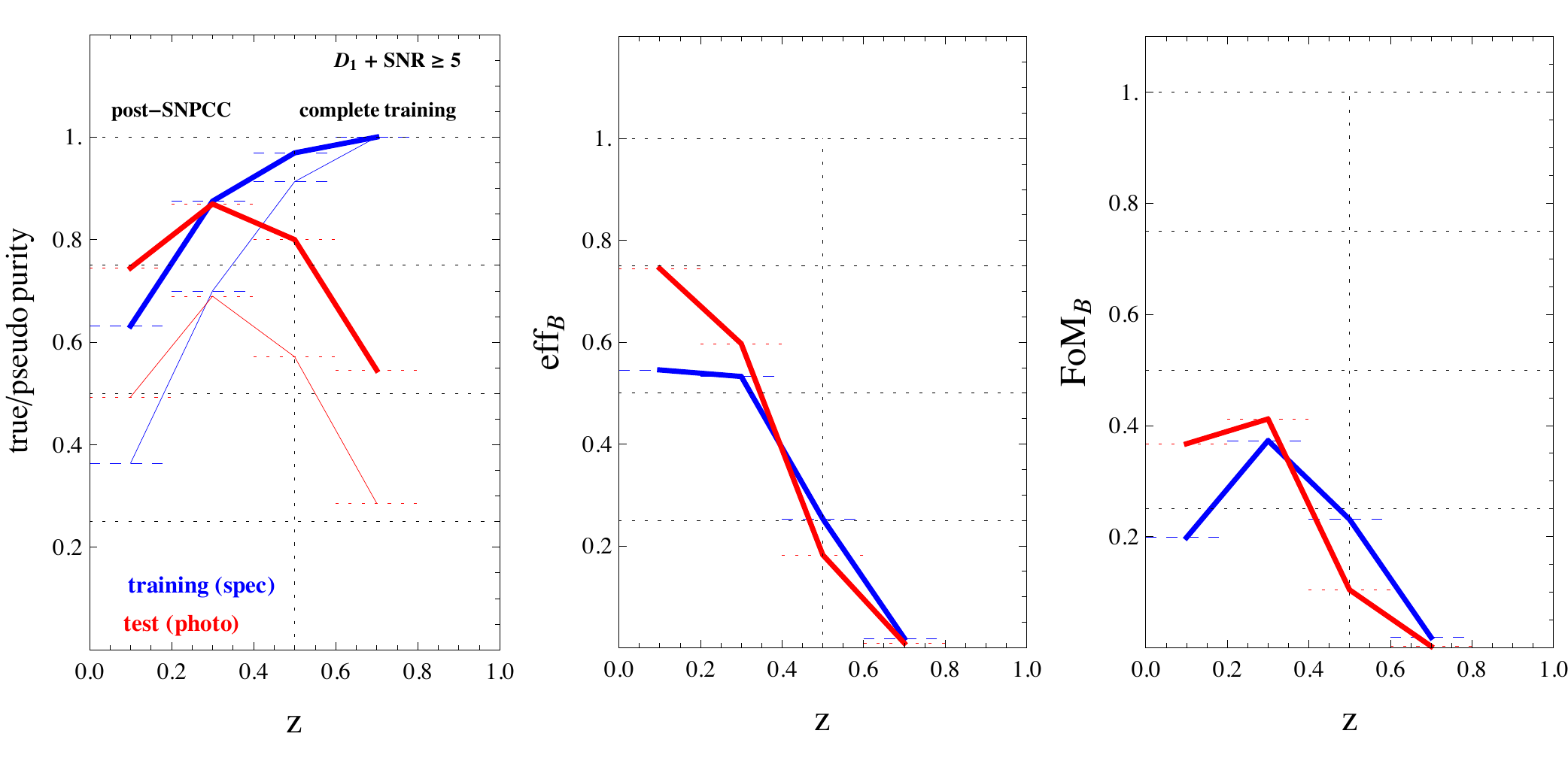}
\caption{Results from the post-SNPCC data for  pur (left), eff$_{\rm B}$ (middle) and FoM$_{\rm B}$ (right) as a function of redshift for $D_1+$SNR5 (alternative view of results shown in figure \ref{fig:D1SNR5}). 
 The red-thick lines correspond to results found for the test sample (cross-validated) and blue-thick lines show results for the training sample. The right panel also shows values for ppur (thin lines, blue for training and red for test sample). These results were calculated for redshift bins of width $0.2$. Redshift dependent outcomes from SC, eff$_{\rm A}$ and pur$_{\rm A}$ for this sample are shown in figure \ref{fig:z_diag_cad1_SNR5_aft_only}.}
\label{fig:z_diag_cad1_SNR5_bef_only}
\end{figure}


For the sake of completeness, we also present results from applying our method to the SNPCC sample. Although this sample contains the bugs mentioned before, it allow us to coherently compare our method to a broader range of alternatives. Detailed comparison of our results with those reported in \citet{Kessler2010b} is presented in section \ref{sec:SNPCC}.

Our first move is to check if KPCA can correctly classify SNe light curves in a best-case scenario. In order to do so, we generated a high quality data set, hereafter SIM1. This set consists of 2206 SNe, composed by 2 sub-samples (training and test), both with at least 3 observation epochs having SNR$\geq$5 in all filters. SNe types and proportions in each subset are the same as those found in the post-SNPCC training sample. As a consequence, the 2 sub-samples in SIM1 are completely representative of one another. This was done to avoid classification problems found by other studies when the training sample is not representative of the test sample (e.g., \citet{Newling2011} and R2012).  At this moment, the purpose of SIM1 is only to perform a consistency check for the KCPA and light curve preparation prescriptions described above. 

In generating SIM1, we used the input SNANA files provided as part of the post-SNPCC package, and ran the simulator until the required number of each SNe type passing selection cuts was reached.The kernel matrix was constructed considering $t_{\rm low}^{\rm SIM1}=-3$ and $t_{\rm up}^{\rm SIM1}=+24$. After verifying that our algorithm was indeed effective in ideal conditions, we will focus on the analysis of the post-SNPCC itself.

The Phillips relation for type Ia SN can be consider the first SNe Ia standardization procedure \citep{Phillips1993}. It establishes a correlation between the magnitude measured at maximum brightness and the magnitude measured 15 days after that (hereafter \textit{Phillips interval}). For our purposes, this relation highlights a time interval in the light curve where important information are stored. However, at this point we cannot say if a data set sampled solely in this time interval can provide enough information. As a consequence, we considered 8 different sub-sets of post-SNPCC data, whose parameters are described in Table \ref{tab:cad}. This requirements were imposed to training and test samples within a given data set.   

$D_1$ to $D_4$ probe the light curve so to include the Phillips interval. $D_5$ and $D_6$ aim at testing the KPCA+1NN procedure in a region of the light curve that was not explored in the SNPCC: with points only before maximum. Although this kind of classification does not result in cosmological useful SNe Ia, it is very important in pointing candidates for spectroscopic follow-up \citep{Kessler2010b}. $D_7$ and $D_8$ are tailored to include the second maxima in infra-red bands expected to occur after 20 days since maximum brightness \citep{Kasen2006}. 

In Table \ref{tab:cad}, we varied not only the maximum and minimum epoch of observation, but also considered different values for $\Delta$. The purpose of this analysis is to investigate if the classification results are sensitive to the step size between different columns of the kernel matrix.  We expect this result to be correlated with data quality, since the interpolated functions are influenced by errors in flux measurements. To test this hypothesis, we applied the classification procedure to different sub-samples of each data set, according to their SNR. 

\begin{figure}
\centering
\includegraphics[scale=0.55]{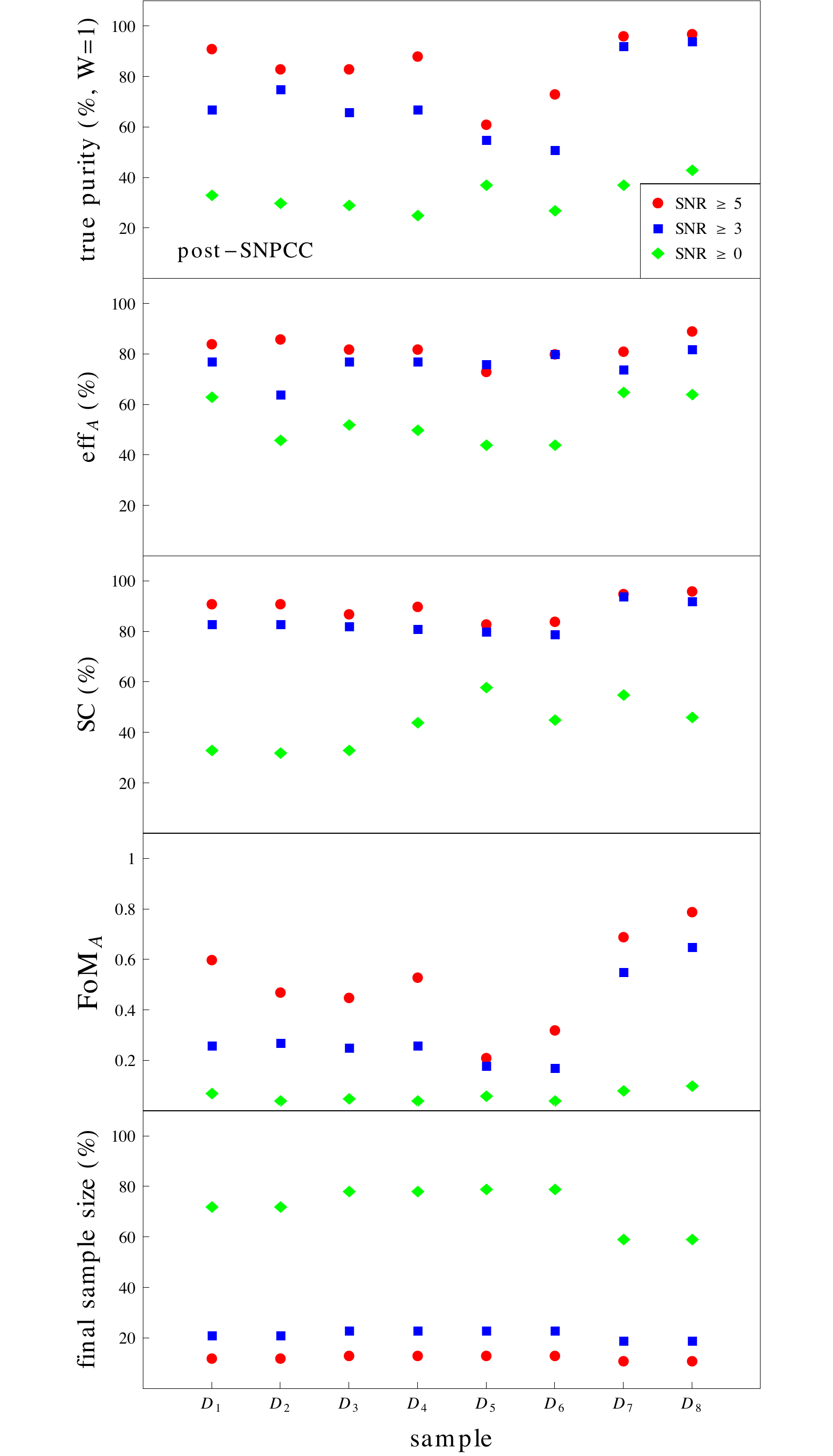}
\caption{Summary of classification results. Panels display pur, eff$_{\rm A}$, SC, FoM$_{\rm A}$ and final sample size, from top to bottom.  Horizontal axis runs through data samples described in table \ref{tab:cad}. Results are displayed for SNR$\geq$5 (red circles), SNR$\geq$3 (blue squares) and SNR$\geq0$ (green diamonds).}
\label{fig:classRES}
\end{figure}

Finally, we only considered SNe with at least 3 observational epochs above a certain SNR threshold in each filter. As the spline fitted functions are supposed to get the overall behaviour of a smooth light curve, this selection cut assures that at least 3 of the points with higher weights in the spline fitting procedure correspond to good quality measurements. We also present results without a SNR selection cut, addressed as SNR$\geq$0.

\subsection{Results}

\begin{figure*}
\centering
\includegraphics[scale=0.625]{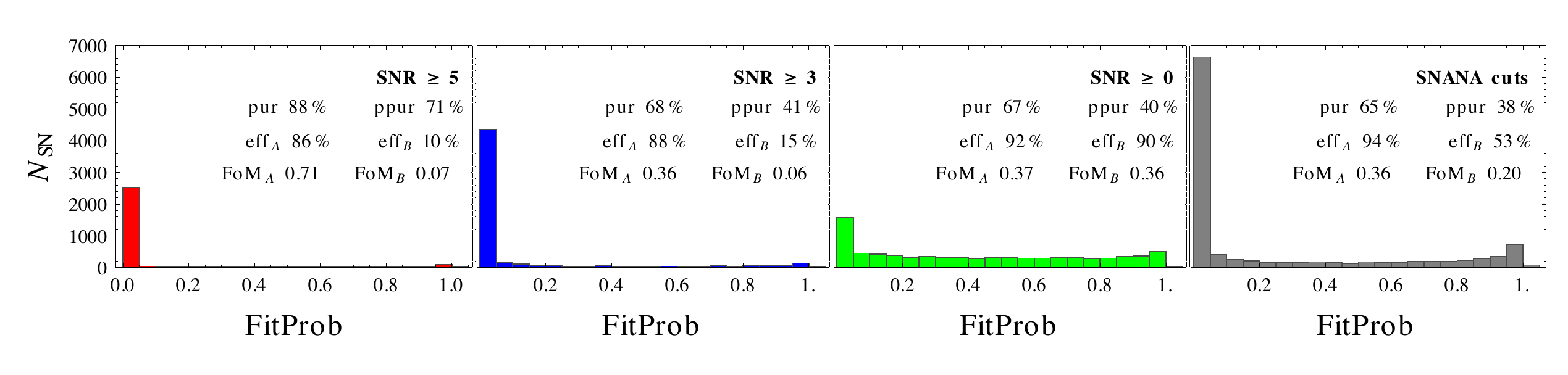}
\caption{Number of SNe as a function of their fit probability calculated from MLCS2k2. Panels show histograms for SNR$\geq$5, SNR$\geq3$, SNR$\geq$0 and SNANA cuts, from left to right. Also shown are the classification outcomes based on FitProb (SNe with FitProb$>$0.1 were tagged as Ia and the remaining ones were tagged as non-Ia).}
\label{fig:fitprob}
\end{figure*}

In order to choose a filter as our reference band, we used the SNe Ia in the training sample of SIM1 and post-SNPCC. As our primary goal is to correctly separate a sample containing only type Ia, our decision was based on the results from SNe Ia in the spectroscopic sample only. Interpolated light curve functions  before normalization were used to determine the time of peak brightness in all bands. The residual between the simulated and derived date of maximum brightness, $\Delta t^{\rm max}$, in each band were computed for all SNe Ia in the training samples. This resulted in a distribution of points whose spread represents our ability (or lack of) in determining this parameter for each filter. The mean values and standard deviations encountered are shown in Table \ref{tab:wfilt}.

From this we realized that the $r$ band is the best choice for determining the time of peak brightness, since it has the less biased mean value with the smallest standard deviation. Such results agree with those found in R2012, also based on SNANA simulations, but with a different argument. All the results presented from now on were calculated using the time of peak brightness in $r$ band as reference. 

The final classification results are reported in terms of efficiency (eff), purity (pur) and successful classification (SC) rates,
\begin{eqnarray}
{\rm eff}&=&\frac{N^{\rm SC}_{\rm Ia}}{N^{\rm tot}_{Ia}}\label{eq:eff}\\
 {\rm pur}&=&\frac{N^{\rm SC}_{\rm Ia}}{N^{\rm WC}_{\rm nonIa}+N^{\rm SC}_{\rm Ia}}\\
 {\rm SC}&=&\frac{N^{\rm SC}_{Ia}+N^{\rm SC}_{\rm nonIa}}{N^{\rm TOT}}
\end{eqnarray}
where $N^{\rm SC}_{\rm Ia}$ ($N^{\rm SC}_{\rm nonIa}$) is the number of successfully classified SNe Ia (nonIa), $N^{\rm tot}_{\rm Ia}$ is the total number of SNe Ia, $N^{\rm WC}_{\rm nonIa}$ is the number of non-Ia wrongly classified as Ia and $N^{\rm TOT}$ is the total number of SNe which survived selection cuts.  

Efficiency values are shown  for two different normalizations: eff$_{B}$ considers $N^{\rm tot}_{\rm Ia}$ the total number of SNe Ia \textit{before any selection cuts}, and eff$_{A}$ was calculated using  the total number of SNe Ia remaining \textit{after selection cuts.}\footnote{By selection cuts we mean the SNR requirement for each subs-sample $+$ the time window constraints of described in table \ref{tab:cad}.} The definition used in the SNPCC corresponds to eff$_{B}$,  and aims at addressing the impact on final sample not only due to the classifier, but also to the  selection cuts used. In our particular case, we chose to display values of eff$_{A}$ in order to isolate the classification power of the algorithm itself. As stated before, our results are mainly influenced by the quality of each observation. Beyond that, we made a specific choice of not extrapolating the light curve where data is not present (table \ref{tab:cad}). As a consequence, we consider our selection cuts as a minimum amount of information necessary to coherently compare different light curves without the need of further \textit{ad hoc} hypothesis. In this scenario, the use of eff$_{A}$ gives a better idea on the classifier performance.  However, when comparing with previous analysis from the literature, eff$_{B}$ should be referred to. From now on, for all our results that can be compared to previous ones,  both quantities are shown. \footnote{For the sake of clarity, when both quantities are present (results that might be compared with others from the literature), outcomes normalized after selection cuts are shown in appendixes \ref{ap:zcad1SNR5} and \ref{ap:SNPCC_comp}.}

By definition, eff measures our capacity in recognizing the SNe Ia, while pur measures the contamination from non-Ia SNe in our final sample. SC values are presented in order to provide an overall picture of our classification results regarding non-Ia as well.

In order to make our results easier to compare with other analysis from the literature, we also report them in terms of the figure of merit (FoM) and pseudo-purity (ppur), used to rank classifiers in the SNPCC,
\begin{eqnarray}
{\rm ppur} &=&\frac{N^{\rm SC}_{\rm Ia}}{N^{\rm SC}_{\rm Ia}+W N^{\rm WC}_{\rm nonIa}}, \\
{\rm FoM}&=&{\rm eff}\times {\rm ppur},
\end{eqnarray}
where $W$ is used to input a stronger penalty on non-Ia contaminating the final SNe Ia sample. Following the SNPCC, we used $W=3$. Given that FoM is a function of efficiency, we report values for FoM$_{A}$ and FoM$_{B}$ for total number of SNe after and before selection cuts, respectively.

\subsubsection{SIM1}

We must now prepare the light curves according the prescription described in section \ref{sec:LC_prep}. We randomly selected one example of type Ia light curve  in SIM1  to illustrate how the fitted functions behave given the data points. This is shown in the left panels of figure \ref{fig:diff_LC}. The right panels show the light curve functions for different types of non-Ia SNe. Panels from top to bottom run over the DES filters $\{g,r,i,z\}$.  In order to facilitate visualization, all curves were normalized as explained in section \ref{sec:LC_prep}.

For the SIM1 data set, the cross-validation procedure returns PCs 1 and 5 along with $\sigma=0.3$ as the most appropriate parameters values. The final geometrical distribution of the training sample in such PCs parameter space, along with the classification results are shown in figure \ref{fig:resSIM1}. In order to facilitate visualization, we show the Ia  and non-Ia SNe in the test sample in two different plots. 

We can see that, in a best case scenario, KPCA+1NN algorithm is efficient enough to separate the two populations in feature space with a minimum loss in the number of SNe Ia (up to 94\% eff$_{A}$) and almost no contamination from non-Ia's in the final sample (up to 99\%  pur).

\subsubsection{Post-SNPCC data}

The analysis of the post-SNPCC data was performed in different steps. We first separate a sub-sample which can be consider the analogous of SIM1 inside post-SNPCC, $D_1$ with SNR$\geq$5 (hereafter $D_1$+SNR5). This data set results from imposing in post-SNPCC data the same selection cuts applied to SIM1. 

Using $D_1$+SNR5, we obtained   $89\%$ (80\%) pur  and SC of $92\%$ (94\%)  in the training (test) sample. The graphical representation of results from $D_1$+SNR5 are shown in the upper and middle panels of Figure \ref{fig:D1SNR5} and the redshift distribution of the diagnostic parameters are displayed in figures \ref{fig:z_diag_cad1_SNR5_bef_only} and \ref{fig:z_diag_cad1_SNR5_aft_only}.

Analysing the geometrical distribution of training sample data points (blue circles and purple squares), the numerical results mentioned above become more clear. There is an obvious  distinction between the preferential \textit{locus} occupied by Ia and non-Ia in this parameter space. However, besides the overlapping area where both species exist, and which was already present in SIM1, we can also spot some contamination of non-Ia points inside the area occupied by Ia. Such ``misplaced" non-Ia probably gave rise to an  important share of the wrong cross-validation classification.  In what follows, we described 2 different approaches aimed at suppressing the influence of these ``problematic" data points. \\

\textit{The Untyped supernova}\\

Let us focus in $D_1$+SNR5 for a moment. Each data point in the training sample is characterized by the SN identification number, its coordinates in PC1 $\times$ PC4 space, the true label and the label  from cross-validation. We identified all points who received a wrong label in the cross-validation process and gathered them in a set $U$. We considered these troubled points, in the sense that, although they are spectroscopically confirmed SNe, their light curve characteristics are not enough to fully distinguish them within the training sample. 

\begin{figure}
\centering
\includegraphics[scale=0.55]{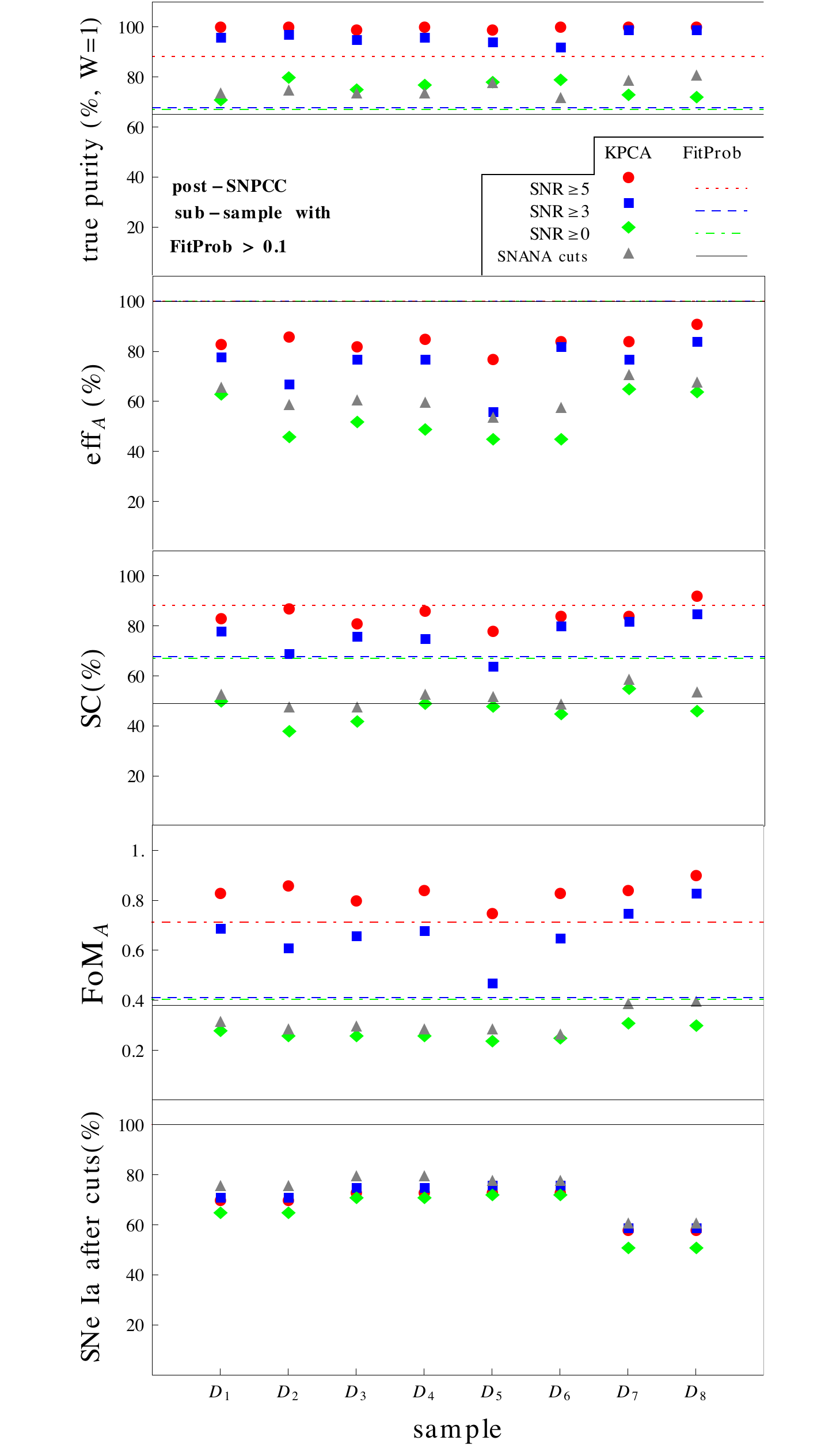}
\caption{Classification results obtained for the sub-sample of SNe with FitProb$>$0.1 using different time windows. Red-circles, blue-squares, green-diamonds and gray-triangles correspond to KPCA+1NN results when SNR$\geq5$, SNR$\geq3$, SNR$\geq$0 and \textit{SNANA cuts} are applied, respectively. Horizontal red (dotted), blue (dashed), green (dot-dashed) and gray (full) lines correspond to the results from FitProb criteria for the same set of cuts. Panels show eff$_{\rm A}$, pur, FoM$_{\rm A}$, SC and the percentage of SNe Ia passing time window requirements from top to bottom.}
\label{fig:KPCAmlcsHF}
\end{figure}

Our first attempt was to remove all points $\in U$ from the training set before classifying the test sample. In doing so, we defined that a new unlabelled test point would be classified according to the region in the parameter space it occupies, since removing the troubled points defines a clear geometrical boundary between Ia and non-Ia regions in PCs parameter space. 
This slightly increased our ratings, leading to 87\% pur, 93\%  eff$_{\rm A}$,  and 96\% SC rates.

Trying to get rid of the remaining contamination as much as possible, we consider the complete training sample with 3 different SNe types: Ia, non-Ia and untyped SNe ($U$). This allows us to take advantage of the information in the troubled points and identify light curves similar to them. An expected consequence of this choice is a decrease in efficiency, since some of the Ia in the test sample will be classified as $U$. On the other hand, as the lost of SNe for the $U$ class happens to non-Ia as well, the pur in our final Ia sample will increase, for $D_1$+SNR$\geq5$ to 91\%. 

The training set divided in 3 sub-samples has its graphical representation shown in the bottom panel of figure \ref{fig:D1SNR5}. For all the cases described here (complete training, excluding $U$ from the training set and including $U$ as a classification type) the distribution of test points will not change, since only the training sample is affected.

We performed the classification for all samples described in table \ref{tab:cad} imposing 3 different SNR cuts (namely SNR$\geq$5, SNR$\geq$3 and SNR$\geq$0). A summary of our finding is detailed in table \ref{tab:final}. 

Figure \ref{fig:classRES} shows results for samples listed in the above mentioned table for the case where the $U$ class was included in the  training sample as a third SNe type\footnote{This plot was constructed with the goal of maximizing SC, however, we also applied the cross-validation process of section \ref{subsec:cross}, aiming at maximum FoM and the results are pretty similar.}.  It is clear from this plot that pur and FoM results become more dependent on time sampling choices as SNR goes higher. The extreme cases being samples $D_5$/$D_6$ (before maximum, worst results) and $D_7$/$D_8$ (wider time sampling, better results).

Finally, we should emphasize that our analysis  was based on the idea that information should be stored somewhere in the light curve function. If this is true, KPCA could easily be able to provide a direction of information clustering in some untouched feature space, which could be accessed through the data points projections in the PCs. That was the main reason why we started our analysis based on SNR requirements. Errors in flux measurements are direct correlated to the SNR of each observation, and higher errors lead to more oscillations in the light curve functions. In the extreme case were we used random number as components of an input data vector (which contains no information), its projections in PCs will always be located very close to the origin. 

Results shown in Table \ref{tab:final} reflects this main idea. Requiring a SNR$\geq$5 in $D_1$ to $D_4$, we obtained  pur, eff$_{\rm A}$, and SC rates higher than 80\% in all 4 cases. These samples contain approximately 5 times more non-Ia than Ia SNe (see Table \ref{tab:number}), which is close to what we expect in a real survey. Beyond that, we did not demand representativeness in redshift or SNe types between the test and training samples.  The training sample inside the  post-SNPCC data have all the biases the organizers were able to predict and which come along with  spectroscopic observational conditions \citep{Kessler2010b}. The selection cuts we applied to SNR, in this context, can be seen as a simple procedure to extract the full potential of a given data set\footnote{We remind the reader that the SNR selection cuts are applied to both, training and test sample.}.

The results presented here are in agreement to those found by R2012, who applied a diffusion map and random forest algorithm to the same data set. Using the spectroscopic sample as given in the post-SNPCC as a training set, they found  56\%/48\% for pur/eff$_{\rm B}$ values. Our analysis for $D_8$+SNR0, which imposes no SNR selection cuts, returns 43\% pur and 35\% eff$_{\rm B}$. In their scenario achieving higher purity, they report 90\% pur  and 8\% eff$_{\rm B}$ from a redshift limited training sample
(R2012, Table 6). For $D_8$+SNR5, our method achieved 98\% pur and 7\% eff$_{B}$. However, we emphasize that while R2012 uses a different prescription for constructing the training sample, our results were reached using a subset of the spectroscopic sample \textit{as it is presented in the SNPCC}.

Focusing in sample $S_{m,25}$ of R2012 and cad1+SNR5 of our method, the first feature to call attention is the exponential decay in our results for eff$_{\rm B}$. It will be clear in what follows that this is a consequence of SNR cuts (figure \ref{fig:SNPCC}). In this particular case, we imposed each filter should observe at least 3 epochs with SNR$\geq$5 and, with higher redshift, SNe fulfilling this requirement  become rare. Also, in the present analysis, we keep only SNe with observations in all available filters, which prevent us from classifying any object with  $z\geq$0.8 (see upper redshift end of our results in figure \ref{fig:z_diag_cad1_SNR5_bef_only}). Obviously these are not intrinsic characteristics of the method, or the data, but choices we made in order to keep results in a conservative perspective. Nevertheless, our values for eff$_{\rm B}$ are comparable to those of R2012 up to $z\approx0.4$ (see figure 10 of R2012).

As a consequence, despite the loss in efficiency for the reasons cited above, the local maximum in FoM$_{\rm B}$ achieved by both groups, us and R2012, are FoM$_{\rm B}\approx0.5$, with our method providing higher results up to $z\approx0.5$.

It was not our purpose to construct a different observation strategy, but instead, to show that if a photometric survey was able to provide a sample similar to post-SNPCC today, it is possible to extract a photometric classified set containing approximately 15\% of the entire sample (more than 2000), with SC$\geq$90\% . Beyond that, such results can be achieved with minimum astrophysical input and no \textit{a priori} hypothesis about light curve shape, colour, SNe host environment or redshift. \\

\textit{Results from Linear PCA}\\

Given the wide spread use of linear PCA in astronomy, we also verified how the standard linear version of PCA performs in the SNe photometric classification problem. The method described in section \ref{subsec:PCA} was applied to the post-SNPCC data. Once the PCs  and projections were calculated, we used a cross-validation algorithm similar to that presented in section \ref{subsec:cross}. The main difference being that, in the linear case, there is no parameter $\sigma$ to determine. 

We present results for $D_1$ in appendix \ref{ap:linearPCA}. As expected, when no SNR cut is applied, linear and KPCA achieved similar rates of  pur and eff$_{\rm A}$ (table \ref{tab:linearPCA}). However, when data quality increases, linear PCA is not able to take advantage of the small details introduced in the light-curve function. Results from linear PCA applied to $D_1$+SNR5 and including U in training achieved maximum values of  73\% pur, 56\% eff$_{\rm A}$, and 79\% SC.  Comparing tables \ref{tab:linearPCA} and \ref{tab:final}, we find that  using KPCA for such a case improves results of pur, eff$_{\rm A}$, and SC by 25\%, 50\% and 15\% respectively, over the linear PCA outcomes. The dependence of these results with redshift are displayed in figures  \ref{fig:z_diag_cad1_SNR5_bef_only} (for KPCA applied to $D_1+$SNR5) and \ref{fig:zlinear} (for the linear case).

\begin{figure*}
\begin{minipage}[t]{0.45\linewidth}
\centering
\includegraphics[scale=0.65]{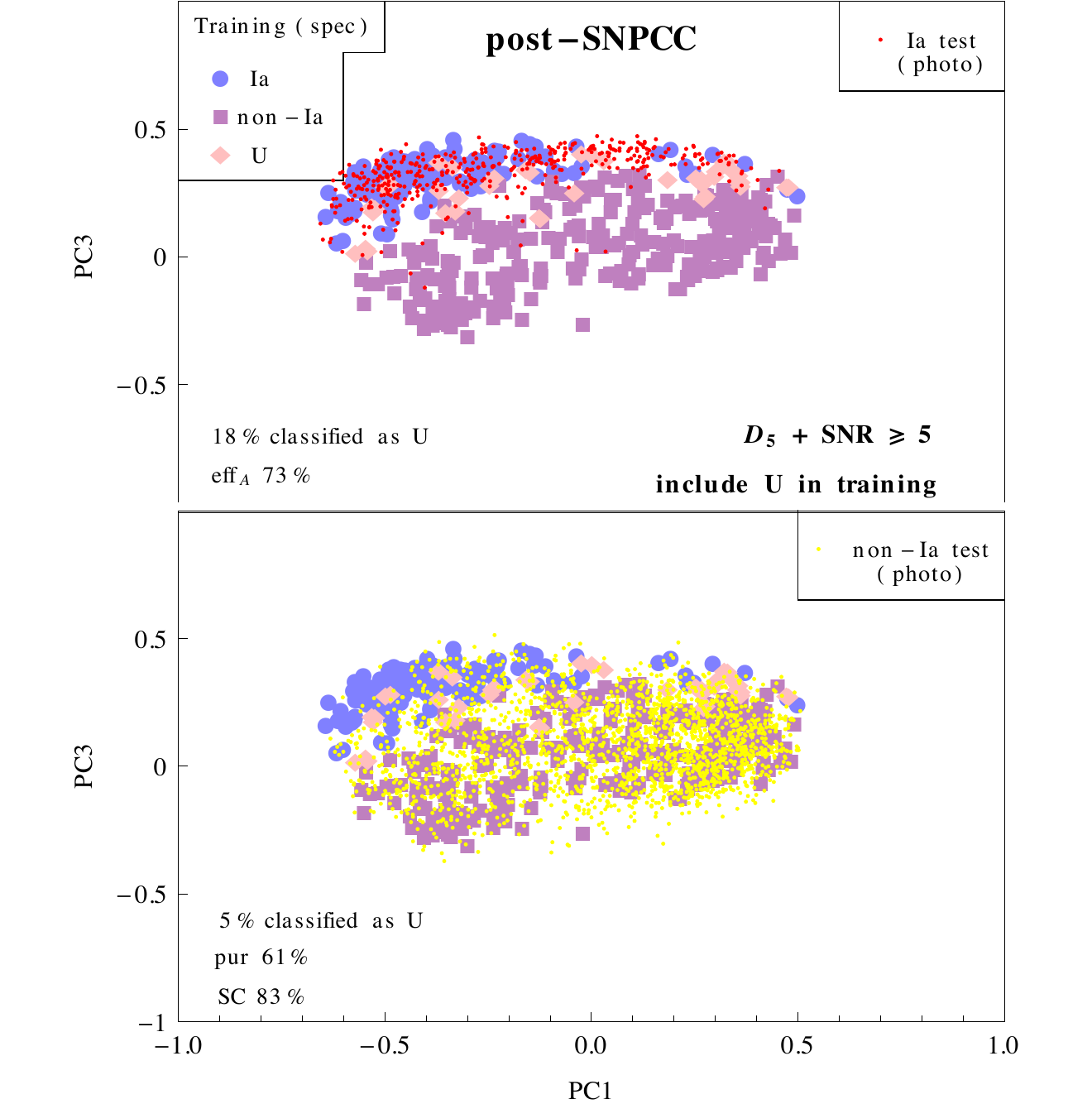}
\caption{Classification results from pre-maximum observations with SNR$\geq$5 ($D_5$+SNR5) and considering $U$ as a classification type. The colour code is the same used in figure \ref{fig:D1SNR5}.}
\label{fig:D5SNR5}
\end{minipage}
\hspace{0.5cm}
\begin{minipage}[t]{0.45\linewidth}
\centering
\includegraphics[scale=0.65]{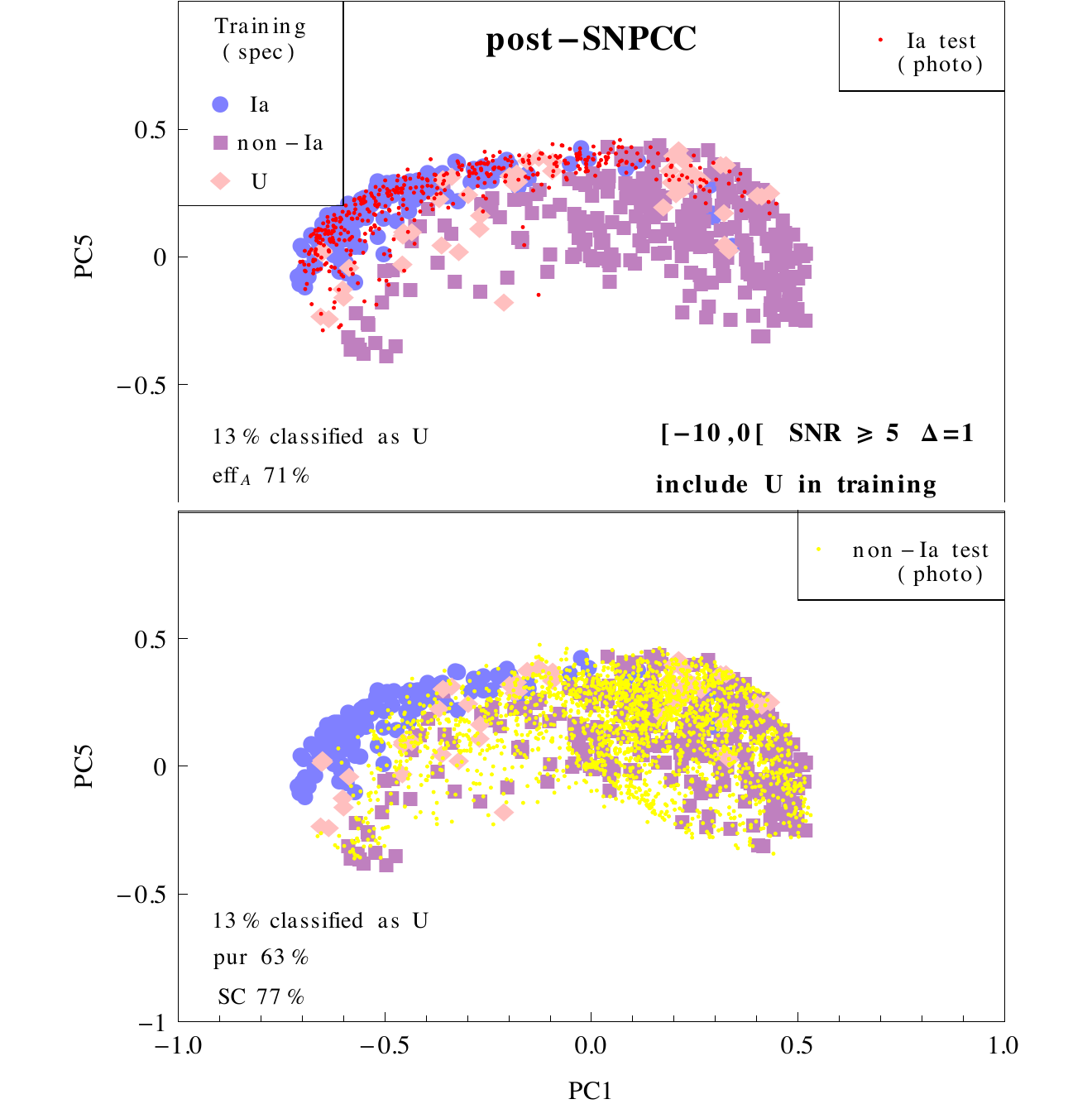}
\caption{Classification results for $t_{\rm low}=-10$, $t_{up}$ being the last point before maximum brightness ($[-10,0[$+SNR5) and considering $U$ as a classification type. The colour code is the same used in figure \ref{fig:D1SNR5}.}
\label{fig:classPMonly}
\end{minipage}
\hspace{1cm}
\begin{minipage}[t]{0.45\linewidth}
\centering
\includegraphics[scale=0.5]{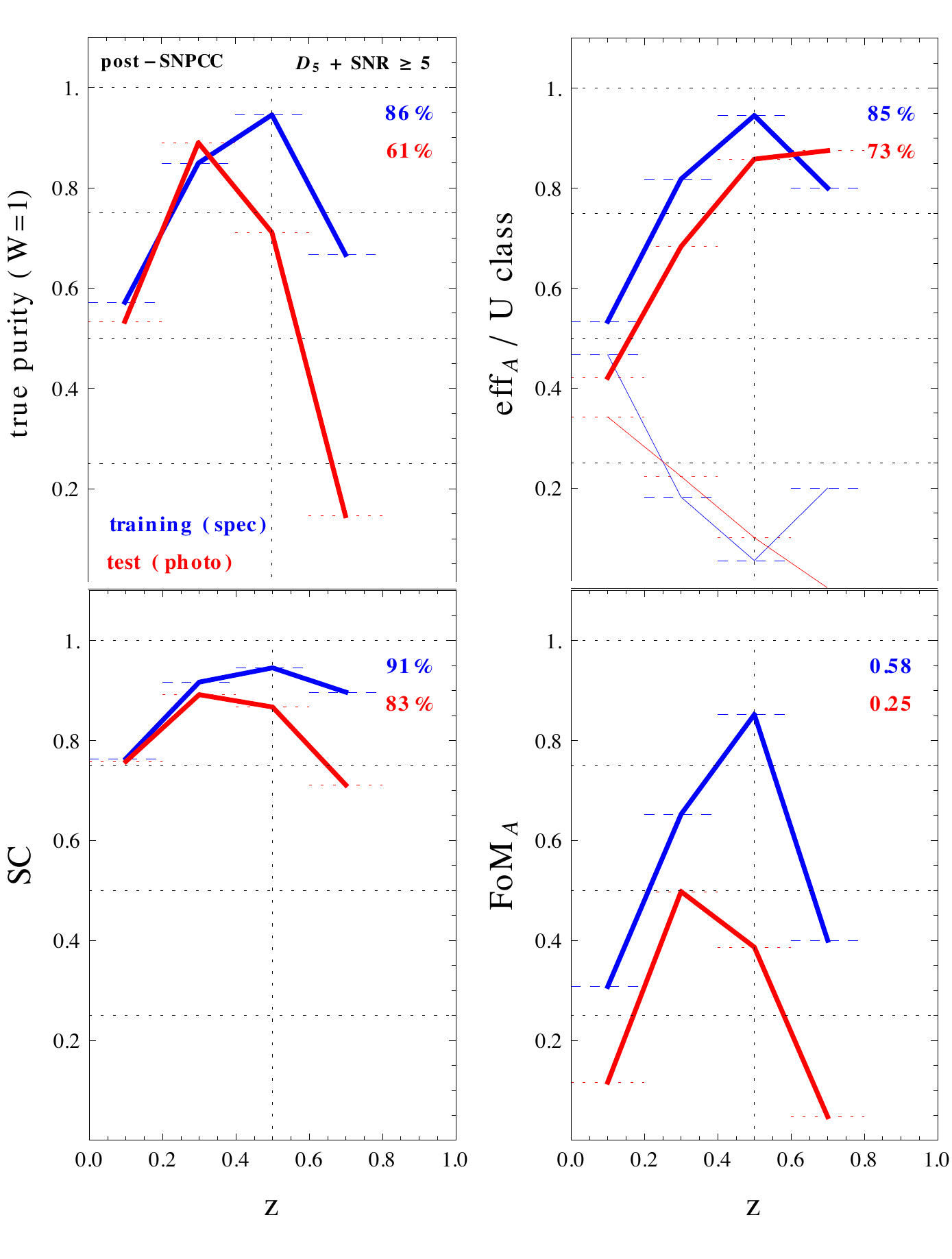}
\caption{Results for pur, eff$_{\rm A}$, SC and FoM as a function of redshift for pre-maximum data ($D_5+$SNR5) and including U class in the training sample. The top right panel also shows the fraction of SNe classified as U. The colour code is the same used in figure \ref{fig:z_diag_cad1_SNR5_bef_only}.}
\label{fig:zcad5}
\end{minipage}
\hspace{0.5cm}
\begin{minipage}[t]{0.45\linewidth}
\centering
\includegraphics[scale=0.5]{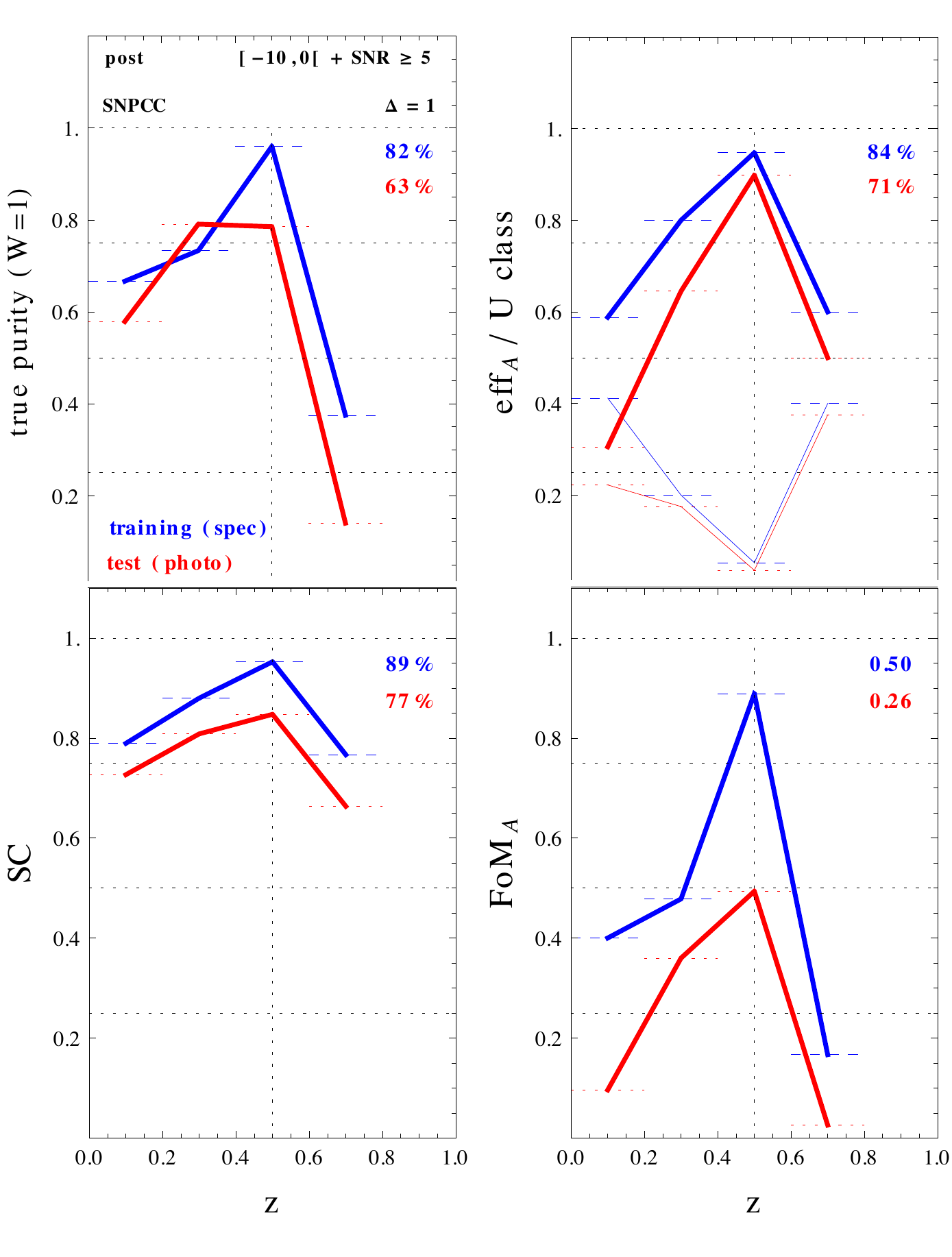}
\caption{Redshift dependence results for $[-10,0[$+SNR5 and including U class in the training sample. The panels show the same quantities described in figure \ref{fig:zcad5}. The colour code is the same used in figure \ref{fig:z_diag_cad1_SNR5_bef_only}.}
\label{fig:zPMonly}
\end{minipage}
\end{figure*}

\subsection{A tougher scenario}
\label{subsec:mlcs}

In order to make a harder test in the classification power of KPCA+1NN, we used MLCS2k2 light curve fitter within SNANA to exclude easily recognizable non-Ia light curves from the test sample. Once the ``obviously" non-Ia are eliminated from the test sample, we were left with a data set containing light curves more similar between each other. If we are able to improve the MLCS2k2 successful classification rates within this sub-sample, we can be sure that the algorithm is doing more than just identifying very strange light curves. We shall see this is the case\footnote{We emphasized that the reader should not consider the classification results using our method and those based on MLCS2k2 in the same grounds. The procedure used to obtain  FitProb values uses information about spectroscopic redshift, and as a consequence, it cannot be considered a photometric classification method.} 
.

We begin by choosing a selection cut. For each light curve surviving this cut we calculated the fit probability of being a SNe Ia (FitProb) as implemented in SNANA. Those with FitProb$>$0.1 were tagged as Ia and the remaining ones were classified as non-Ia. Figure \ref{fig:fitprob} shows the number of SN according to the calculated FitProb for 4 different selection cuts. Beyond the 3 SNR cuts mentioned previously, we also analysed the outcomes of those used by the \textit{SNANA cuts} entry submitted to the SNPCC \citep{Kessler2010b}. These are defined as: at least 1 observation epoch before maximum brightness, at least 1 epoch after +10 days, at least 1 epoch with SNR$\geq$10 and filters \{$r$,$i$\} should have maximum SNR$\geq$5. Panels also show results for pur, ppur, eff$_{\rm A}$, eff$_{\rm B}$, FoM$_{\rm A}$ and FoM$_{\rm B}$ obtained from classifying the entire samples according to FitProb. In this plot, it is evident that, no matter which selection cut we choose, there is a high concentration of SNe with FitProb$<$0.1. This reflects the fact that such group of high quality non-Ia light curves are most obviously  different from standard SNe Ia, and was responsible for a significant part of our SC rates in the previous sub-section.  Analysing the efficiency values, we see that only $\approx 10\%$ of type Ia SNe are wrongly classified as non-Ia according to the FitProb criteria.

We now separate only the SNe classified as Ia according to the FitProb criteria for each  selection cut and consider these our entire test sample. After that, we re-calculated the FitProb results and ran  the KPCA+1NN classifier. For the \textit{SNANA cuts} entry, no extra SNR cuts were applied. Results for different time windows are shown in figure \ref{fig:KPCAmlcsHF}. From this plot, it is evident that, when no SNR cuts are applied, both methods return very similar results for pur. The FoM$_{\rm A}$ obtained from the FitProb criteria is higher than those obtained with our method, due the their maximum efficiency in this context (all SNe tagged as Ia). The main difference appears when results for higher SNR are compared. For SNR$\geq3$, our method is able to increase pur results from pur$\approx 70\%$ to pur$>90\%$ without using any kind of astrophysical information.  

In order to have a better idea of how demanding the time sampling is on the SNe Ia sample which already passed the selection cuts, we show in the bottom panel of figure \ref{fig:KPCAmlcsHF} the fraction of SNe Ia that fulfils such requirements. These results are quite similar and almost independent of selection cuts. For $D_1$ to $D_6$ around $70\%$ of SNe Ia were classifiable  and for $D_7$ and $D_8$ around $60\%$.

\subsection{Pre-maximum observations}
\label{sec:pm}

We also explored the ability of KPCA+1NN to classify light curves given only observation epochs before maximum brightness.
A proposal that was submitted to the participants of the SNPCC but did not received any reply. Although such kind of analysis do not produce a SN sample useful for cosmology, it is extremely important in pointing candidates for spectroscopic follow-up. 

In a first approach, the light curves were treated as described in section \ref{sec:LC_prep}. Once the spline fitted functions and time of maximum were obtained, we constructed the data matrix, $\pmb{G}$, with time sampling between -10 e 0 days since maximum brightness ($D_5$ and $D_6$ in table \ref{tab:cad}). We emphasize that this scenario uses points after maximum in order to determine $t_{\rm max}$, but not in the construction of matrix $\pmb{G}$. The more realistic situation,  where the points after maximum are not used in any step of the process is also analysed bellow.

For $D_5$+SNR5 and $D_5$+SNR0, results are shown in figure \ref{fig:D5SNR5} and \ref{fig:D5SNR0}, respectively. 
Figure \ref{fig:D5SNR5} is similar to figure \ref{fig:D1SNR5}, in the sense that both present a clear separation between Ia and non-Ia points in the training sample and the Ia in the test sample seem to obey that boundary (upper panels). On the other hand, when non-Ia points from the test sample are superimposed, they occupy almost the entire populated region of the parameter space.

In figure \ref{fig:D5SNR0} the situation changes completely. The effect mention previously, describing data vectors corresponding to low information content localized close to the origin in PC space, is translated into an over-density of points in this area. Beyond that, we also see that the difference between the Ia and non-Ia distributions are not that clear any more. There is a slightly tendency of the non-Ia points agglomerate along the vertical axis, but this entire area is also occupied by Ia. The plot also states that the amount of relevant information contained in Ia input vectors is larger than that in non-Ia, since the spread in the first is much larger than the second.  Classification results for $D_5$+SNR5 ($D_5$+SNR0) achieved 61\% (38\%) pur, 73\% (44\%) eff$_{\rm A}$ and 83\% (58\%) SC\footnote{It is important to emphasize that, given the training sample contains much more non-Ia than Ia, a 50\% SC does not correspond to the outcomes of a random decision making process.}, which leads to a FoM$_{\rm A}$ of 0.25 (0.07).

We now turn to a more restrict situation. Although very promising, results for $D_5$ and $D_6$ were not obtained using strictly only pre-maximum data, since the entire light curve was used to determine $t_{\rm max}$ (section \ref{sec:LC_prep}). 
In order to analyse a more realistic scenario, we also studied the classification outcomes when points after maximum are removed from the process of determining $t_{\rm max}$.

For each light curve in the post-SNPCC we took just epochs observed before the simulated time of maximum brightness\footnote{SNANA variable: SIM\_PEAKMJD.}. The spline fit was then applied to these data points and the time of maximum is defined by the \textit{r}-band as before. If in any other filter the last observed data point correspond to an  earlier epoch them $t_{\rm max}$ in $r$-band, we extrapolated the light curve function until it reaches $t_{\rm max}$. We performed classification for $\Delta=1,3$ and in both cases $t_{\rm low}$ was kept as $-10$. After  the curves were obtained, we followed the construction of the data matrix $\pmb{G}$ and the KPCA$+$1NN algorithm as explained before. In what follows, these data sample is tagged as $[-10,0[$.

\begin{figure}
\centering
\includegraphics[scale=0.65]{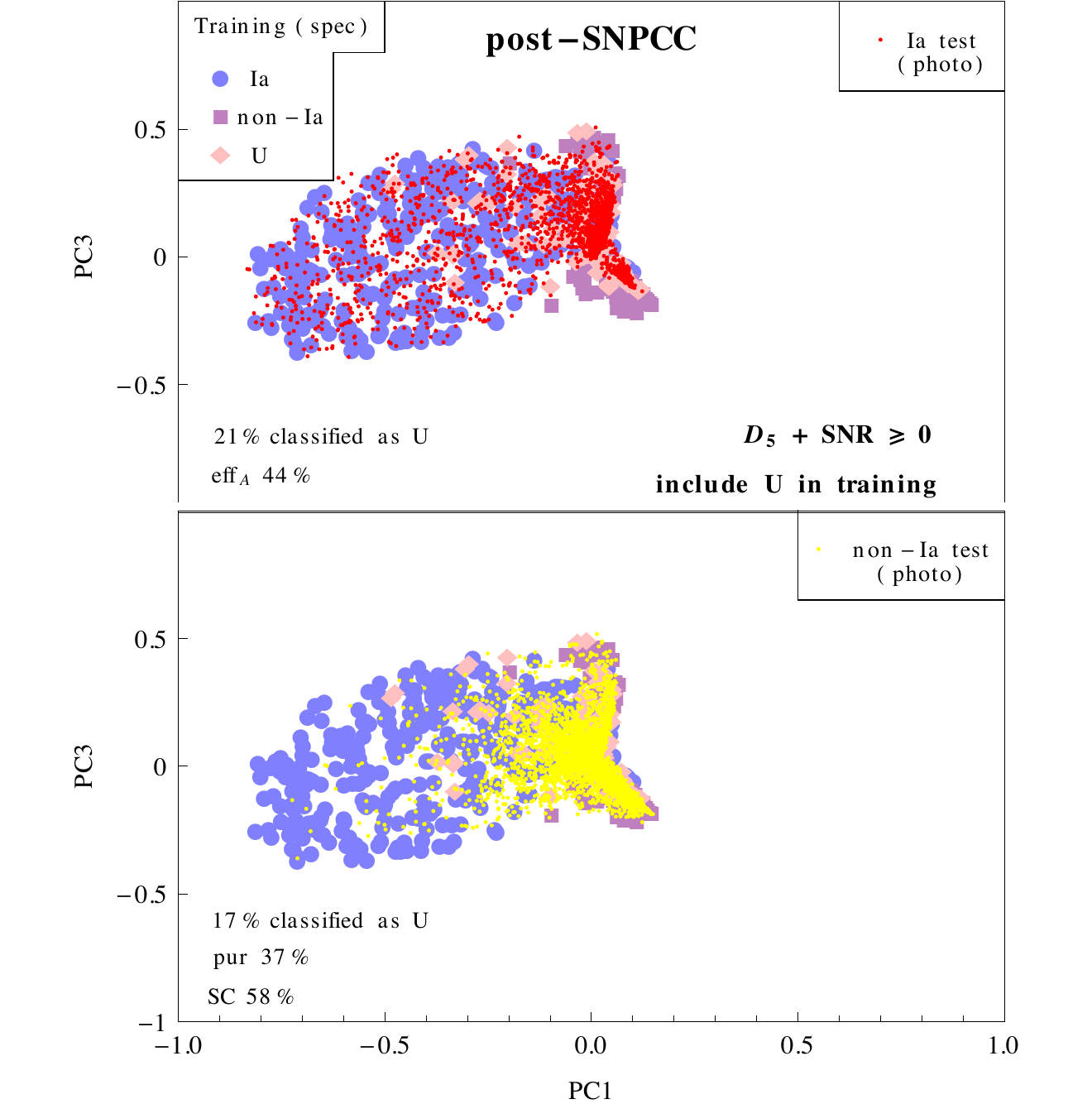}
\caption{Classification results from pre-maximum observations for $D_5$+SNR0. The colour code is the same used in figure \ref{fig:D1SNR5}.}
\label{fig:D5SNR0}
\end{figure}

\begin{figure}
\centering
\includegraphics[trim = 0mm 0mm 0mm 3.5mm, clip, width=1\columnwidth]{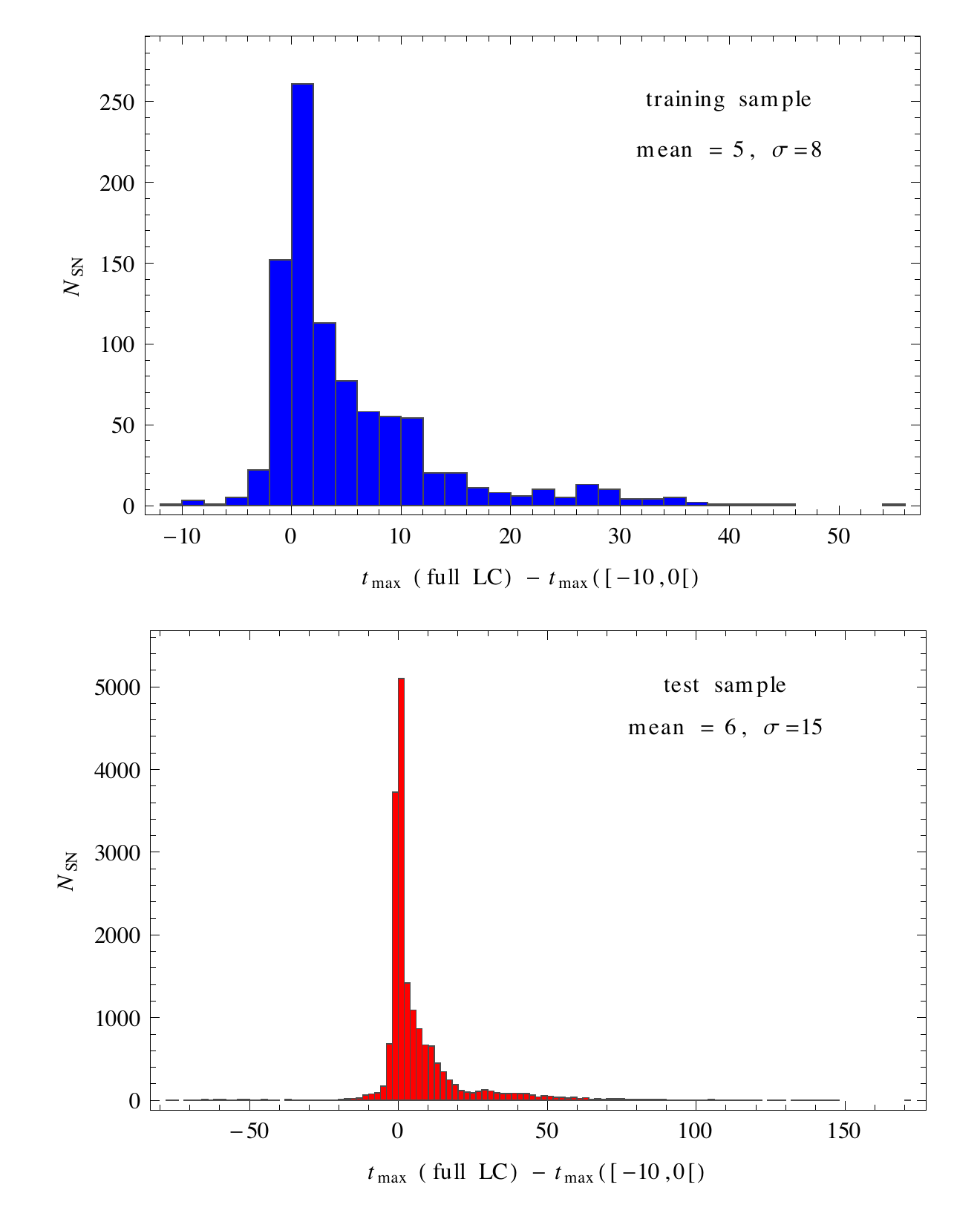}
\caption{Number of SNe as a function of the difference between the time of maximum brightness determined using the full light-curve (samples $D_i$) and using only points before maximum brightness ($[-10,0[$). The upper panel shows histogram for the training sample and the lower panel corresponds to test sample outcomes. }
\label{fig:diffPKMJD}
\end{figure}

Differences between the time of maximum brightness determined using the entire light curve and using only pre-maximum data are shown in figure \ref{fig:diffPKMJD}. Classification results for $[-10,0[$+SNR5 are shown in figure \ref{fig:classPMonly} ($\Delta=1$) and \ref{fig:pmD3class} ($\Delta=3$) and numerical results for other cases are displayed in table \ref{tab:final}. Comparison with results from $D_5$+SNR5 (figure \ref{fig:D5SNR5}) shows that, although  pur and efficiency remain almost unchanged, there is a larger number of non-Ia classified as $U$. The $U$ type SNe, in this case, acts like a barrier between Ia and non-Ia regions, such that expanding non-Ia cover area (adding data a little more noisy) makes them being classified as $U$ before pur levels are diminished. However, this barrier only works up to a certain point.

Classification results for  $D_6$ (figure \ref{fig:D6SNR5}) and $[-10,0[$ with $\Delta=3$ (figure \ref{fig:pmD3class}), both satisfying SNR$\geq$5, reflect this point. The determination of the time of maximum brightness is the only difference between these two data sets, and yet, it is already enough to lower the classification results significantly. A feature that was not verified among the $D_i$ samples (figure \ref{fig:classRES}). This demonstrates the importance of a correct determination of the time of maximum brightness. The redshift dependent results for these 2 instances of the data are displayed in figures \ref{fig:zcad6SNR5} ($D_6$+SNR5) and \ref{fig:zpm2SNR5} ($[-10,0[$+SNR5, with $\Delta=3$). 

\begin{figure*}
\centering
\begin{minipage}[t]{0.45\linewidth}
\includegraphics[scale=0.65]{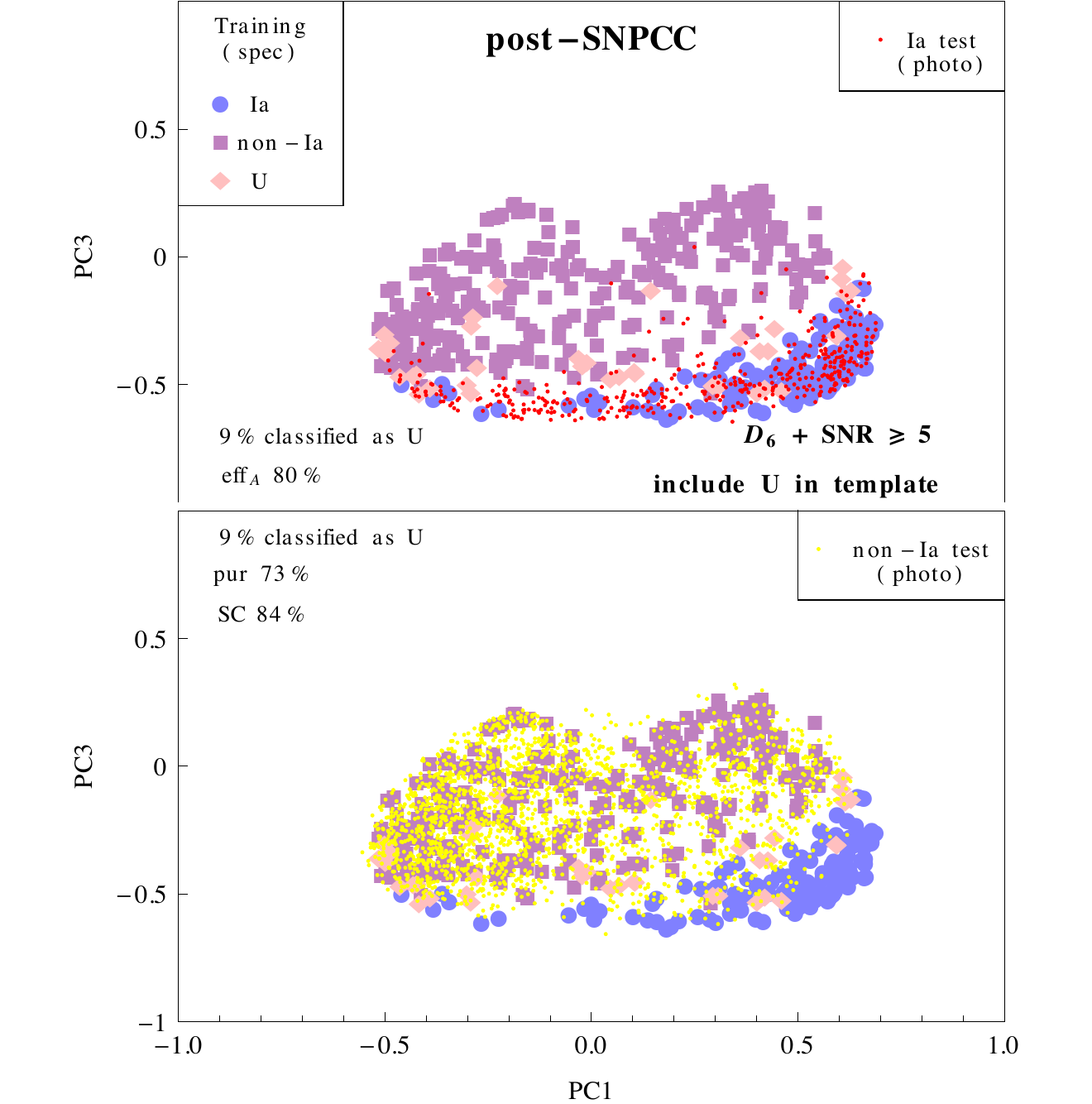}
\caption{Classification results for $D_6$+SNR5. The colour code is the same used in Figure \ref{fig:D1SNR5}. }
\label{fig:D6SNR5}
\end{minipage}
\hspace{0.5cm}
\begin{minipage}[t]{0.45\linewidth}
\includegraphics[scale=0.65]{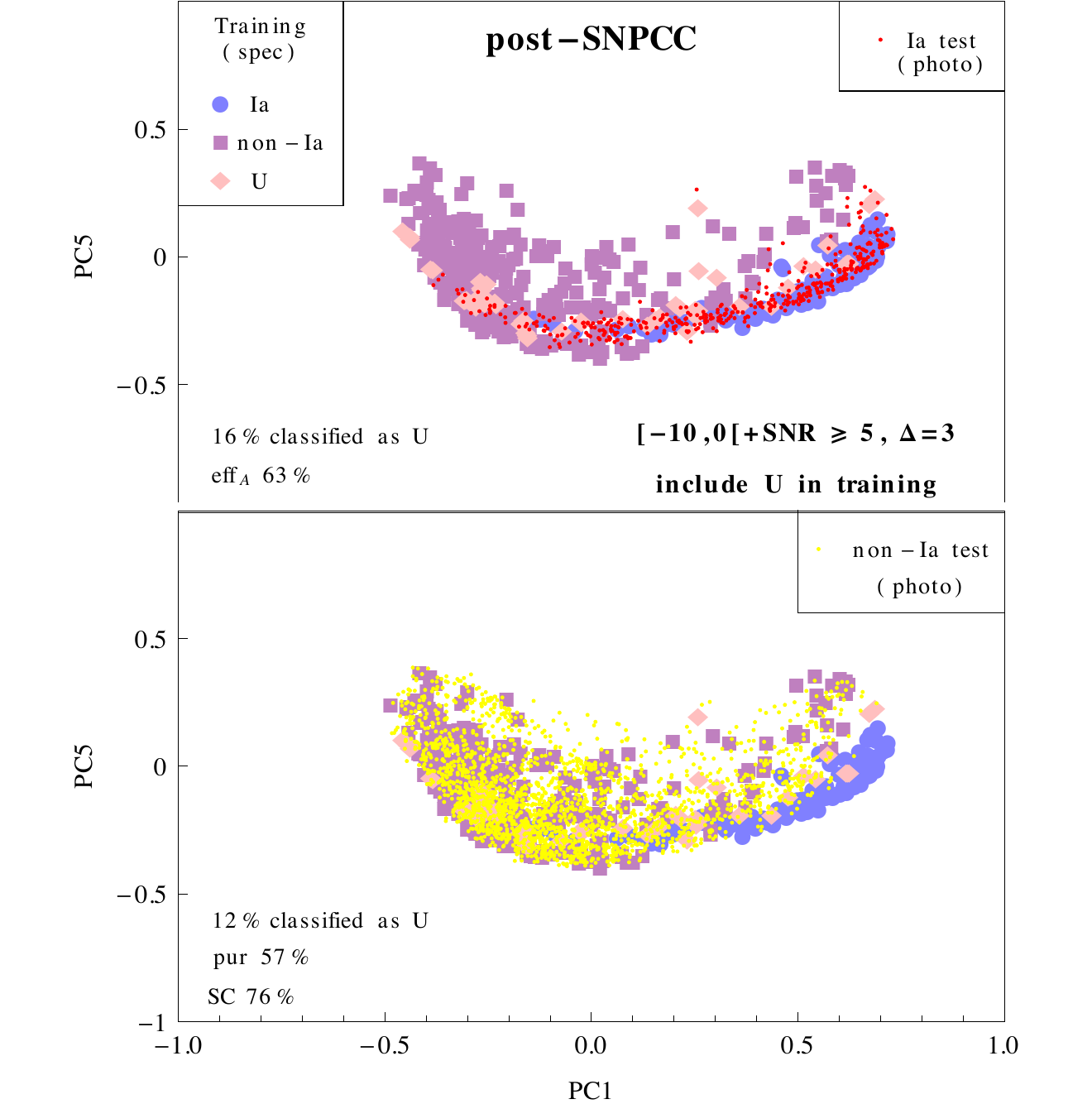}
\caption{Classification results for $[-10,0[$+SNR5 with $\Delta=3$. The colour code is the same used in Figure \ref{fig:D1SNR5}. }
\label{fig:pmD3class}
\end{minipage}
\hspace{1cm}
\begin{minipage}[b]{0.45\linewidth}
\includegraphics[scale=0.55]{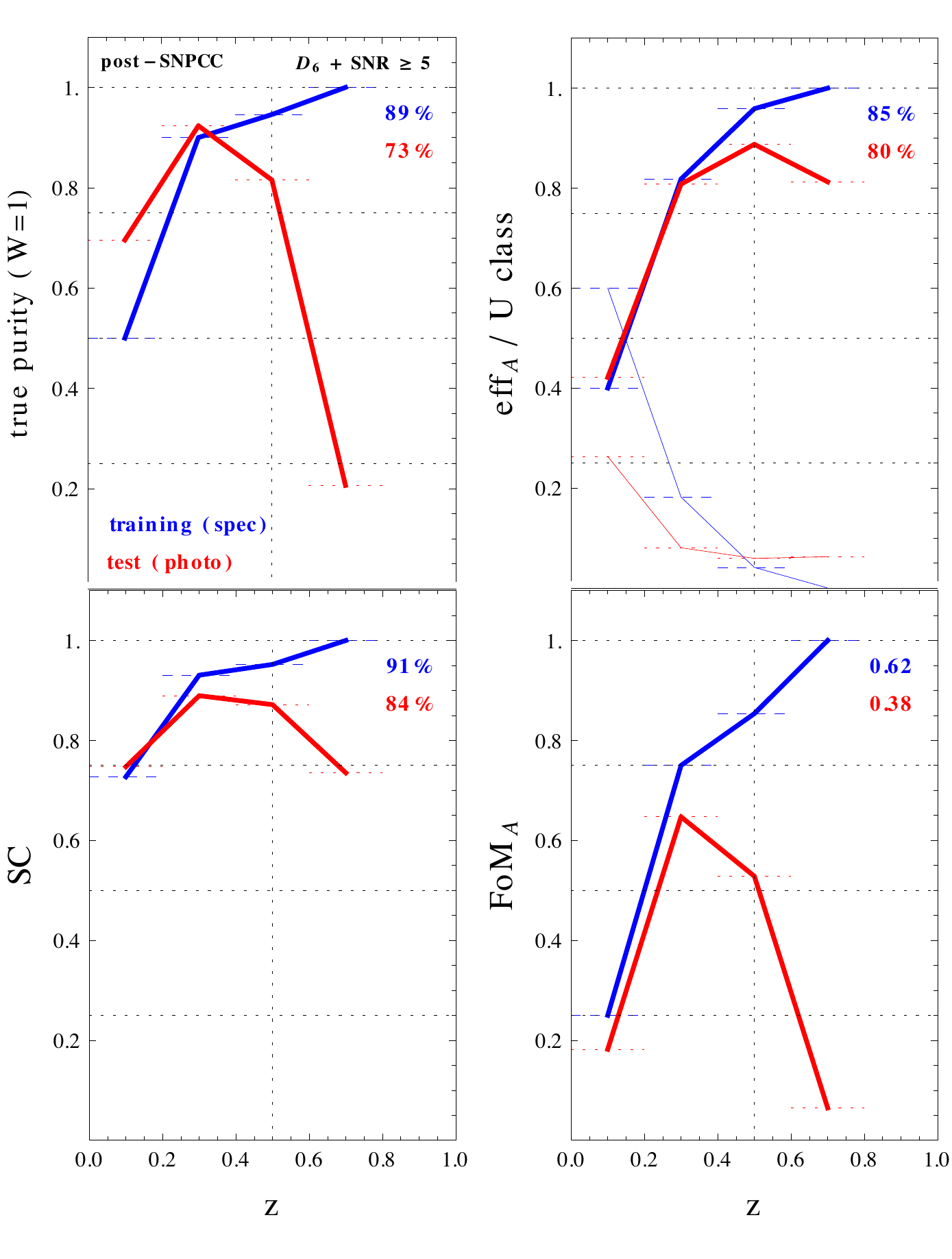}
\caption{Results for eff$_{\rm A}$, pur, FoM and SC as a function of redshift for $D_6+$SNR5. The colour code is the same used in Figure \ref{fig:z_diag_cad1_SNR5_bef_only}.}
\label{fig:zcad6SNR5}
\end{minipage}
\hspace{0.5cm}
\begin{minipage}[b]{0.45\linewidth}
\includegraphics[scale=0.55]{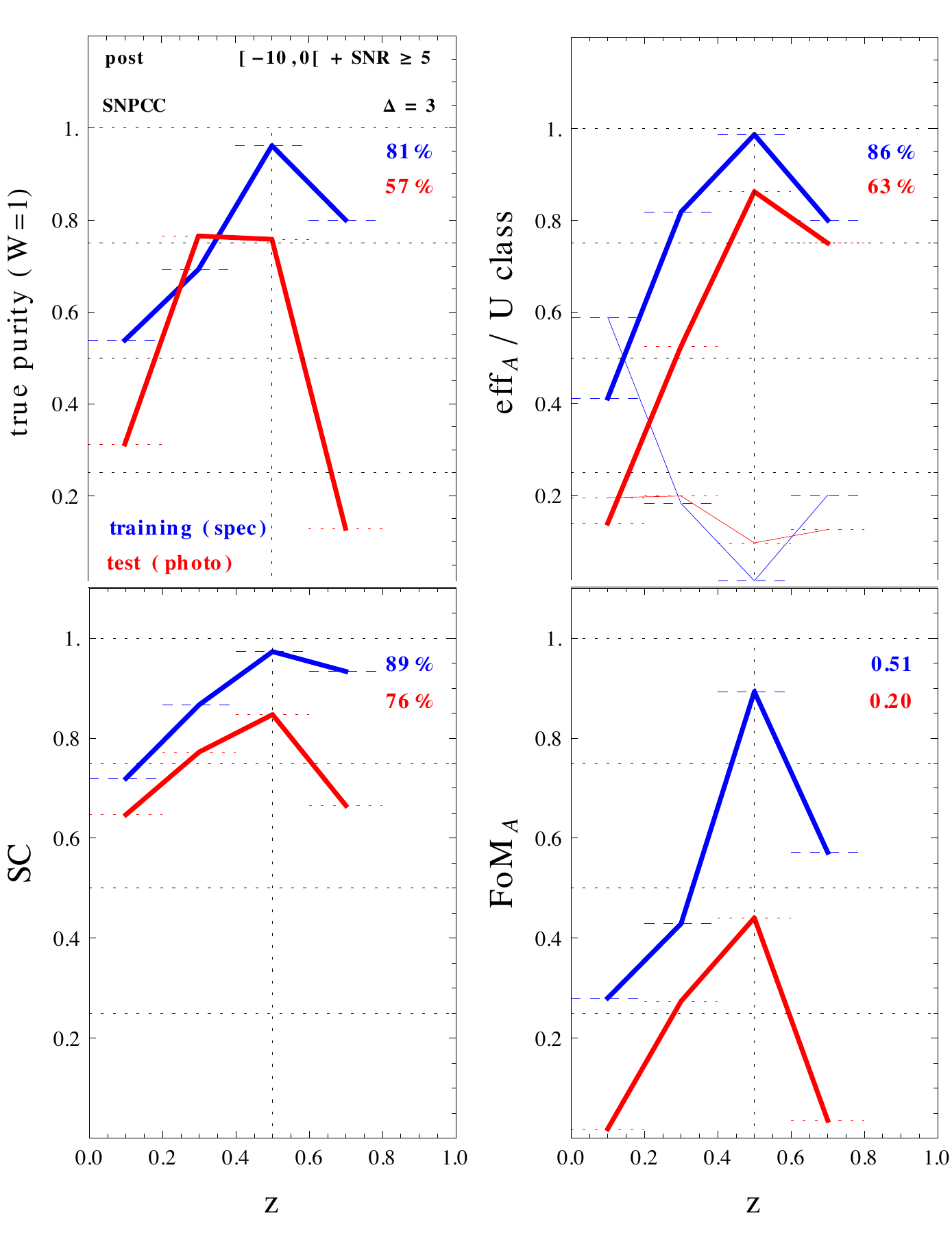}
\caption{Results for eff$_{\rm A}$, pur, FoM and SC as a function of redshift $[-10,0[+$SNR5 and $\Delta=3$. The colour code is the same used in Figure \ref{fig:z_diag_cad1_SNR5_bef_only}.}
\label{fig:zpm2SNR5}
\end{minipage}
\end{figure*}

These results are very encouraging. It means that, in the context of future DES data, the algorithm can correctly classify approximately $75\%$ of the initial data sample using only pre-maximum data, if the entire data set was given at once. But in a real situation this can be improved. Suppose that initially, our training sample is composed by the spectroscopic SNe sample available today. As time goes by and  pre-maximum light curves are observed, they are automatically classified. An example strategy would be to target with spectroscopic observations the light curves whose projections in PC feature space lay in the boundaries of the SNe Ia/non-Ia regions. Once the SNe type is confirmed, it can be added to the training sample, improving future classification results.
\\

\section{SNPCC sample}
\label{sec:SNPCC}

In order to allow a direct comparison of our results with those reported in the SNPCC,  we also applied the KPCA+1NN algorithm to the data set used in the competition. This consists of 20216 simulated light curves of which 1105 represent the spectroscopic sample. This data can be consider less likely to represent the future DES data, given that all bugs listed as fixed ``after SNPhotoCC" in table 4 of \citet{Kessler2010b} are still part of this sample. However,  the application is instructive to have an idea of how our method performs when faced to other algorithms. 

Results for FoM$_{\rm B}$, eff$_{\rm B}$, ppur (W=3) and pur (W=1) are shown in figure \ref{fig:SNPCC} for different SNPCC sub-samples and SNR cuts. This should be compared to figure 5 of \citet{Kessler2010b}, which reports results from different classifiers without using host galaxy photometric redshift. A detailed analysis of the multiple panels in figure \ref{fig:SNPCC} is presented in appendix \ref{ap:SNPCC_comp}.

Our findings from this sample can be summarized through the items bellow:
\begin{itemize}
\item There is a weak dependence of the overall classification results with particular time sampling choices. The only eye-catching difference comes from time window including the second maximum in  the infrared (D$_7$).
\item Results are highly dependent on SNR cuts, specially efficiency and consequently, FoM.
\item D$_7$+SNR5 achieved FoM$_{\rm B}$ $>$ 0.25 for 0.2$\leq$ z $\leq$ 0.4. A result only achieved by 3 of the entries participating on the SNPCC (namely Sako, JEDI-KDE and SNANA).
\item Our method achieved outstanding pur and ppur results for z$\geq$0.2. In this redshift range, all samples with SNR$\geq$5 reported pur values larger than 75\%: a result that was not obtained by \textit{none} of the SNPCC entries. Particularly, in 0.2$\leq$z$<$0.4,  D$_7$+SNR$\geq$5 obtained 94\% $\leq$ pur $\leq$ 97\%, while keeping a moderate FoM$_{\rm B}$. The redshift dependence of these results are displayed in figure \ref{fig:cad7_SNR5}.
\end{itemize}

\begin{figure*}
\centering
\includegraphics[scale=0.6]{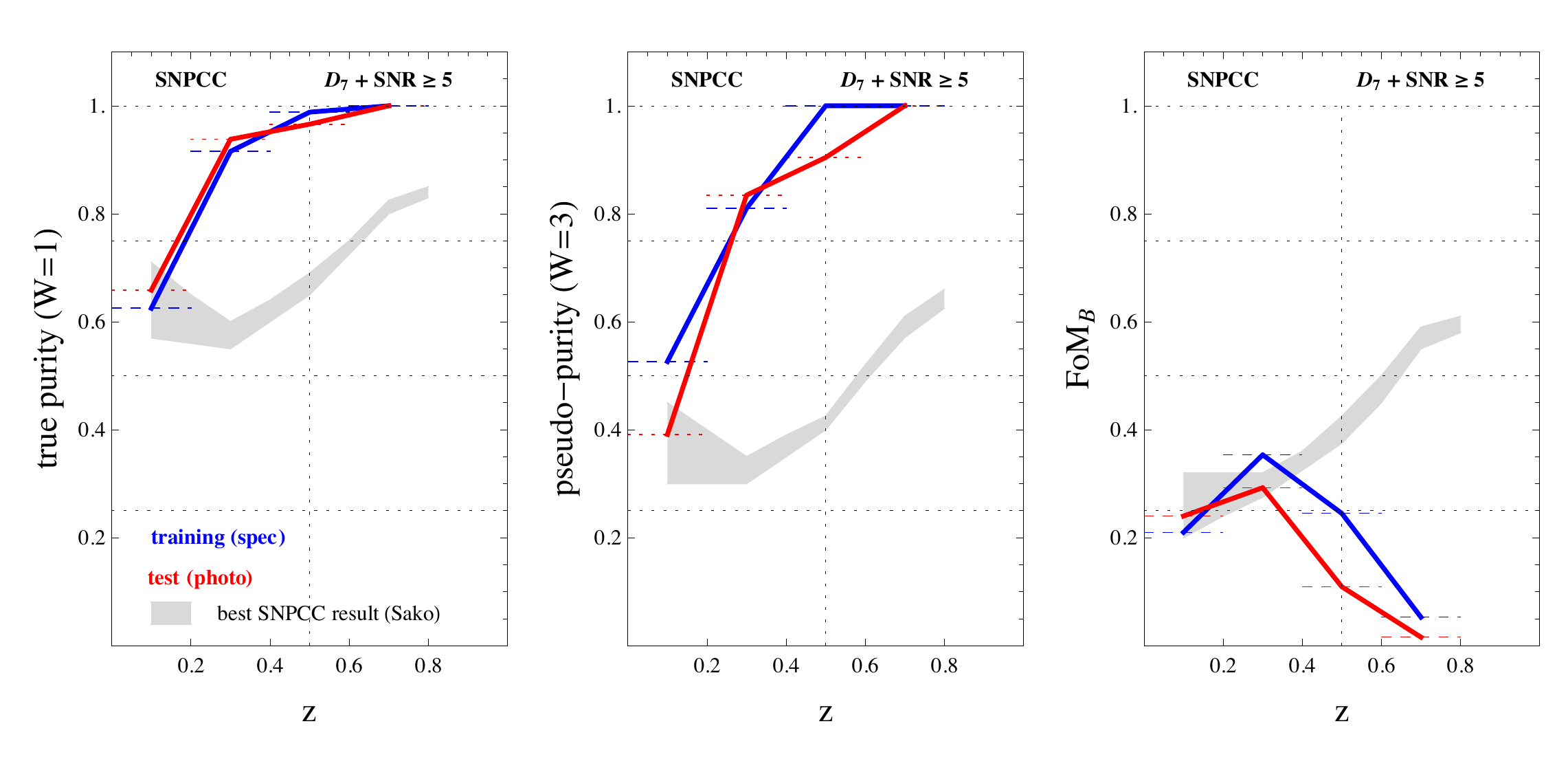}
\caption{Classification results for D$_{7}$+SNR5 from the SNPCC sample (original SNPCC data set) compared to results reported by the group achieving highest FoM in the SNPCC (Sako). Panels show true purity, pseudo-purity and FoM$_{\rm B}$ from left to right. Blue (red) lines correspond to results from KPCA+1NN when applied to spectroscopic/training (photometric/test) samples. Gray region correspond to results reported by the group which achieved the best overall classification results in the SNPCC, without using host galaxy  photometric redshift information \citep{Kessler2010b}.  }
\label{fig:cad7_SNR5}
\end{figure*}

\section{Conclusion}	
\label{sec:discuss}

Current SNe surveys already have at hand  much more SNe light curves than it is possible to spectroscopically confirm. This situation will increase tremendously in the next decade, which makes SNe Ia photometric identification a crucial issue. In this work, we propose the use of KPCA combined with
k = 1 nearest neighbour algorithm (KPCA+1NN) as a framework
for SNe photometric classification. 

Lately, a large effort has been applied to the SNe photometric classification problem. An up to date compilation of those efforts is reported in  \citet{Kessler2010}, known as the \textit{SuperNova Photometric Classification Challenge} (SNPCC). It consisted of a blind simulated light curve sample as expected for the \textit{Dark Energy Survey} (DES) to be used as a test ground for different classifiers. Although there were some fundamental differences between the algorithms submitted, none of the entries performed obviously better than all the others. After the results were reported, the organizers made public an updated version of the simulated data (post-SNPCC). Both samples, SNPCC and post-SNPCC were analysed in this work.

Our method fit in
the class of statistical inference algorithms, according to the SNPCC nomenclature. 
All calculations are done in the observer frame. There is no corrections
due to reddening, local environment, redshift or observation conditions and all available spectroscopically
confirmed data surviving quality selection cuts should be used to shape the PCs feature space. The
dimensionality reduction is performed using only spectroscopically
confirmed SNe (training sample) and each new unlabelled light curve (test sample) is
classified one at a time. This allow us to avoid introducing noisy information from non-confirmed SNe in the classifier training. The algorithm is built so that once a new spectroscopic light curve is available or we have total confidence in a photometric one, it can easily be included in the training process, but it is not necessary to redefine the PC feature space every time a new point is to be classified.

In designing our method, we prioritize purity in the final SNe Ia sample, once it is the most important characteristic of a data set to be use for cosmology. We also decided to take a conservative approach towards the unknown features of the data. As a consequence, no extrapolation on time or wavelength domain was used and we demanded that each SNe was observed in all available filters. As expected, these choices have a great impact in our efficiency results. However, we believe that the high purity levels achieved justifies our choices (figure \ref{fig:SNPCC}), specially in a context where there are already observed light curves not being used for cosmology due to lack of classification \citep{Sako2011}.  

We highlight that we chose not to include high complexity
in the different steps along the process in order to
keep focus in the KPCA performance. Although, as remarked
before, there is plenty of room for improvement. For
example, in choosing the kernel function, the nearest neighbour
algorithm degree and studying more flexible selection cuts. Such developments are worth pursuing, but one should also be aware not to fine tuning the procedure too much, so the results will apply only to one specific data set. Quantifying the dependence of our results with such change of choices is out of the scope of this work.  

Results presented in this work show that KPCA+1NN algorithm provide excellent purity in the final SNe Ia sample. Although a time window since maximum brightness needs to be defined, its width does not have a large impact in final classification results. On the other hand, SNR of each observation epoch plays a crucial role. As a consequence, our best results are mainly concentrated in the intermediate range, 0.2$\leq$z$\leq0.4$. From the SNPCC sample analysis in these redshifts, our method returned FoM$_{\rm B}$ $>$ 0.25, using D$_7$+SNR5 (figure \ref{fig:cad7_SNR5}). A result only achieved by 3 of the entries participating on the SNPCC (namely Sako, JEDI-KDE and SNANA).

We also found  outstanding purity and pseudo-purity results. All samples with SNR$\geq$5 reported purity values larger than 75\% for z$\geq$0.2: a result that was not obtained by \textit{none} of the SNPCC entries. Particularly, for 0.2$\leq$ z $\leq$ 0.4, D$_7$+SNR$\geq$5 obtained 94\% $\leq$ pur $\leq$ 97\%, while keeping a moderate FoM (figure \ref{fig:cad7_SNR5}).

Among the entries submitted to the SNPCC, only the InCA group used a similar approach, although by means of completely different techniques. The results they reported to the competition provide purity rates similar the ones we get for SNR$\geq0$.

We stress that, although the comparison with the SNPCC results is important, it cannot be considered exactly in the same grounds as our results. First because since they were built with different purposes (the SNPCC aimed at maximum FoM$_{B}$ and our goal was to achieve the highest possible purity while maintaining a reasonable FoM), second because we were not time constrained as the groups taking the challenge and finally, we had access to the answer key before hand. Something the competitors did not have. However, a strictly direct comparison with other results in the literature is possible through the post-SNPCC sample.

Recently, the InCA group made public a detailed analysis of the results achieved by their method when applied to the post-SNPCC data set \citep{Richards2011} (R2012). The two algorithms provided similar classification results. Both achieving local maximum of FoM$_{\rm B}$ around 0.5, with our method giving better results at lower and theirs at higher redshifts. Averaging over the entire redshift range, we achieve FoM$_{\rm B}$ of 0.06 and R2012 reported 0.35. R2012 also provides results with different spectroscopic samples, constructed by re-distributing DES available follow-up time. In their result with highest purity, they reported 90\% purity, 8\% eff$_{\rm B}$ and 0.08 FoM$_{\rm B}$ using a redshift limited spectroscopic sample. Our method provides 96\% purity, 6\% eff$_{\rm B}$ and 0.06 FoM$_{\rm B}$ for $D_7$+SNR$\geq$5.

\citet{karpenka2012} also present results from post-SNPCC data. In their analysis, results from a parametric fit to the spectroscopic light curves are used to train a neural network which subsequently returns the probability of a new object being a Ia.  Using 50\% of the initial sample as a training set ($\approx$10000 objects considered spectroscopically confirmed), they found  80\% purity, 85\% eff$_{\rm B}$ and 0.51 FoM$_{\rm B}$. 

It is important to emphasize that the results we report above
were achieved using a sub-set of the spectroscopic sample \textit{as it is given within the post-SNPCC data}.
This means that it is not necessary to tailor the spectroscopic
sample \textit{a priori} in order to get high purity results,
making our method ideal as a first approach to a large photometric data set.

In order to test the algorithm in a more restrictive scenario, we present results obtained from the post-SNPCC sub-sample with 
MultiColor Light-curve Shape
(MLCS2k2) fit probability, FitProb$>0.1$. This sample contains light curves very similar between each other, and represents a more difficult classification challenge than the complete SNPCC data. We show
that our method is not able to do more than identifying the obviously non-Ia light curves when no SNR cuts are applied. However, when we compare results from data samples with SNR$\geq$3, KPCA+1NN  can boost purity levels to $>95\%$ independently of time window sampling.

Finally, we report the first attempt in classifying the post-SNPCC data using only pre-maximum epochs. This study is very
important in selecting candidates for spectroscopic follow
up. Using only data between -10 and 0 days since maximum brightness,
we obtained 63\% purity, 71\% eff$_{\rm A}$,  77\% SC and FoM$_{\rm A}$ of 0.26.
This is a very enthusiastic result and reflects the
vast room for improvement this kind of analysis may provide
in different stages of the pipeline.

We stress that the application proposed here is merely an example of how the KPCA$+$kNN algorithm might be applied in astronomy. Beyond the specific problem of SNe Ia photometric classification, the same procedure can be used to identify other expected transient sources and even to spot still non-observed objects among a large and heterogeneous data set. The projection of such objects in PCs feature space would occupy a previously non-populated \textit{locus}, what would give us a hint to further investigate that  particular object. In the more ideal scenario, when synthetic light curves from a non-observer object is available, a synthetic target can be included in the training sample, leading to a detection tailored according to our expectations. This provides still another advantage over template fitting techniques, which deserve further investigation.

From what was presented here, we conclude that the decision of choosing one method over the other is not a straightforward one, but must be balanced by the characteristics of the data available and our  goal in classifying it. Given that SNe without spectroscopic confirmation  is not a future issue of large surveys, but a problem that is already present in the SDSS data \citep{Sako2011}, KPCA+1NN algorithm proved to be the ideal choice to quickly increase the number of SNe Ia available for cosmology with minimum contamination. Alternatively, it can also be used as a complement to other techniques in helping to increase the number of SNe Ia in the training sample. Either way, we have enough evidence to trust the competitiveness of our algorithm within the current status of the SNe photometric classification field.

\section*{Acknowledgements}

We thank Masaomi Tanaka, Naoki Yoshida, Takashi Moriya, Laerte Sodr\'e Jr, Andrea Ferrara, Andrei Mesinger and Rick Kessler for fruitful discussions and suggestions. We also thank the anonymous referee for comments that highly improved the quality of the paper. The authors are happy to thank the Institute for the Physics and Mathematics of the Universe (IPMU), Kashiwa, Japan, Scuola Normale Superiore (SNS), Pisa, Italy, Centro Brasileiro de Pesquisas Fisicas (CBPF), Rio de Janeiro, Brazil and the Asia Pacific Center for Theoretical Physics (APCTP), Pohang, South Korea, for hosting during the development of this work. RSS thanks the Excellence Cluster Universe Institute, Garching, Germany, for hosting while this work was developed. The authors acknowledge  financial support from the Brazilian financial agency FAPESP through grants number 2011/09525-3 (EEOI) and 2009/05176-4 (RSS). EEOI thanks the Brazilian agency CAPES for financial support (1313-10-0). RSS thanks the Brazilian agency CNPq for financial support (200297/2010-4).

\appendix
\section{Basic proofs}
\label{ap:proof}

This appendix contain basic proofs for the statements used throughout the text. These are common to machine learning theory field, but may not be as such for the astronomy community. They follow closely Max Welling's notes \textit{A first encounter with Machine Learning} and \citet{Scholkopf1996}, which the reader is advised to check for a comprehensible introduction to the basic concepts used here.

\begin{enumerate}
\item \textit{All the vectors in the eigenvector space $V$ lie in the space spanned by the data vectors contained in $X$}

Consider $\mathbf{v}_a \in V$,
\begin{eqnarray}
\lambda_a \mathbf{v}_a&=&C \mathbf{v}_a=\frac{1}{N}\sum_{i=1}^N\mathbf{x}_{i}\mathbf{x}_{i}^T\mathbf{v}_a=\frac{1}{N}\sum_{i=1}^N\left(\mathbf{x}_{i}^T\mathbf{v}_a\right)\mathbf{x}_{i}\nonumber\\
&\Rightarrow&\nonumber\\
\mathbf{v}_a&=&\sum_{i=1}^{N}\left[\frac{\left(\mathbf{x}_{i}^T\mathbf{v}_a\right)}{N\lambda_a}\right]\mathbf{x}_{i}=\sum_{i=1}^N\alpha_i\mathbf{x}_{i}.
\end{eqnarray} 

In other words, any eigenvector can be written as a linear combination of the vectors in $X$ and, as a consequence, must lie in the space spanned by them.\\

\item  \textit{Determining equation (\ref{eq:eigenK})}

Consider the projected eigenvalue equations,
\begin{equation}
\mathbf{x}_{i}^T C \mathbf{v}_a=\lambda_a\mathbf{x}_{i}^T\mathbf{v}_a.
\end{equation}
Using equations (\ref{eq:covmatrix}) and (\ref{eq:vlambda}), we have
\begin{eqnarray}
\mathbf{x}_{i}^T\frac{1}{N}\sum_{j=1}^N\mathbf{x}_{j}\mathbf{x}_{j}^T\sum_{k=1}^N\alpha_k^a\mathbf{x}_{k}&=&\lambda_a\mathbf{x}_{i}^T
\sum_{l=1}^N\alpha_l^a\mathbf{x}_{l}\\
\frac{1}{N}\sum_{j,k}\alpha_k^a\left[\mathbf{x}_{i}^T\mathbf{x}_{j}\right]\left[\mathbf{x}_{j}^T\mathbf{x}_{k}\right]&=&\lambda_a
\sum_{l=1}^N\alpha_l^a\left[\mathbf{x}_{i}^T\mathbf{x}_{l}\right].\nonumber
\end{eqnarray}
Addressing $K_{ij}=\left[\mathbf{x}_{i}^T\mathbf{x}_{j}\right]$, we can write
\begin{equation}
K\pmb{\alpha}^a=\tilde{\lambda_a} \pmb{\alpha}^a \qquad {\rm where} \qquad \tilde{\lambda}=N\lambda_a.
\end{equation}\\

\item \textit{Determination of $||\pmb{\alpha}^a||$}\\
The norm of parameters $\pmb{\alpha}^a$ is a consequence of the normalization of the eigenvectors in $V$.Using equation (\ref{eq:vlambda}),
\begin{eqnarray}
\mathbf{v}_a^T\mathbf{v}_a=1 \qquad &\Rightarrow& \qquad \sum_{i,j}\alpha_i^a\alpha_j^a\left[\mathbf{x}_{i}^T\mathbf{x}_{j}\right]=\left(\mathbf{\alpha}^a\right)^T K \mathbf{\alpha^a}=1\nonumber\\
&\Rightarrow & \qquad N\lambda_a\left(\pmb{\alpha}^a\right)^T\pmb{\alpha}^a=1\nonumber\\
&\Rightarrow & \qquad ||\pmb{\alpha}^a||=\frac{1}{\sqrt{N\lambda_a}}.
\end{eqnarray}\\

\item \textit{Obtaining $K_F$ and $\pmb{\alpha}_{\Phi}$}

We begin with the definition of the covariance matrix in feature space
\begin{equation}
C_{\rm F} = \frac{1}{N}\sum_{i=1}^N\Phi(x_i)\Phi(x_i)^T,
\end{equation}
we have to find the eigenvalues, $\lambda_{\Phi}$, and eigenvectors, $\mathbf{v}_{\Phi}$, which satisfy
\begin{equation}
\lambda_{\Phi}\mathbf{v}_{\Phi}=C_{\rm F}\mathbf{v}_{\Phi}.
\end{equation}
Using item (ii) above, we have that all $\mathbf{v}_{\Phi}$ can be written as a linear combination of the $\Phi$'s. This means that we are allowed to consider the equivalent equations
\begin{equation}
\lambda_{\Phi}\left(\Phi(\mathbf{x}_k)\cdot \mathbf{v}_{\Phi}\right)=\left(\Phi(\mathbf{x}_k)\cdot  C_{F}\mathbf{v}_{\Phi}\right), \qquad \forall k,
\label{eq:projeigen}
\end{equation}
with the prescription that
\begin{equation}
\mathbf{v}_{\Phi}=\sum_{i=1}^N\alpha_{\Phi}^i\Phi(\mathbf{x}_i).
\label{eq:expphi}
\end{equation}
Using equations (\ref{eq:projeigen}) and (\ref{eq:expphi}), 
\begin{eqnarray}
&&\lambda_{\Phi}\sum_{i=1}^N\alpha_{\Phi}^i\left(\Phi(\mathbf{x}_k)\cdot\Phi(\mathbf{x}_i)\right)=\nonumber\\
&=&\frac{1}{N}\sum_{i=1}^N\alpha_{\Phi}^i\left(\Phi(\mathbf{x}_k)\cdot\sum_{j=1}^N\Phi(\mathbf{x}_j)\right)\left( \Phi(\mathbf{x}_j)\cdot\Phi(\mathbf{x}_i)\right).
\end{eqnarray}
Calling
\begin{equation}
\left(K_F\right)_{ij}:=\left(\Phi(\mathbf{x}_i)\cdot\Phi(\mathbf{x}_j)\right),
\end{equation}
leads to
\begin{equation}
N\lambda_{\Phi}K_F\pmb{\alpha}=K_F^2\pmb{\alpha},
\end{equation}
where $\pmb{\alpha}$ is a column vector. As $K_F$ is symmetric, 
\begin{equation}
K_F\pmb{\alpha}=\widetilde{\lambda_{\Phi}}\pmb{\alpha},
\label{eq:eigphi}
\end{equation}
with $\widetilde{\lambda_{\Phi}}=N\lambda_{\Phi}$. In order to obtain $\pmb{\alpha}_{\Phi}$, we only need to diagonalize $KF$.

The normalization of $\pmb{\alpha}_{\Phi}$ is achieved by requiring
\begin{equation}
\left(\mathbf{v}_{\Phi}^k\cdot\mathbf{v}_{\Phi}^k\right)=1, \qquad \forall k.
\end{equation}
Through equations (\ref{eq:expphi}) and (\ref{eq:eigphi}) this converts into
\begin{eqnarray}
1&=&\sum_{i,j=1}^N\left[\alpha_{\Phi}^k\right]_i\left[\alpha_{\Phi}^k\right]_j\left(\Phi(\mathbf{x}_i)\cdot \Phi(\mathbf{x}_j)\right)\nonumber\\
&=&\sum_{i,j=1}^N\left[\alpha_{\Phi}^k\right]_i\left[\alpha_{\Phi}^k\right]_j {K_F}_{ij}\nonumber\\
&=&\left(\pmb{\alpha}^k\cdot K_F \pmb{\alpha}_{\Phi}^k\right)\nonumber\\
&=&\lambda_{\Phi}^k\left(\pmb{\alpha}_{\Phi}^k\cdot\pmb{\alpha}_{\Phi}^k\right).
\end{eqnarray}\\

\item \textit{Centralization in feature space}

Considered the centred vectors in feature space
\begin{equation}
\widetilde{\Phi}(\mathbf{x}_i):=\Phi(\mathbf{x}_i)-\frac{1}{N}\sum_{i=1}^N\Phi(\mathbf{x}_i),
\label{eq:centfeat}
\end{equation}
our goal now is to define the dot product matrix 
\begin{equation}
\widetilde{K_F}_{ij}=\widetilde{\Phi}(\mathbf{x}_i)^T\widetilde{\Phi}(\mathbf{x}_j).
\label{eq:centeigen}
\end{equation}
In a procedure similar to (v) above, we arrive at the eigenvalue equation
\begin{equation}
\widetilde{\lambda}_{\Phi}\widetilde{\pmb{\alpha}}_{\Phi}=\widetilde{K_F}\widetilde{\pmb{\alpha}}_{\Phi},
\end{equation}
which has eigenvectors $\widetilde{\mathbf{v}_{\Phi}}$ and
\begin{equation}
\widetilde{\mathbf{v}_{\Phi}}=\sum_{i=1}^N\widetilde{\pmb{\alpha}}_i\widetilde{\Phi}(\mathbf{x}_i).
\end{equation}

In this case, we do not have the centered data points represented by equation (\ref{eq:centfeat}), so we need to write $\widetilde{K_F}$ in terms of $K_F$. In what follows, consider $1_{ij}=1, \forall i,j$. 

Using equations (\ref{eq:centfeat}) and (\ref{eq:centeigen}),
\begin{eqnarray}
\widetilde{K_F}_{ij}&=&\widetilde{\Phi}(\mathbf{x}_i)^T\widetilde{\Phi}(\mathbf{x}_j)\nonumber\\
&=&\left(\Phi(\mathbf{x}_i)-\frac{1}{N}\sum_{m=1}^N\Phi(\mathbf{x}_m)\right)^T \times \nonumber \\
&&\times\left(\Phi(\mathbf{x}_j)-\frac{1}{N}\sum_{n=1}^N\Phi(\mathbf{x}_n)\right)\nonumber\\
&=&\Phi(\mathbf{x}_i)^T\Phi(\mathbf{x_j})-\frac{1}{N}\sum_{m=1}^N\Phi(\mathbf{x}_m)^T\Phi(\mathbf{x}_j)\nonumber\\
&&-\frac{1}{N}\sum_{n=1}^N\Phi(\mathbf{x}_i)^T\Phi(\mathbf{x}_n)\nonumber\\
&&+\frac{1}{N^2}\sum_{n,m=1}^N\Phi(\mathbf{x}_m)^T\Phi(\mathbf{x}_n)\nonumber\\
&=&{K_F}_{ij}-\frac{1}{N}\sum_{i=1}^M 1_{im}{K_F}_{mj}\nonumber\\
&&-\frac{1}{N}\sum_{n=1}^N {K_F}_{in}1_{nj}+\frac{1}{N^2}\sum_{n,m=1}^N1_{im}{K_F}_{mn}1_{nj}.\nonumber\\
\end{eqnarray}
Considering $\left(1_N\right)_{ij}:=1/N, \quad \forall \quad \{i,j\}$, we have the shorter version,
\begin{equation}
\widetilde{K_F}=K_F-1_N K_F - K_F 1_N+1_N K_F 1_N.
\end{equation}

\end{enumerate}

\section{Linear PCA}
\label{ap:linearPCA}
\begin{figure}
\centering
\includegraphics[scale=0.65]{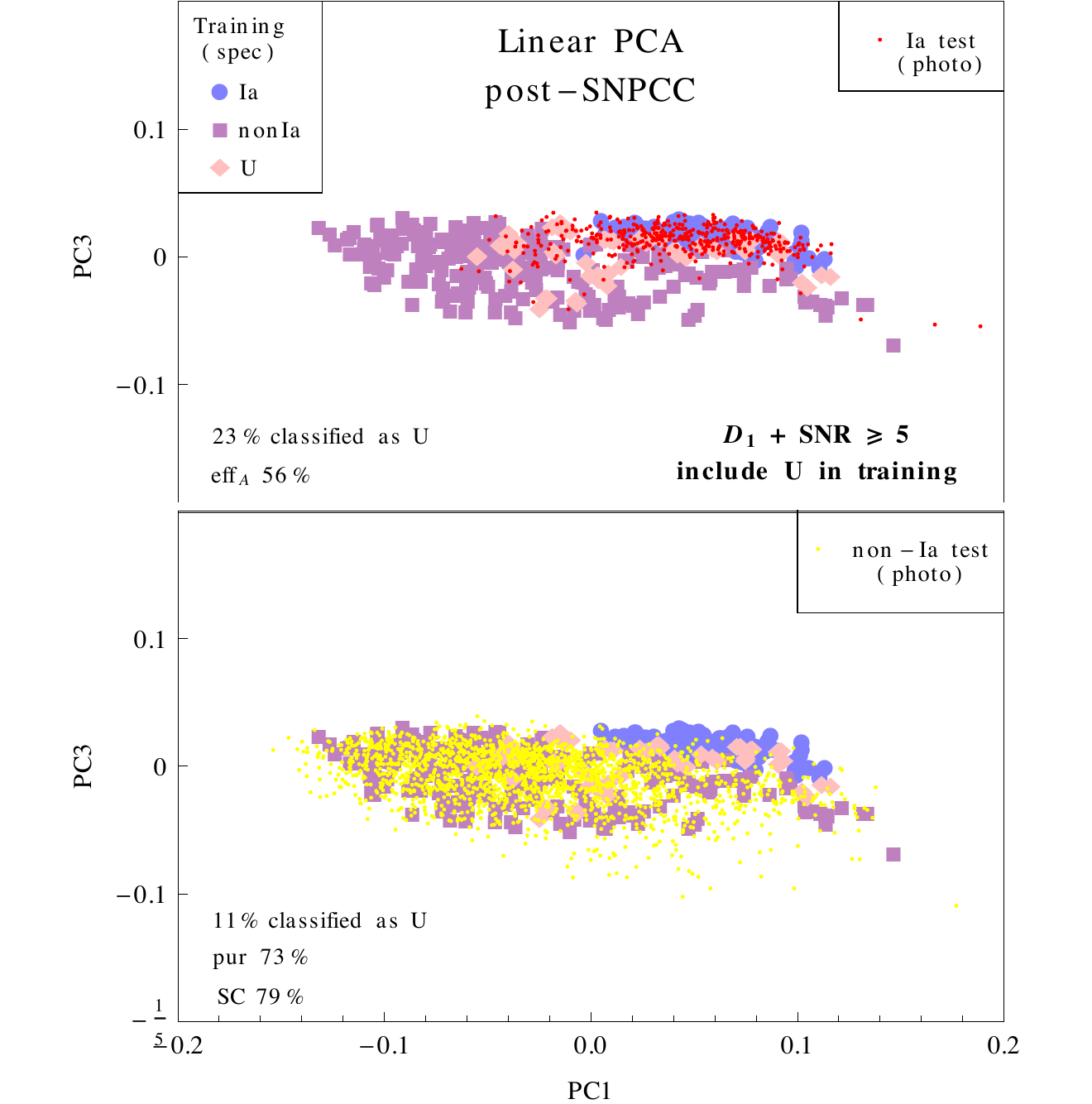}
\caption{Classification results using linear PCA for $D_1$+SNR5. The colour code is the same used in Figure \ref{fig:D1SNR5}.}
\label{fig:lPCAD1SNR5}
\end{figure}

We present here the results we achieved from applying linear PCA  to the post-SNPCC data. The procedure for deriving the PCs are described in subsection \ref{subsec:PCA}. The 2 PCs that best separate Ia and non-Ia data points were identified by using a cross-validation algorithm similar to the one described in subsection \ref{subsec:cross}. The only difference is that, in the linear case, there is no parameter $\sigma$ to adjust. The outcomes for sample $D_1$ using different SNR cuts are displayed in table \ref{tab:linearPCA}. The graphical representation of data points projections for the SNR$\geq$5 case is shown in figure \ref{fig:lPCAD1SNR5} and the redshift dependence of the classification results are displayed in figure \ref{fig:zlinear}.

Comparing results for $D_1$+SNR5 when U class is included in the training, presented in Tables \ref{tab:linearPCA} and \ref{tab:final}, the reader can verify that the using KPCA raises the efficiency levels from 56\% to 84\% and the purity levels from 73\% to 91\%. This corresponds to approximately  50\% increase in efficiency and 25\% increase in purity. 
\begin{table}
\caption{Results from applying linear PCA+1NN to the post-SNPCC data,  $D_1$ sample. Ratios of efficiency (eff), purity (pur) and successful classification (SC) are reported in percentages (\%).}
\centering
\begin{tabular}{ |c | c | c c c |c c c c |}
\cline{3-9}
 \multicolumn{2}{c|}{} & \multicolumn{3}{c|}{\tiny{Training sample}} & \multicolumn{4}{c|}{\tiny{Test sample}}\\
 \multicolumn{2}{c|}{} & \multicolumn{3}{c|}{\tiny{cross-validated}} & \multicolumn{4}{c|}{\tiny{including U}}\\
\cline{2-9}
\multicolumn{1}{c|}{} & \tiny{PC pair} & \tiny{eff$_{A}$} & \tiny{pur} & \tiny{SC} & \tiny{eff$_{A}$} & \tiny{pur} & \tiny{SC} & \tiny{FoM$_{A}$}\\
\hline
\tiny{SNR$\geq$5} & \tiny{1-3} & \tiny{77} & \tiny{80} & \tiny{86} & \tiny{56} & \tiny{73} & \tiny{79} & \tiny{0.27}\\
\hline
\tiny{SNR$\geq$3} & \tiny{1-3} & \tiny{85} & \tiny{83} & \tiny{87} & \tiny{64} & \tiny{63} & \tiny{78} & \tiny{0.23}\\
\hline
\tiny{SNR$\geq$0} & \tiny{1-3} & \tiny{84} & \tiny{84} & \tiny{84} & \tiny{48} & \tiny{32} & \tiny{50} & \tiny{0.07}\\
\hline
\end{tabular}\label{tab:linearPCA}
\end{table}

\begin{figure}
\includegraphics[scale=0.55]{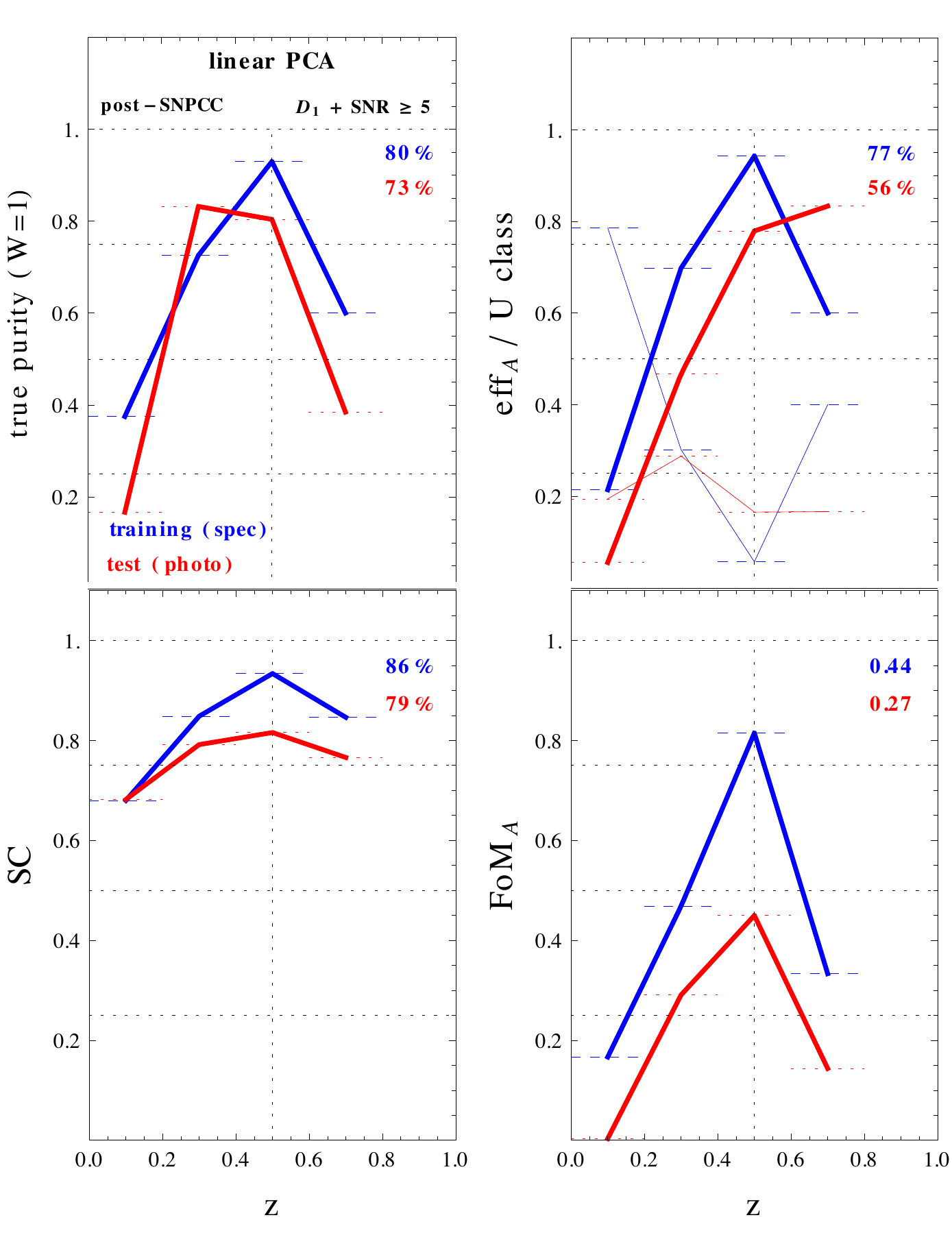}
\caption{Classification results for $D_1+$SNR5 as a function of redshift using linear PCA. The color code is the same used in figure \ref{fig:z_diag_cad1_SNR5_bef_only}.}
\label{fig:zlinear}
\end{figure}


\section{Results for $D_1$ as a function of redshift and SNR cuts}
\label{ap:zcad1SNR5}

\begin{figure}
\centering
\includegraphics[trim = 0mm 0mm 0mm 3.5mm, clip, width=1\columnwidth]{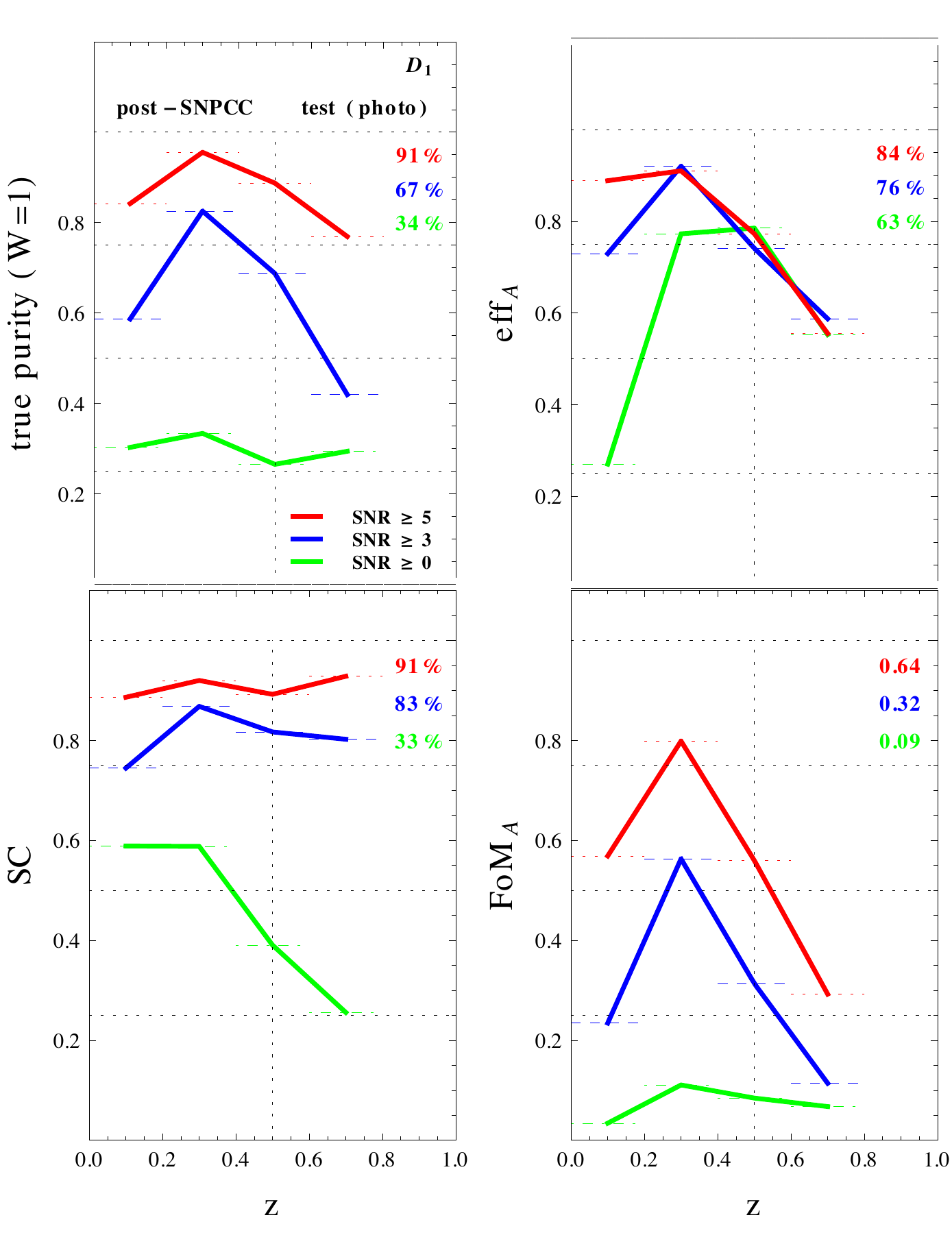}
\caption{Test sample classification results of efficiency, purity, FoM and SC for $D_1$ as a function of redshift. The orange (dot-dashed), brown (dashed) and green (dotted) lines correspond to SNR$\geq$5, SNR$\geq$3 and SNR$\geq$0, respectively.}
\label{fig:zcad1SNR}
\end{figure}

\begin{figure}
\centering
\includegraphics[scale=0.4]{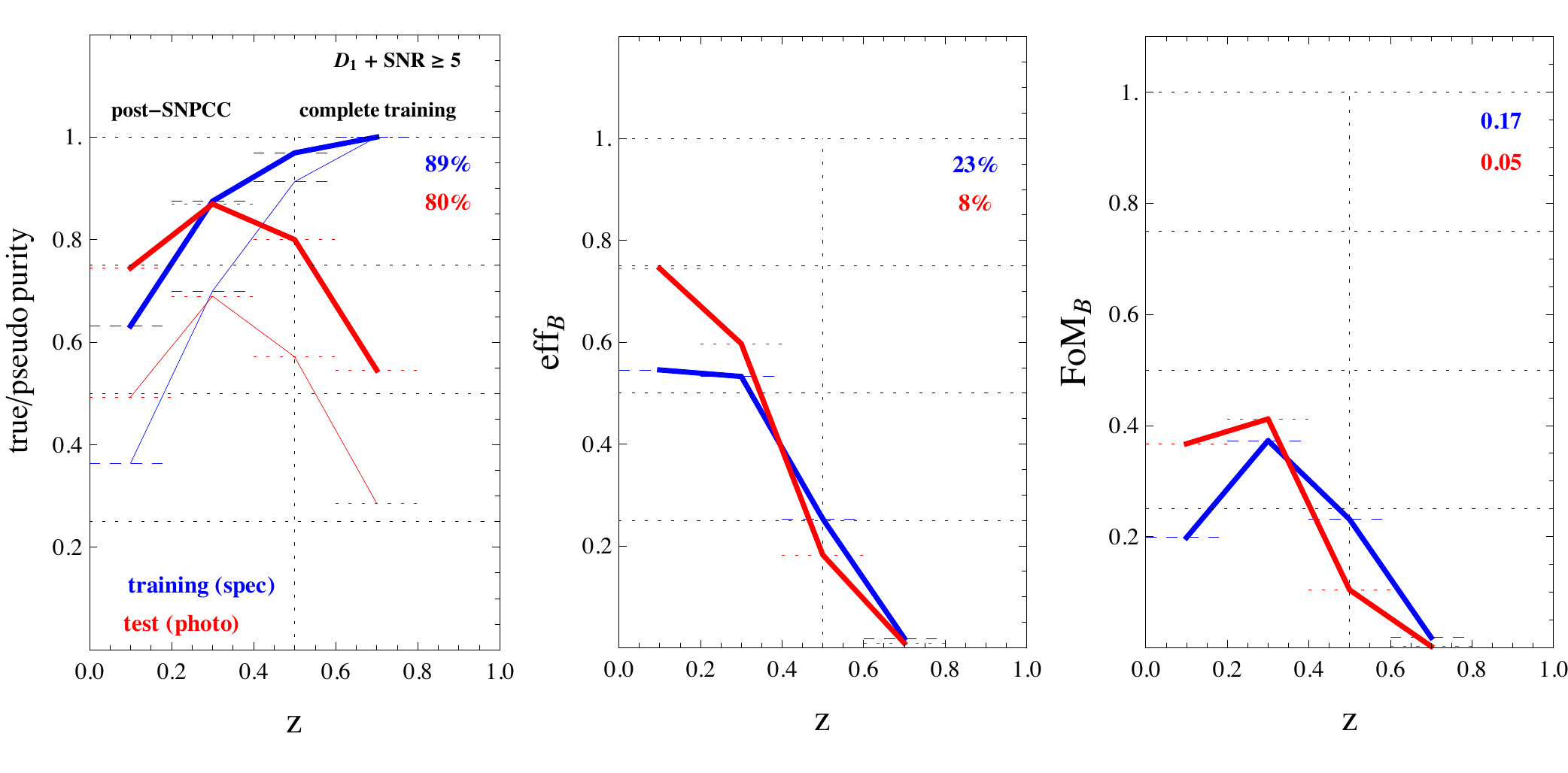}
\caption{Results from the post-SNPCC data for  SC (left), eff$_{\rm A}$ (middle) and FoM$_{\rm A}$ as a function of redshift for $D_1+$SNR5. The color code is the same used in figure \ref{fig:z_diag_cad1_SNR5_bef_only}. Top-middle panel also shows values of the percentage of SNe classified as $U$ (thin lines, blue for training and red for test sample).  }
\label{fig:z_diag_cad1_SNR5_aft_only}
\end{figure}

\begin{figure}
\centering
\includegraphics[scale=0.4]{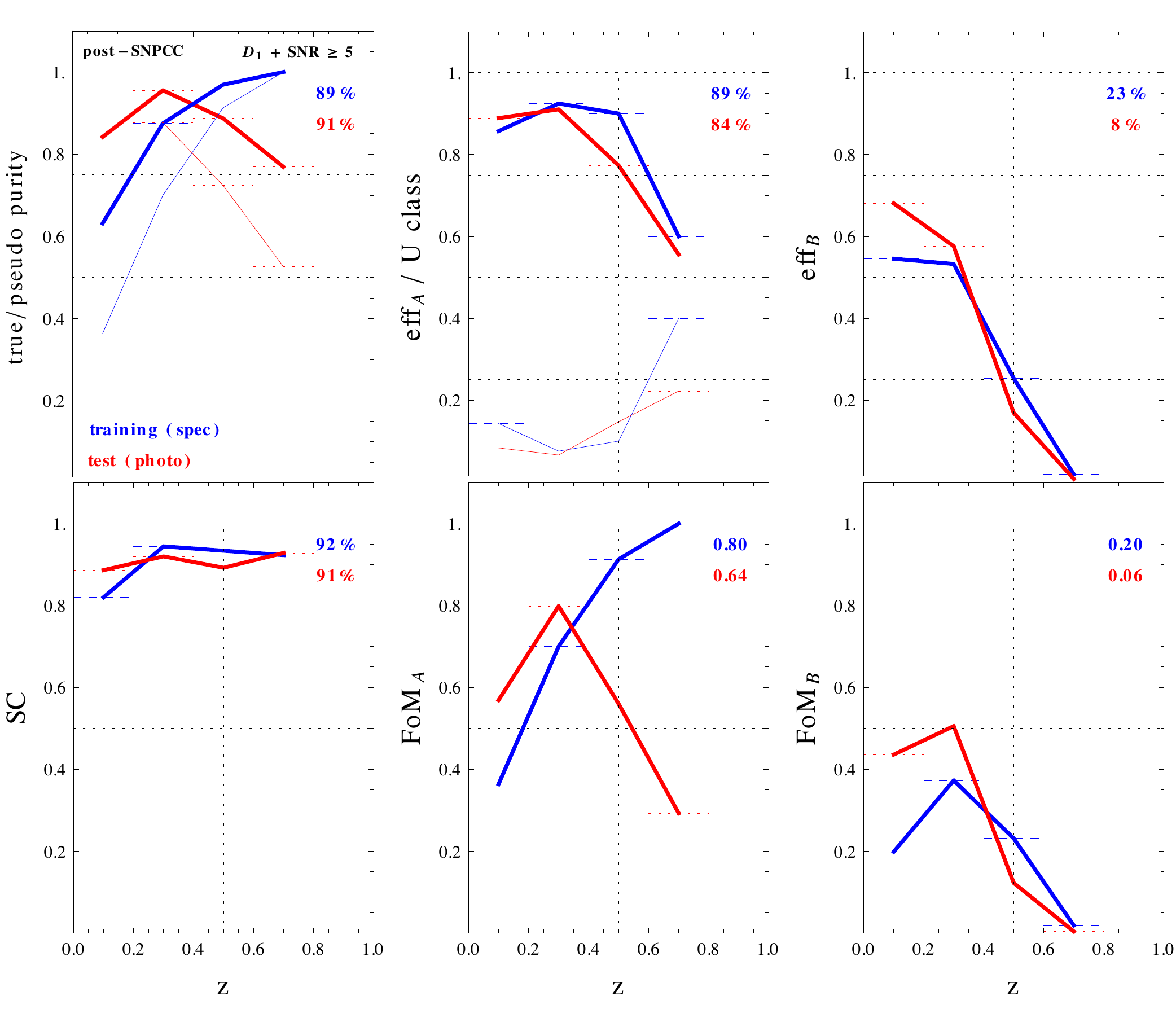}
\caption{Results from the post-SNPCC data for  pur (top-right), eff$_{\rm A}$ (top-middle), eff$_{\rm B}$ (top-right), SC (bottom-left), FoM$_{\rm A}$ (bottom-middle) and FoM$_{\rm B}$ (bottom-right) as a function of redshift for $D_1+$SNR5 and including U class in the training sample. The color code is the same used in figure \ref{fig:z_diag_cad1_SNR5_bef_only}. Top-left and top-middle panels also show values of pseudo-purity and the percentage of SNe classified as $U$ (thin lines, blue for training and red for test sample), respectively.  }
\label{fig:z_diag_cad1_SNR}
\end{figure}


Figure \ref{fig:zcad1SNR} shows how the classification results for $D_1$ (test sample) behave as a function of redshift and SNR selection cuts. Figure \ref{fig:z_diag_cad1_SNR5_aft_only} shows SC, efficiency and FoM results normalized after election cuts.

Examining the top-middle panel of figure \ref{fig:z_diag_cad1_SNR}, we see that eff$_{\rm A}$ also suffers in high redshift due to SNe classified as $U$ (thin lines). This was another choice we made in order to preserve purity. Although a few  SNe Ia are lost to the $U$ class (which is bad for efficiency), so are non-Ia that would easily be mistaken with SNe Ia (which is good for purity). This effect becomes clear if we compare figures \ref{fig:z_diag_cad1_SNR5_bef_only}  and \ref{fig:z_diag_cad1_SNR5_aft_only} to figure \ref{fig:z_diag_cad1_SNR}. From these we see that eff$_{\rm A}$ gets from 89\% (without $U$ type) to 84\% (with $U$ type) but at the same time purity increased from 80\% to 91\%, staying above 75\% for the entire redshift range.

\section{$D_8$+SNR0 classifications}
\label{ap:d8}
\begin{figure}
\centering
\includegraphics[scale=0.65]{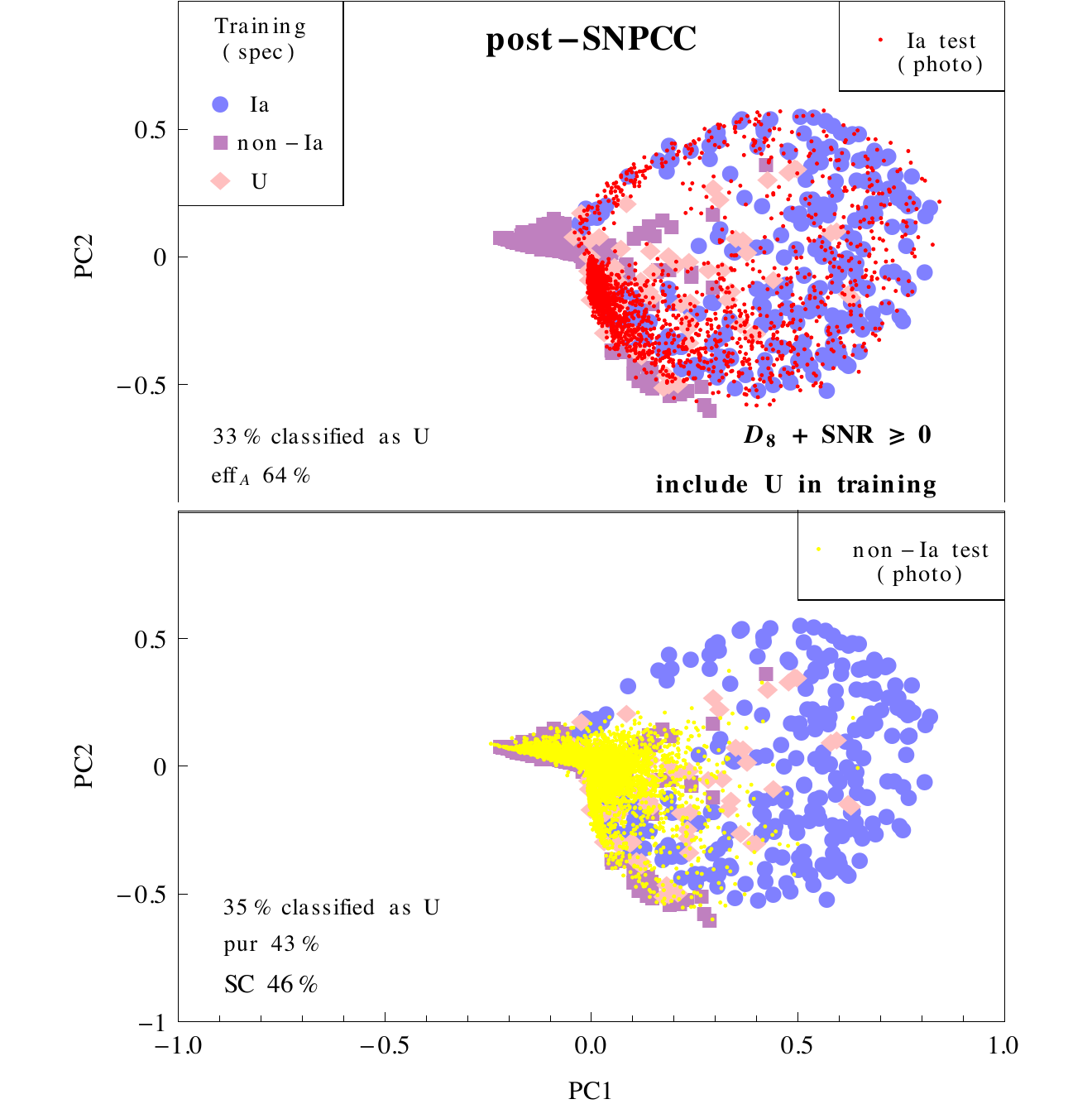}
\caption{Classification results for $D_8$+SNR0,  including U class in the training sample. The color code is the same used in figure \ref{fig:D1SNR5}.}
\label{fig:rescad8SNR0}
\end{figure}
We present in figures \ref{fig:rescad8SNR0}, \ref{fig:zcad8SNR0} and \ref{fig:znonIacad8SNR0} the classification results for $D_8$+SNR0. This is shown in order to facilitate comparison with other methods from the literature which do not apply SNR cuts.
However, we emphasize that, for a given time sampling, this is the worst case scenario for our method. As shown in figure \ref{fig:zcad1SNR}, the classification potential of the method is highly increased with better quality data (higher SNR).

\begin{figure*}
\centering
\begin{minipage}[t]{0.45\linewidth}
\includegraphics[scale=0.375]{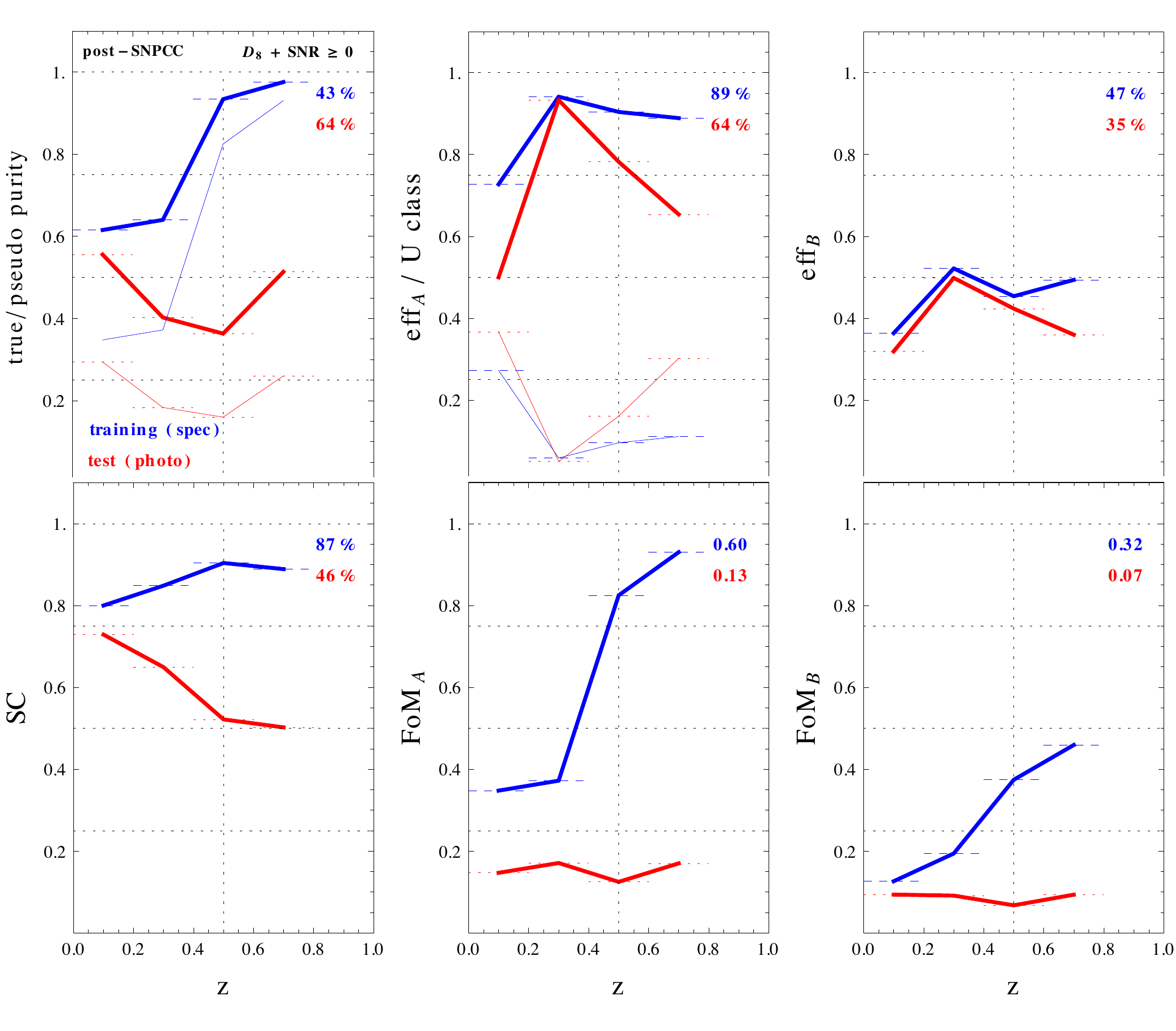}
\caption{Classification results as a function of redshift for Ia ($D_8$+SNR0), including U class in the training sample. The panels show efficiency, purity, FoM and SC from top to bottom. The colour code is the same used in figure \ref{fig:z_diag_cad1_SNR5_bef_only}.}
\label{fig:zcad8SNR0}
\end{minipage}
\hspace{0.5cm}
\begin{minipage}[t]{0.45\linewidth}
\includegraphics[scale=0.375]{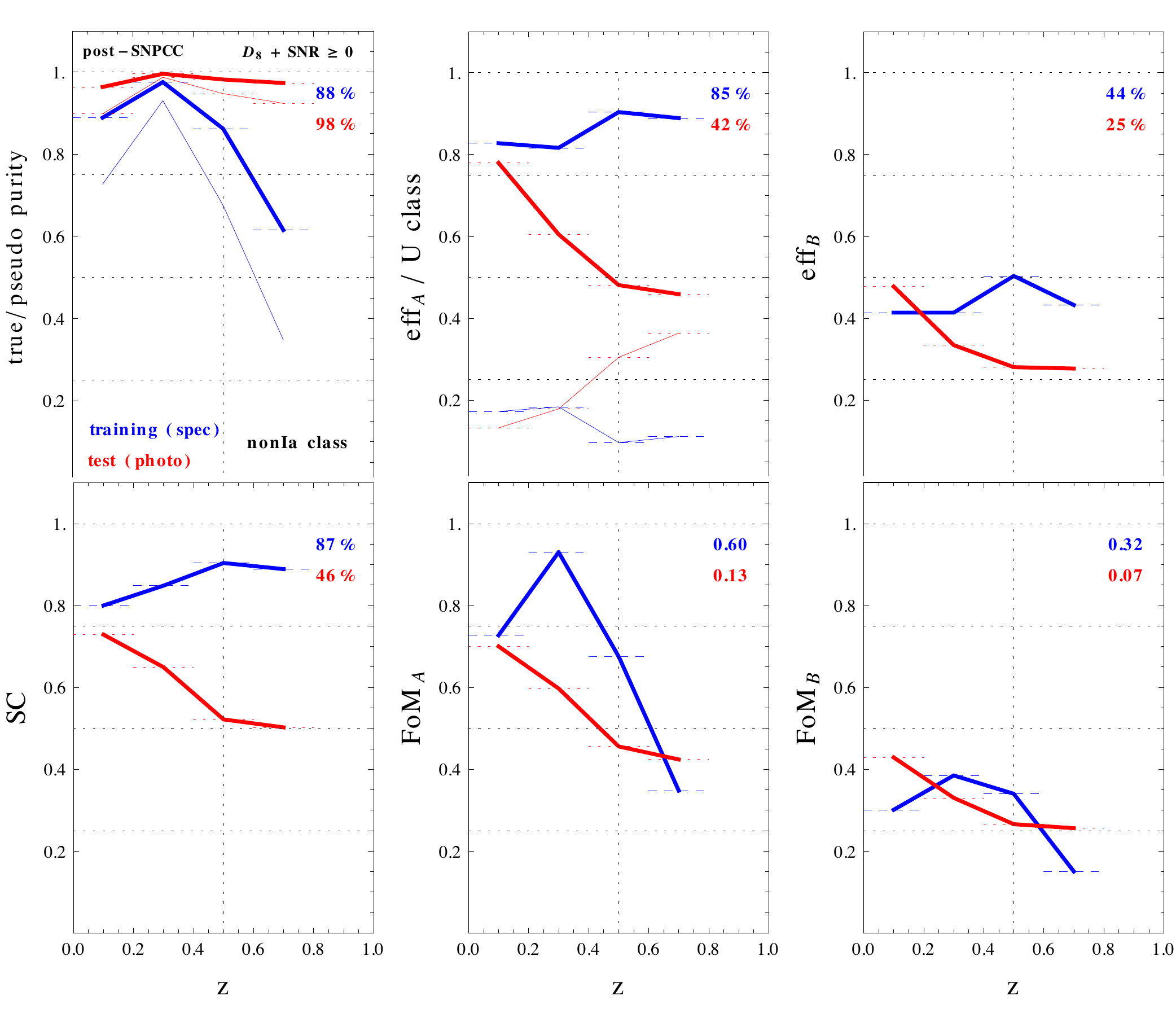}
\caption{Analogous of figure \ref{fig:zcad8SNR0} for non-Ia classifications. }
\label{fig:znonIacad8SNR0}
\end{minipage}
\end{figure*}

\section{SNPCC complete results}
\label{ap:SNPCC_comp}

Figure \ref{fig:SNPCC} shows detailed results obtained from the SNPCC sample for different time window samplings. It is composed by 4 big panels, each one containing plots for a diagnostic parameter, organized in 3 rows and 4 columns. The rows run through SNR$\geq$5, SNR$\geq$3 and SNR$\geq0$, from top to bottom. The left-most column in each panel show results for SNR cuts only. Meaning that all SNe surviving the corresponding SNR cut were classified as Ia.  Other columns represent $D_1$, $D_3$ and $D_7$, from left to right. Outcomes from $D_2$, $D_4$ and $D_8$ are similar to the ones presented in the plot, so we decided not to show them.

The first thing to notice from this figure is that the time window sampling leads to small differences in the overall classification results. Obviously higher purity results  comes from $D_7$, the only sub-sample which includes the second maximum in the infra-red, for SNe Ia in $z\leq0.8$. However, discrepancies between results from different SNR cuts are much larger. This shows that, despite the need to define a time window, the specific choice is not crucial in the determination of final results.

The same argument does not hold for SNR selection cuts. We see the crucial role played by the quality of each observation, no matter which diagnostic we analyse. Although this effect is noticeable in all of them, it is more evident in outcomes from eff$_{\rm B}$ and FoM$_{\rm B}$, due to reasons already discussed in section \ref{sec:application}. Nevertheless, our method achieved FoM$_{\rm B}>0.25$ for $z\leq0.25$. In this redshift range, only SNPCC entries Sako, JEDI-KDE and SNANA cuts reported comparable results. The behaviour of our eff$_{B}$ plots is almost opposite to what is reported from the SNPCC. In those, the efficiency is almost always very high, what frequently comes accompanied by a low purity result.  

On the other hand, our results for purity and pseudo-purity are very good, specially for redshifts within $[0.2,0.5]$. For all sub-samples with  SNR$\geq$5, we achieved purity values larger than 75\% in this redshift range, a result that is not present in \textit{none} of the entries in the SNPCC. Beyond that, $D_7$+SNR$\geq$5 gives  good results for purity and pseudo-purity for $z\geq0.5$, confirming the importance of observing the second maximum in the infra-red. 


\begin{figure*}
\begin{center}
\subfigure{%
\includegraphics[trim = 0mm 0mm 0mm 3.5mm, clip, width=0.45\linewidth]{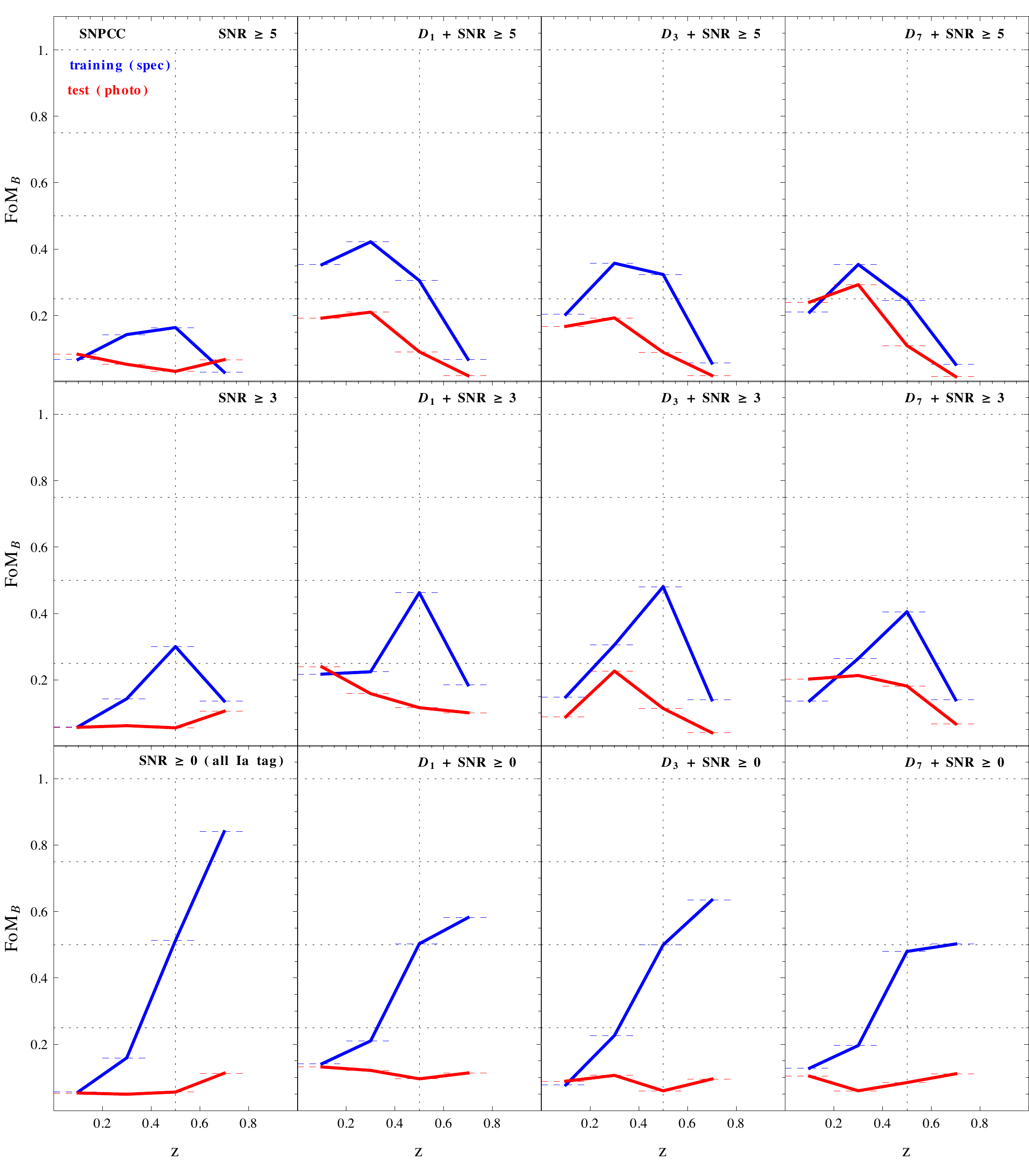}
}
\subfigure{
\includegraphics[trim = 0mm 0mm 0mm 3.5mm, clip, width=0.45\linewidth]{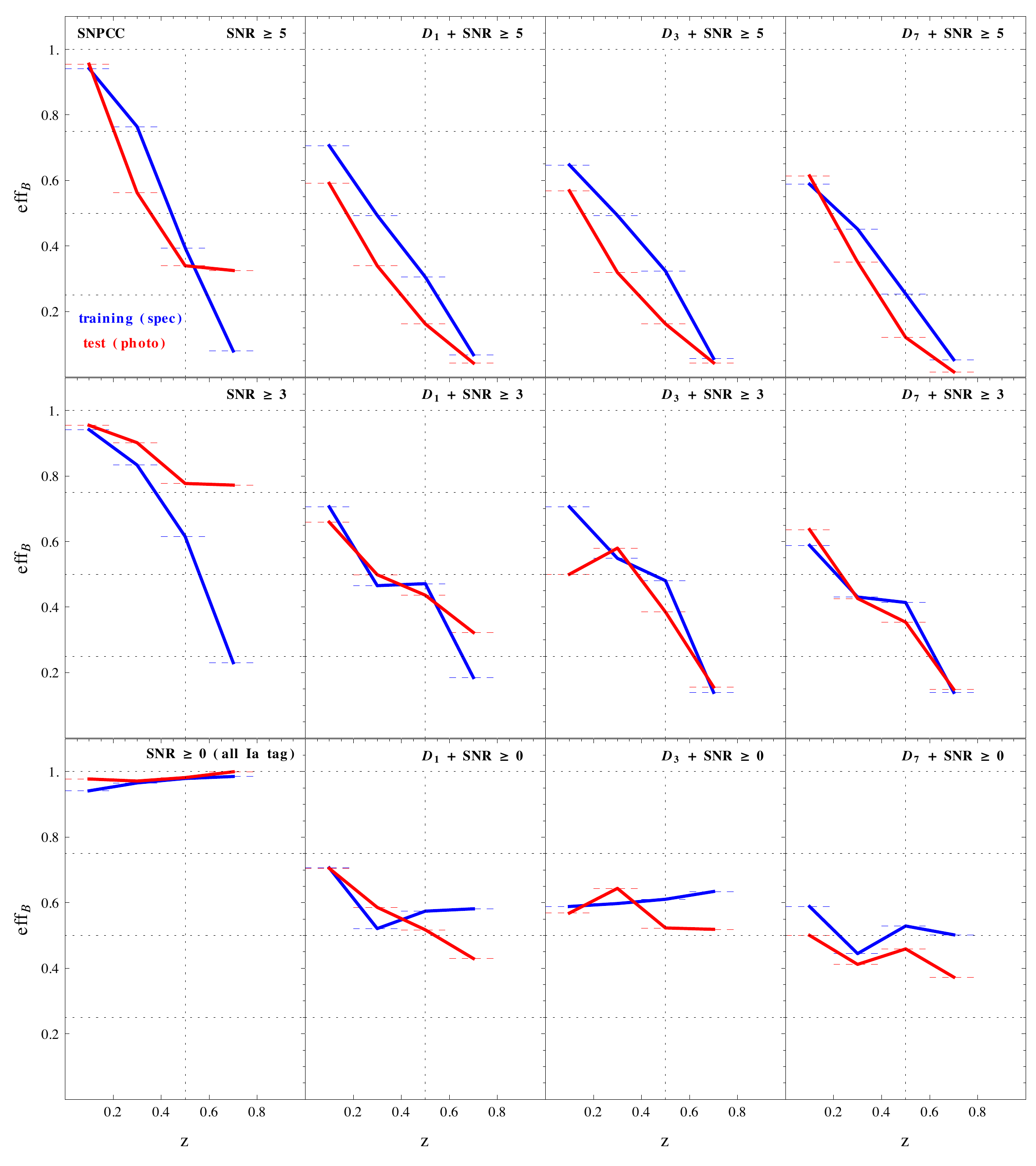}
}\\
\subfigure{
\includegraphics[trim = 0mm 0mm 0mm 3.5mm, clip, width=0.45\linewidth]{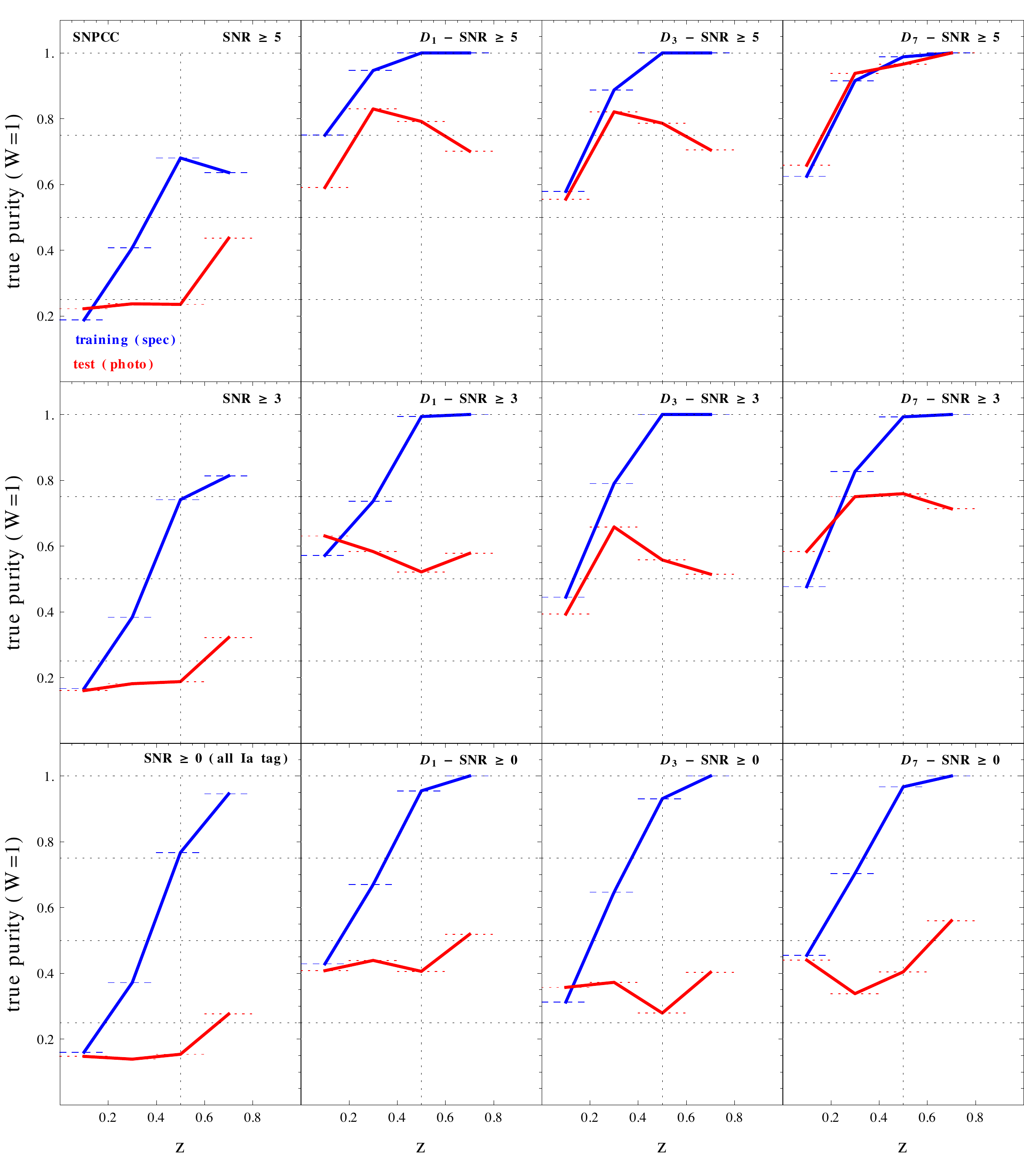}
}
\subfigure{
\includegraphics[trim = 0mm 0mm 0mm 3.5mm, clip, width=0.45\linewidth]{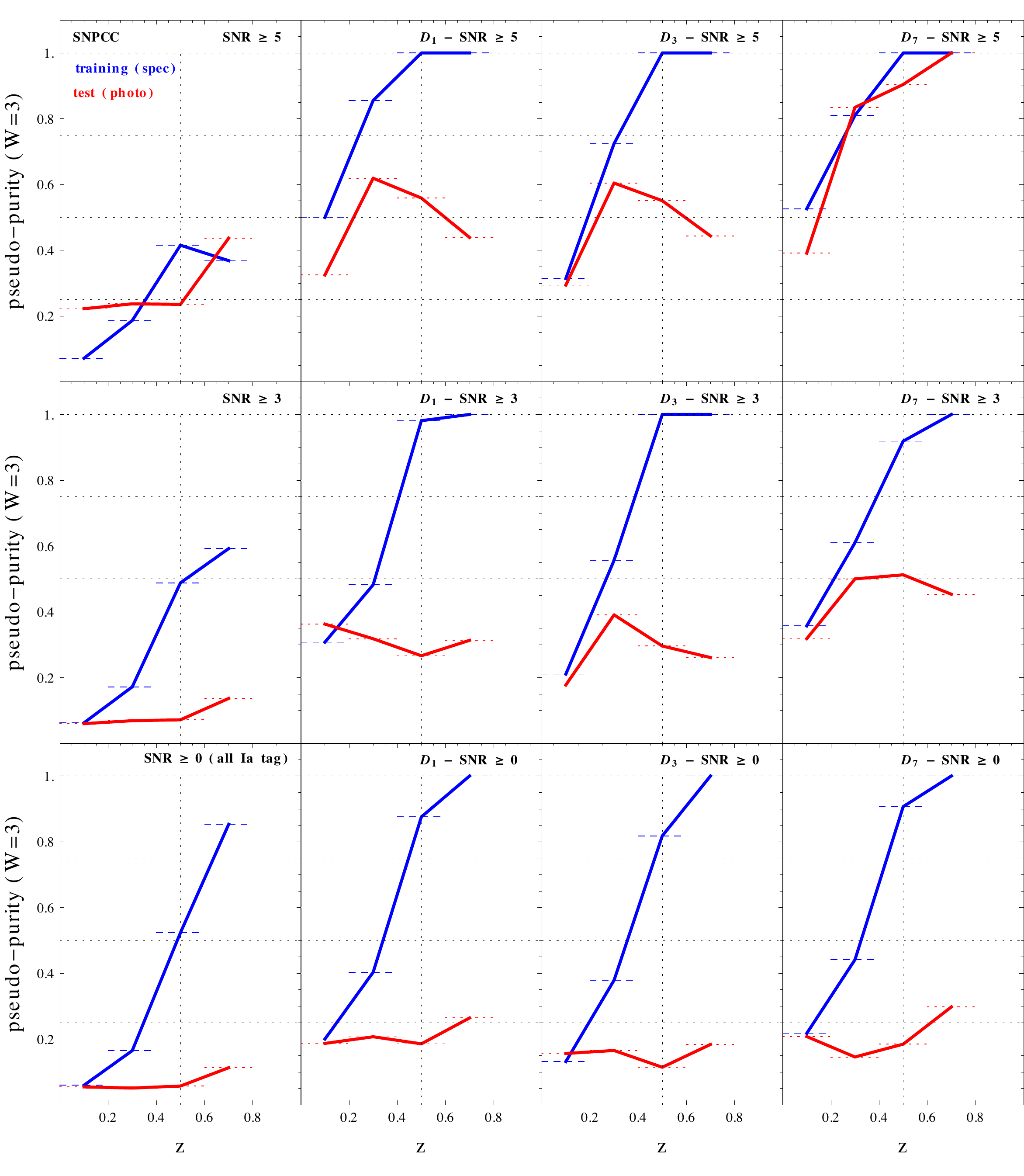}
}
\end{center}
\caption{Redshift dependence results from the SNPCC data set for FoM$_{B}$ (top-left), eff$_{B}$ (top-right), pseudo-purity (bottom-left) and true purity (bottom-right), including $U$ as an extra class in the training sample. The blue-thick lines correspond to results from the training (spectroscopic) sample and the red-thick to results from the test (photometric) sample. The left-most columns in each big panel show results where all SNe satisfying the SNR selection cuts were tagged as Ia. Rows run through SNR cuts: SNR$\geq$5, SNR$\geq$3 and SNR$\geq$0 from top to bottom. Columns 2 to 4 show results for $D_1$, $D_3$ and $D_5$, from left to right.  }
\label{fig:SNPCC}
\end{figure*}

%
\section{Summary tables}
\label{ap:tables}

We present bellow complete tables describing our results for different light curve time samplings and SNR cuts.

\begin{table*}
\caption{Number of SNe in each post-SNPCC subset. The table also shows subsamples of the $D_i$ and $[-10,0[$ according to SNR cuts.}
\centering
\begin{tabular}{| c | c *{14}{ c|}}
\multicolumn{1}{c}{ }& \multicolumn{1}{c }{ }&\multicolumn{12}{c}{Training sample} \\
\cline{4-15}
\multicolumn{2}{c}{ }& &\multicolumn{3}{|c|}{$D_1$} & \multicolumn{3}{c|}{$D_3$} & \multicolumn{3}{c|}{$D_5$} & \multicolumn{3}{c|}{$D_7$} \\
\cline{2-2} \cline{4-15}
\multicolumn{1}{c}{ } &\multicolumn{1}{||c||}{SIM1} & &\multicolumn{1}{|c|}{\tiny{SNR$\geq$0}} & \tiny{SNR$\geq$3} & \tiny{SNR$\geq$5}&\tiny{SNR$\geq$0} & \tiny{SNR$\geq$3} & \tiny{SNR$\geq$5}& \tiny{SNR$\geq$0} & \tiny{SNR$\geq$3} & \tiny{SNR$\geq$5} & \tiny{SNR$\geq$0} & \tiny{SNR$\geq$3} & \tiny{SNR$\geq$5}\\
\cline{1-2} \cline{4-15}
\multicolumn{1}{|c}{Ia} & \multicolumn{1}{|| c ||}{559}& & 
\multicolumn{1}{|c|}{374} & 213& 142 &  409& 225 & 145 & 418 & 232 &  148 & 297 & 173 & 119\\
\multicolumn{1}{|c}{non-Ia} & \multicolumn{1}{||c||}{544}& & 
\multicolumn{1}{|c|}{355} & 315 &  273&  397& 347 & 303 & 412&350& 303 & 282 & 257 & 222\\
\multicolumn{1}{|c}{total} & \multicolumn{1}{||c||}{1103}& & 
\multicolumn{1}{|c|}{729} &  528& 415&  806&  572&  448& 830 & 582&415 & 579 & 430 & 341\\
\cline{1-2} \cline{4-15}
\multicolumn{15}{c}{ }\\
\multicolumn{3}{c}{ } & \multicolumn{12}{c}{Test sample}\\
 \cline{4-15}
\multicolumn{2}{c}{ }& &\multicolumn{3}
{|c|}{$D_1$} & \multicolumn{3}{c|}{$D_3$} & \multicolumn{3}{c|}{$D_5$}  & \multicolumn{3}{c|}{$D_7$}\\
\cline{2-2} \cline{4-15}
\multicolumn{1}{c}{ } &\multicolumn{1}{||c||}{SIM1} & &\multicolumn{1}{|c|}{\tiny{SNR$\geq$0}} & \tiny{SNR$\geq$3} & \tiny{SNR$\geq$5}&\tiny{SNR$\geq$0} & \tiny{SNR$\geq$3} & \tiny{SNR$\geq$5}& \tiny{SNR$\geq$0} & \tiny{SNR$\geq$3} & \tiny{SNR$\geq$5} & \tiny{SNR$\geq$0} & \tiny{SNR$\geq$3} & \tiny{SNR$\geq$5} \\
\cline{1-2} \cline{4-15}
\multicolumn{1}{|c}{Ia}&\multicolumn{1}{||c||}{559} & & 
\multicolumn{1}{|c|}{3181} &633	&431	&3480	&666	&453	&3575	&673	&448	& 2525& 520 & 354\\
\multicolumn{1}{|c}{non-Ia}&\multicolumn{1}{||c||}{544}& &
\multicolumn{1}{|c|}{11346}&3716	&1993	&12255	&3926	&2100	&12413	&3900	&2096	& 9340 & 3241 & 1776\\
\multicolumn{1}{|c}{total}  &\multicolumn{1}{||c||}{1103}& &
\multicolumn{1}{|c|}{14527}	&4349	&2424	&15735	&4592	&2553	&15988	&4573	&2544	&11865 & 3761 & 2130\\
\cline{1-2} \cline{4-15}
\multicolumn{15}{c}{ }\\
\multicolumn{4}{c}{ } &  \multicolumn{3}{c }{Test sample}& \multicolumn{1}{c}{ } &\multicolumn{3}{c}{Training Sample} &\multicolumn{4}{c}{ }\\
\cline{5-7} \cline{9-11}
\multicolumn{4}{c}{ } & \multicolumn{3}{ | c |}{$[-10,0[$} & \multicolumn{1}{c}{ } & \multicolumn{3}{ | c |}{$[-10,0[$}&  \multicolumn{4}{c}{}\\
\cline{5-7} \cline{9-11}
\multicolumn{4}{c}{ } & \multicolumn{1}{ | c |}{\tiny{SNR$\geq$0}} & \multicolumn{1}{c|}{\tiny{SNR$\geq$3}} & \multicolumn{1}{c| }{\tiny{SNR$\geq$5}} & \multicolumn{1}{c}{ } & \multicolumn{1}{ |c |}{\tiny{SNR$\geq$0}} & \multicolumn{1}{c|}{\tiny{SNR$\geq$3}} & \multicolumn{1}{c |}{\tiny{SNR$\geq$5}}&\multicolumn{4}{c}{ } \\
\cline{4-7} \cline{9-11}
\multicolumn{3}{c}{ } &  \multicolumn{1}{ |c }{Ia} &\multicolumn{1}{|c|}{444} &\multicolumn{1}{c|}{238} &\multicolumn{1}{c|}{153} &\multicolumn{1}{c}{} &\multicolumn{1}{|c|}{3555} &\multicolumn{1}{c|}{661} &\multicolumn{1}{c|}{437} &\multicolumn{4}{c}{ } \\
\multicolumn{3}{c}{ } &  \multicolumn{1}{ |c }{nonIa} &\multicolumn{1}{|c|}{440} &\multicolumn{1}{c|}{361 } &\multicolumn{1}{c|}{312} &\multicolumn{1}{c}{} &\multicolumn{1}{|c|}{12544 } &\multicolumn{1}{c|}{3926} &\multicolumn{1}{c|}{2125} &\multicolumn{4}{c}{ } \\
\multicolumn{3}{c}{ } &  \multicolumn{1}{ |c }{total} &\multicolumn{1}{|c|}{884} &\multicolumn{1}{c|}{599} &\multicolumn{1}{c|}{465} &\multicolumn{1}{c}{} &\multicolumn{1}{|c|}{ 16099} &\multicolumn{1}{c|}{4587} &\multicolumn{1}{c|}{2562} &\multicolumn{4}{c}{ } \\
\cline{4-7} \cline{9-11}
\end{tabular}\label{tab:number}
\end{table*}

\begin{table*}
 \caption{Summary of classifications results for post-SNPCC data. Ratios of efficiency (eff$_{\rm A}$/eff$_{\rm B}$), purity (pur) and successful classification (SC) are reported in percentages (\%).}
 \centering
 \begin{tabular}{| c| c | c | c || c c c ||c c c | c c c | c c c c c c|}
 \cline{8-19}
 \multicolumn{7}{c|}{ } & \multicolumn{12}{c|}{Test sample}\\
  \cline{5-19}
  \multicolumn{4}{c||}{\multirow{2}{*}{ }} &  \multicolumn{3}{c||}{Training sample}& \multicolumn{3}{c|}{\multirow{2}{*}{complete}} & \multicolumn{3}{c|}{\multirow{2}{*}{exclude U}} &  \multicolumn{6}{c|}{\multirow{2}{*}{include U}}\\
 \multicolumn{4}{c||}{}  & \multicolumn{3}{c||}{cross validated} & \multicolumn{3}{c|}{ }& \multicolumn{3}{c|}{ }&  \multicolumn{6}{c|}{ }\\            
\hline
 data set & SNR & $\sigma$ & PCs & eff$_{\pmb A}$ & pur &  SC & eff$_{\pmb A}$  &  pur & SC & eff$_{\pmb A}$ & pur & SC & eff$_{\pmb A}$ &  pur & SC & FoM$_{\pmb A}$ &  eff$_{\pmb B}$ &  FoM$_{\pmb B}$\\
 \hline
\multicolumn{1}{|c|}{\multirow{3}{*}{$D_1$}}& 
	$\geq5$ & 0.9 & $1\quad 4$ & 89 & 89 & 92 & 89 & 80 & 94 & 93 & 87 & 96 & 84 & 91 & 91 & 0.64 &  8 & 0.06\\
  & $\geq3$ & 0.9 & $1\quad 2$ & 87 & 86 & 89 & 83 & 56 & 88 & 89 & 64 & 91 & 76 & 67 & 83 & 0.32 & 11 & 0.04\\
  & $\geq0$ & 0.7 & $1\quad 5$ & 88 & 87 & 87 & 73 & 30 & 57 & 81 & 24 & 41 & 63 & 34 & 33 & 0.09 & 44 & 0.06\\
\hline
\multicolumn{1}{|c|}{\multirow{3}{*}{$D_2$}}&
	$\geq5$ & 0.4 & $1\quad 3$ & 90 & 90 & 92 & 90 & 77 & 94 & 93 & 80 & 95 & 86 &  83 & 91 & 0.54 & 8 & 0.05\\  
 &  $\geq3$ & 0.7 & $1\quad 5$ & 88 & 88 & 90 & 77 & 62 & 90 & 78 & 70 & 92 & 64 & 75 & 83 & 0.32 &  9 & 0.04\\
 &  $\geq0$ & 0.4 & $1\quad 5$ & 86 & 88 & 87 & 65 & 32 & 63 & 87 & 27 & 46 & 32 & 30 & 32 & 0.06 & 46 & 0.04\\
\hline
\multicolumn{1}{|c|}{\multirow{3}{*}{$D_3$}}&  
	$\geq5$ & 1.0 & $1\quad 2$ & 84 & 86 & 90 & 89 & 71 & 92 & 91 & 75 & 93 & 82 & 83 & 87 & 0.51 &  8 & 0.05\\  
  & $\geq3$ & 1.0 & $1\quad 2$ & 85 & 87 & 89 & 84 & 52 & 86 & 88 & 58 & 89 & 77 & 67 & 82 & 0.31 & 11 & 0.05\\   
  & $\geq0$ & 0.6 & $1\quad 4$ & 85 & 84 & 84 & 69 & 30 & 58 & 79 & 26 & 45 & 52 &  29 & 34 & 0.06 & 40 & 0.05\\  
\hline
\multicolumn{1}{|c|}{\multirow{3}{*}{$D_4$}}&
	$\geq5$ & 0.3 & $1\quad 4$ & 85 & 90 & 92 & 87 & 78 & 93 & 88 & 86 & 95 & 82 & 88 & 91 & 0.58 & 8 & 0.06\\   
  & $\geq3$ & 0.6 & $1\quad 2$ & 83 & 86 & 88 & 85 & 53 & 87 & 88 & 57 & 89 & 77 & 67 & 81 & 0.31 & 11 & 0.05\\
  & $\geq0$ & 1.7 & $1\quad 3$ & 85 & 85 & 85 & 60 & 26 & 53 & 62 & 25 & 51 & 50 & 25 & 44 & 0.05 & 38 & 0.04\\
\hline
\multicolumn{1}{|c|}{\multirow{3}{*}{$D_5$}}&
    $\geq5$ & 1.9 & $1\quad 3$ & 85 & 86 & 91 & 82 & 59 & 87 & 88 & 60 & 87 & 73 & 61 & 83 & 0.25 &  - &  
-\\
  & $\geq3$ & 1.1 & $1\quad 3$ & 87 & 91 & 91 & 85 & 48 & 84 & 87 & 53 & 87 & 76 & 56 & 80 & 0.22 &  - & 
-\\    
  & $\geq0$ & 0.4 & $1\quad 3$ & 82 & 85 & 84 & 56 & 34 & 65 & 57 & 38 & 69 & 44 & 38 & 58 & 0.07 &  - & 
-\\
\hline
\multicolumn{1}{|c|}{\multirow{3}{*}{$D_6$}}& 
    $\geq5$ & 0.8 & $1\quad 3$ & 85 & 89 & 91 & 86 & 62 & 88 & 87 & 69 & 91 & 80 & 73 & 84 & 0.38 &  - &
- \\  
  & $\geq3$ & 1.0 & $1\quad 3$ & 88 & 90 & 91 & 86 & 45 & 83 & 91 & 48 & 84 & 81 & 51 & 79 & 0.21 &  - &
- \\
  & $\geq0$ & 1.3 & $1\quad 3$ & 86 & 85 & 85 & 60 & 28 & 57 & 52 & 25 & 55 & 44 & 27 & 45 & 0.05 &  - & 
- \\         
\hline         
\multicolumn{1}{|c|}{\multirow{3}{*}{$D_7$}}& 
	$\geq5$ & 0.5 & $1\quad 2$ & 93 & 92 & 95 & 86 & 95 & 97 & 86 & 96 & 97 & 81 & 96 & 95 & 0.72 &  6 & 0.06\\  
  & $\geq3$ & 0.6 & $1\quad 2$ & 91 & 90 & 92 & 80 & 86 & 96 & 84 & 90 & 97 & 74 & 92 & 95 & 0.59 &  9 & 0.07\\
  & $\geq0$ & 1.3 & $1\quad 2$ & 90 & 87 & 88 & 72 & 36 & 66 & 77 & 35 & 64 & 65 & 37 & 55 & 0.11 & 36 & 0.06\\         
\hline       
\multicolumn{1}{|c|}{\multirow{3}{*}{$D_8$}}& 
	$\geq5$ & 0.4 & $1\quad 2$ & 92 & 90 & 94 & 90 & 92 & 97 & 92 & 96 & 98 & 89 & 98 & 96 & 0.83 & 7 & 0.06\\  
  & $\geq3$ & 0.5 & $1\quad 2$ & 91 & 89 & 92 & 84 & 76 & 94 & 88 & 92 & 97 & 82 & 94 & 92 & 0.69 & 9 & 0.08\\
  & $\geq0$ & 0.7 & $1\quad 2$ & 89 & 86 & 87 & 74 & 41 & 71 & 94 & 33 & 58 & 64 & 43 & 46 & 0.13 & 35 & 0.07\\         
\hline         
\hline
\multicolumn{1}{|c|}{\multirow{2}{*}{$[-10,0[$}}& 
    $\geq5$ & 1.2 & $1\quad 5$ & 84 & 82 & 89 & 78 & 53 & 85 & 78 & 60 & 87 & 71 & 63 & 77 & 0.26 & - & -\\
  & $\geq3$ & 1.4 & $1\quad 5$ & 81 & 81 & 85 & 76 & 35 & 76 & 77 & 40 & 80 & 67 & 43 & 67 & 0.13 &  - & - \\
$\Delta=1$   
  & $\geq0$ & 1.6 & $1\quad 5$ & 80 & 81 & 81 & 65 & 28 & 56 & 67 & 29 & 56 & 52 & 30 & 45 & 0.06 & - & -\\
\hline
\multicolumn{1}{|c|}{\multirow{2}{*}{$[-10,0[$}}& 
    $\geq5$ & 0.9 & $1\quad 5$ & 86 & 81 & 89 & 69 & 51 & 83 & 74 & 51 & 84 & 63 & 57 & 76 & 0.20 & - & -\\
  & $\geq3$ & 1.6 & $1\quad 3$ & 84 & 87 & 89 & 85 & 41 & 80 & 86 & 46 & 84 & 78 & 50 & 76 & 0.19 & - & -\\
$\Delta=3$   
  & $\geq0$ & 1.2 & $1\quad 3$ & 81 & 82 & 81 & 74 & 37 & 67 & 75 & 40 & 70 & 58 & 44 & 54 & 0.12 &- & -\\
\hline
 \end{tabular}
 \label{tab:final}
\end{table*}

\begin{table*}
\caption{Number of SNe in each SNPCC subset. The table also shows sub-samples of the $D_i$ according to SNR cuts.}
\centering
\begin{tabular}{| c | c *{9}{ c|}}
\multicolumn{1}{c}{ }& \multicolumn{9}{c}{Training sample} \\
\cline{2-10}
\multicolumn{1}{c}{ }&\multicolumn{3}{|c|}{$D_1$} & \multicolumn{3}{c|}{$D_3$} &  \multicolumn{3}{c|}{$D_7$} \\
 \cline{2-10}
\multicolumn{1}{c}{ } &\multicolumn{1}{|c|}{\tiny{SNR$\geq$0}} & \tiny{SNR$\geq$3} & \tiny{SNR$\geq$5}&\tiny{SNR$\geq$0} & \tiny{SNR$\geq$3} & \tiny{SNR$\geq$5}& \tiny{SNR$\geq$0} & \tiny{SNR$\geq$3} & \tiny{SNR$\geq$5} \\
\cline{1-10}
\multicolumn{1}{|c}{Ia} & \multicolumn{1}{|c|}{546} & 311 & 216 & 601 & 330 & 601 & 455 & 272 &  188 \\
\multicolumn{1}{|c}{non-Ia} & \multicolumn{1}{|c|}{278} & 254 & 225 & 312 & 283 & 312 & 242 & 221& 199 \\
\multicolumn{1}{|c}{total} & \multicolumn{1}{|c|}{824} & 565 & 441 & 913 & 613 &  913& 697 &493 & 388 \\
\cline{1-10}
\multicolumn{10}{c}{ }\\
\multicolumn{1}{c}{ } & \multicolumn{9}{c}{Test sample}\\
 \cline{2-10}
\multicolumn{1}{c}{ }& \multicolumn{3}
{|c|}{$D_1$} & \multicolumn{3}{c|}{$D_3$} & \multicolumn{3}{c|}{$D_7$}\\
\cline{2-10}
\multicolumn{1}{c}{ } &\multicolumn{1}{|c|}{\tiny{SNR$\geq$0}} & \tiny{SNR$\geq$3} & \tiny{SNR$\geq$5}&\tiny{SNR$\geq$0} & \tiny{SNR$\geq$3} & \tiny{SNR$\geq$5}& \tiny{SNR$\geq$0} & \tiny{SNR$\geq$3} & \tiny{SNR$\geq$5} \\
\cline{1-10}
\multicolumn{1}{|c}{Ia}&\multicolumn{1}{|c|}{2713} 		&	2201	&	972	& 2942	& 2366	& 1019	& 2303	&	1904 & 956	\\
\multicolumn{1}{|c}{non-Ia}&\multicolumn{1}{|c|}{9785}	&	6651	&	2262	& 10372	& 6969	& 2338	& 8411	&	6139 & 2185	\\
\multicolumn{1}{|c}{total}  &\multicolumn{1}{|c|}{12498}	&	8852	&	3234	& 13314	& 9335	& 3357	& 10714	&	8043& 3141	\\
\cline{1-10}
\end{tabular}\label{tab:number}
\end{table*}

\begin{table*}
 \caption{Summary of classifications results for SNPCC sub-samples. Results for efficiency before (eff$_{B}$) and after (eff$_{A}$) selection cuts, purity (pur) and successful classification (SC) are reported in percentages (\%).}
 \centering
 \begin{tabular}{| c| c | c | c || c c c || c c c c c c|}
 \cline{8-13}
 \multicolumn{7}{c|}{ } & \multicolumn{6}{c|}{Test sample}\\
  \cline{5-13}
  \multicolumn{4}{c||}{\multirow{2}{*}{ }} &  \multicolumn{3}{c||}{Training sample}& \multicolumn{6}{c|}{\multirow{2}{*}{include U}}\\
 \multicolumn{4}{c||}{}  & \multicolumn{3}{c||}{cross validated} & \multicolumn{6}{c|}{ }\\            
\hline
 data set & SNR & $\sigma$ & PCs & eff$_{A}$ & pur &  SC & eff$_{A}$ &  pur & SC & FoM$_{A}$ &  eff$_{B}$ & FoM$_{B}$\\
 \hline
\multicolumn{1}{|c|}{\multirow{3}{*}{$D_1$}}& 
	   $\geq5$ & 1.0 & $1\quad 5$ & 94 & 96 & 95 & 32 &   75 & 75 & 0.16 & 7 &0.03\\
     & $\geq3$ & 0.9 & $1\quad 5$ & 91 & 89 & 89 & 50 &  55 & 66 & 0.15 & 24 &0.07\\
     & $\geq0$ & 0.8 & $1\quad 4$ & 87 & 88 & 84 & 58 &  50 & 63 & 0.15 & 35 &0.09\\
     \hline
\multicolumn{1}{|c|}{\multirow{3}{*}{$D_2$}}&
		$\geq5$ & 0.6 & $1\quad 5$ & 94 & 96 & 95 & 32 &   77 & 75 & 0.17 & 7 &0.04\\  
     &  $\geq3$ & 0.6 & $1\quad 5$ & 91 & 91 & 90 & 47 &  53 & 69 & 0.13 & 23 &0.06\\
     &  $\geq0$ & 1.0 & $1\quad 2$ & 88 & 89 & 85 & 59 &  34 & 49 & 0.09 & 36 &0.05\\
      \hline
\multicolumn{1}{|c|}{\multirow{3}{*}{$D_3$}}&  
		$\geq5$ & 0.7 & $1\quad 5$ & 90 & 92 & 92 & 30 &   73 & 71 & 0.14 & 7 &0.03\\  
      & $\geq3$ & 0.6 & $1\quad 2$ & 87 & 89 & 87 & 35 &  56 & 41 & 0.10 & 18 &0.05\\   
      & $\geq0$ & 0.8 & $1\quad 3$ & 86 & 86 & 81 & 64 &  36 & 39 & 0.10 & 41 &0.07\\  
       \hline
\multicolumn{1}{|c|}{\multirow{3}{*}{$D_4$}}&
		$\geq5$ & 0.5 & $1\quad 5$ & 91 & 92 & 92 & 30 &   72 & 72 & 0.14 & 7 &0.03\\   
      & $\geq3$ & 0.4 & $1\quad 4$ & 87 & 88 & 87 & 42 &  33 & 56 & 0.06 & 22 &0.05\\
      & $\geq0$ & 1.7 & $1\quad 3$ & 87 & 89 & 86 & 32 &  32 & 48 & 0.09 & 40 &0.06\\
       \hline         
\multicolumn{1}{|c|}{\multirow{3}{*}{$D_7$}}& 
		$\geq5$ & 0.8 & $1\quad 4$ & 92 & 93 & 93 & 27 & 91 & 72 & 0.21 &  6 & 0.04\\  
      & $\geq3$ & 0.8 & $1\quad 2$ & 91 & 91 & 90 & 34 & 68 & 74 & 0.14 & 15 & 0.06\\
      & $\geq0$ & 1.0 & $1\quad 5$ & 93 & 90 & 89 & 55 & 48 & 64 & 0.13 & 28 & 0.07\\         
\hline       
\multicolumn{1}{|c|}{\multirow{3}{*}{$D_8$}}& 
		$\geq5$ & 1.2 & $1\quad 4$ & 93 & 93 & 93 & 54 & 74 & 64 & 0.26 & 11 & 0.06\\  
      & $\geq3$ & 1.1 & $1\quad 4$ & 91 & 93 & 91 & 59 & 51 & 66 & 0.15 & 26 & 0.07\\
      & $\geq0$ & 0.8 & $1\quad 5$ & 89 & 88 & 86 & 63 & 43 & 42 & 0.12 & 32 & 0.06\\         
\hline         
 \end{tabular}
 \label{tab:final_SNCC}
\end{table*}

\bibliographystyle{mn2e}
\bibliography{refkernel}

\label{lastpage}
\end{document}